\documentclass[pdflatex,sn-nature]{sn-jnl}

\usepackage{graphicx}%
\usepackage{multirow}%
\usepackage{amsmath,amssymb,amsfonts}%
\usepackage{amsthm}%
\usepackage{mathrsfs}%
\usepackage[title]{appendix}%
\usepackage{xcolor}%
\usepackage{textcomp}%
\usepackage{manyfoot}%
\usepackage{booktabs}%
\usepackage{algorithm}%
\usepackage{algorithmicx}%
\usepackage{algpseudocode}%
\usepackage{listings}%
\usepackage{ulem}

\newcommand{\araa}{Annu. Rev. Astron. Astrophys.}   
\newcommand{\aj}{Astron. J.}   
\newcommand{\apj}{Astrophys. J.}   
\newcommand{\apjl}{Astrophys. J. Lett.}   
\newcommand{\apjs}{Astrophys. J. Suppl. Ser.}   
\newcommand{\aap}{Astron. Astrophys.}   
\newcommand{\mnras}{Mon. Not. R. Astron. Soc.}   
\newcommand{\nat}{Nature} 

\newcommand{\xmm}{\textit{XMM-Newton}}

\newcommand{\ztba}{{\sc zTBabs}}

\newcommand{\cms}{{cm$^{-2}$}}
\newcommand{\nh}{{$N_{H}$}}
\newcommand{\nhc}{{$N_{H,X,neutral}$}}
\newcommand{\nhx}{{$N_{H,X,TEPID}$}}

\newcommand{\grb}{{lGRB}}
\newcommand{\logn}{{log($n_0$)}}

\newcommand{\tepi}{{TEPID-only}}
\newcommand{\tepiO}{{TEPID+optical}}
\newcommand{\nho}{{$N_{H,optical}$}}
\DeclareRobustCommand{\ion}[2]{\textup{#1\,\textsc{\lowercase{#2}}}}

\theoremstyle{thmstyleone}%
\theoremstyle{thmstyletwo}%

\theoremstyle{thmstylethree}%

\unnumbered

\begin{document}

\title[Dense gas linked to star-forming regions photoionised by embedded gamma-ray bursts]{Dense gas linked to star-forming regions photoionised by embedded gamma-ray bursts}


\author*[1]{\fnm{Aishwarya Linesh} \sur{Thakur}}\email{aishwarya.thakur@inaf.it}

\author[1]{\fnm{Luigi} \sur{Piro}}\email{luigi.piro@inaf.it}

\author[1,2]{\fnm{Alfredo} \sur{Luminari}}

\author[2]{\fnm{Fabrizio} \sur{Nicastro}}

\author[3,4,5]{\fnm{Sandra} \sur{Savaglio}}

\author[6]{\fnm{Yair} \sur{Krongold}}

\author[7,8]{\fnm{Bruce} \sur{Gendre}}

\affil*[1]{\orgdiv{Istituto di Astrofisica e Planetologia Spaziali}, \orgname{INAF}, \orgaddress{\street{Via Fosso del Cavaliere, 100}, \city{Rome}, \postcode{I-00133}, \country{Italy}}}

\affil[2]{\orgdiv{Osservatorio Astronomico di Roma}, \orgname{INAF}, \orgaddress{\street{Via Frascati, 33}, \city{Monte Porzio Catone}, \postcode{I-00040}, \country{Italy}}}

\affil[3]{\orgdiv{Physics Department}, \orgname{University of Calabria}, \orgaddress{\street{Via P. Bucci}, \city{Rende}, \postcode{I-87036}, \country{Italy}}}

\affil[4]{{\orgdiv{Osservatorio di Astrofisica e Scienza dello Spazio}, \orgname{INAF}, \orgaddress{\street{Via Piero Gobetti 93/3}, \city{Bologna}, \postcode{I-40129}, \country{Italy}}}}

\affil[5]{\orgdiv{Laboratori Nazionali di Frascati}, \orgname{INFN}, \orgaddress{\street{Via Enrico Fermi, 54}, \city{Frascati}, \postcode{I-00044}, \country{Italy}}}

\affil[6]{\orgdiv{Instituto de Astronom\'ia}, \orgname{Universidad Nacional Aut\'onoma de M\'exico}, \orgaddress{\street{Circuito Exterior}, \city{Ciudad Universitaria}, \postcode{04510}, \country{Mexico}}}

\affil[7]{\orgdiv{Department of Physics}, \orgname{University of Western Australia}, \orgaddress{\street{35 Stirling Highway} \city{Crawley, WA}, \postcode{6009}, \country{Australia}}}

\affil[8]{\orgdiv{College of Science and Mathematics}, \orgname{University of the Virgin Islands}, \orgaddress{\street{2 John Brewer's Bay} \city{St. Thomas, VI}, \postcode{00802}, \country{USA}}}


\abstract{
 The 1-100 pc region embedding long-duration gamma-ray bursts (lGRBs) has been hitherto unexplored, as extremely high ionisation by the GRB prevents application of optical absorption spectroscopy on such distances. We show that the GRB ionising flux imprints a unique time- and spatially-dependent ionisation structure on the gas, which can be probed by X-ray absorption. Application of this model to a selected sample of 7 bright GRB X-ray afterglow spectra observed by \textit{XMM-Newton} EPIC-pn enables an independent, quantitative estimation of the density (log(n) $\sim$ 2-4) and distances (5-100 pc) of the ionised absorber directly from the GRB X-ray spectrum, thereby allowing us to locate the absorbing medium of this representative sample of long GRBs in the region of the density-size diagram populated by star-forming regions versus other gravitationally bound objects in the Universe. Our results provide one of the most direct links between lGRBs and star formation and open the potential of high-resolution X-ray spectroscopy as a powerful probe of star-forming regions that embed GRBs up to the highest redshifts.}

\maketitle

Gamma-ray bursts (GRBs) are brief bursts of emission detected in $\gamma$ wavelengths. They are the most energetic explosions in the Universe with bright emission composed of two well-defined consecutive segments: the eponymous burst of gamma-ray photons called the prompt emission, and the second visible on a longer timescale called afterglow emission. GRBs have a dichotomous classification in terms of the duration of the prompt emission, those lasting longer than two seconds have been largely associated with massive stars that explode through core-collapse \citep[collapsar,][]{1993ApJ...405..273W, 2006ApJ...637..914W}, whereas short-duration GRBs are associated to compact object mergers (sGRBs, hereinafter)  \citep{2017ApJ...848L..12A, 2017ApJ...848L..15S, 2017ApJ...848L..14G} with a few, notable, exceptions of long duration GRBs that are powered by mergers \citep{2022Natur.612..228T, 2024Natur.626..737L}. Throughout the rest of this work, we use \grb\ to mean canonical long-duration gamma-ray bursts powered by collapsars. 

Spectroscopy in the optical band has enabled in-depth study of the interstellar medium (ISM) within the \grb\ host galaxy in absorption \citep{2006ApJ...648...95P} and the properties of the host itself in emission \cite{2015A&A...581A.125K}. Absorption in the observed optical-UV spectra arises from low- to moderate-ionisation ultraviolet metals (rest-frame E $\simeq$ 3-10 eV, eg. Mg I-II, CIV and NV) in the host ISM. Analysis of time-varying, excited fine-structure lines of these metals places this material at sub-kpc to kpc scales from the \grb\ site \cite[eg.][]{2007Vreeswijk, refId0, 2023A&A...671A..84S}. 

\begin{figure}[h]
    \centering
    \includegraphics[width=\linewidth]{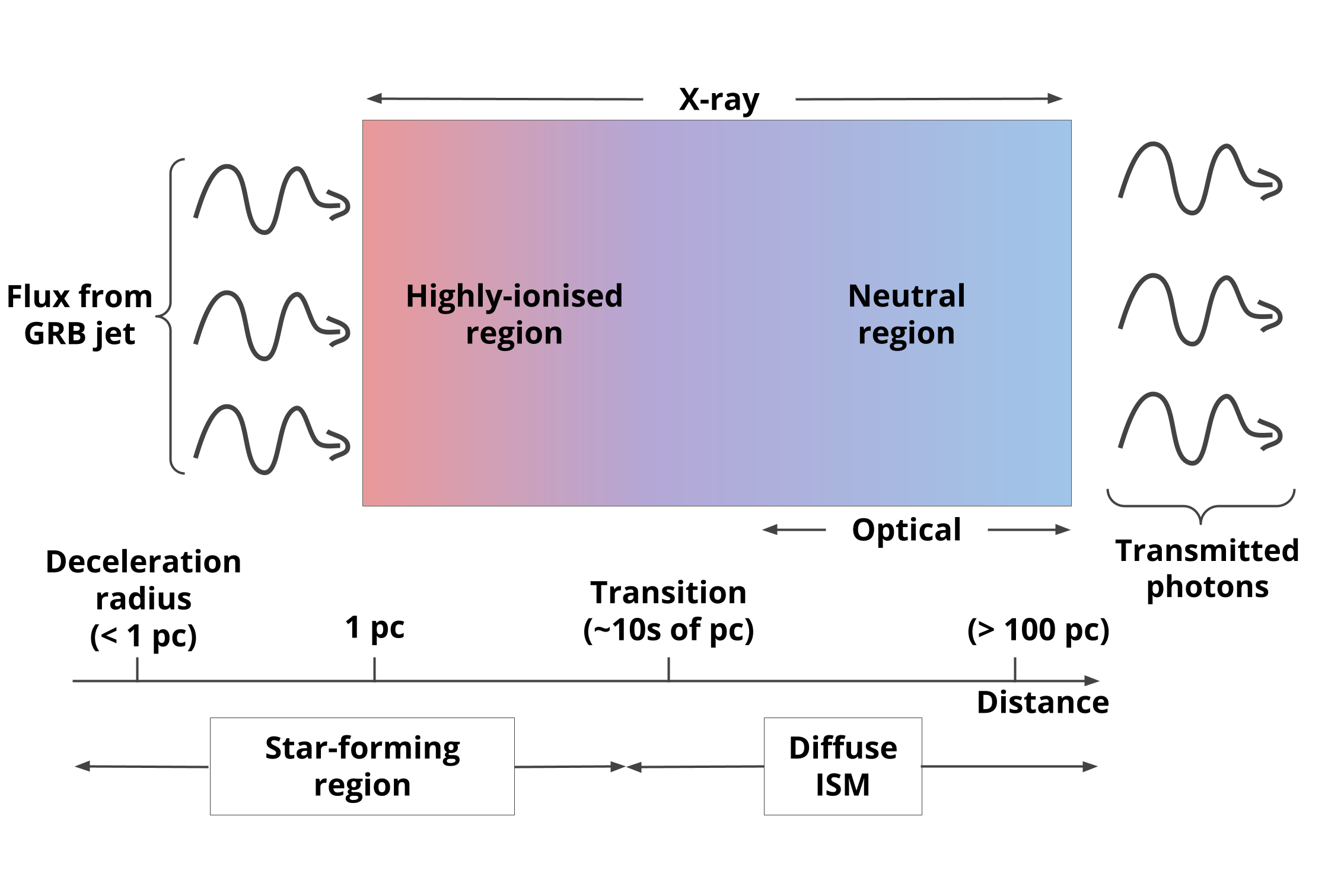}
    \caption{\textbf{Schematic representation of the stratification of the absorbing medium along the line of sight to the GRB} The flux from the GRB jet directly impinges the high-ionisation region (shaded red). This region consists of the star-forming region (denser) in which the GRB is embedded and extends up to several tens of pc from the location of the GRB. The relatively neutral ISM (more tenuous) that is further away ($\gtrsim 10$ pc) is shaded blue. As the text discusses, X-ray absorption is seen from both media, whereas optical absorption is seen only from the low ionisation ISM. The transition region connecting the high and low ionisation media is shaded purple. The black lines on the right show the photons emerging from the host galaxy.}
    \label{fig:schematic}
\end{figure}

The luminous \grb\ emission is expected to have a strong ionising effect on the gas situated within 100 parsec of the \grb\ site. The gas is ionised to a degree that renders it transparent to optical-UV photons, which has so far resulted in this region being largely inaccessible to current techniques. However, this medium does have opacity in the soft X-rays and contributes the observed excess (over the Galactic column) X-ray column density in GRB spectra \citep{2004ApJ...608..846S, 2010MNRAS.402.2429C}. Past studies have inferred that the X-ray absorber is internal to the host galaxy \citep[based on observed correlation of optical extinction and the gas column,][]{2013ApJ...768...23W, 2013MNRAS.432.1231C} and that the X-ray absorbing columns have a distribution consistent with the observed properties of Giant Molecular Clouds (GMCs) \citep{2001ApJ...549L.209G, 2002ApJ...565..174R}, strongly suggesting that the X-ray absorption arises in the photoionised medium of the star-forming region hosting the GRB progenitor. Indeed, systematic comparison of the optical and X-ray column densities have found an indication for highly photoionised material within a few parsec of the lGRB site \citep[see Fig \ref{fig:schematic},][]{2007ApJ...660L.101W, 2011A&A...525A.113S, 2013ApJ...774..115K}. Additional studies investigating the contributions of various elements to the X-ray absorption have inferred that He is likely dominating the soft X-ray opacity \citep{2013ApJ...768...23W}, due to heavy stripping of electrons from metals up to distances $\sim 10$ pc. Furthermore, they find that for such sizes, the observed column densities imply star-forming region-like number densities in the absorbing medium ($\sim 10^{3-4}$ cm$^{-3}$). A similarly qualitative conclusion, by assuming an ad-hox density profile, was obtained by \citep{2013ApJ...774..115K}.

However, contrary to those expectations, photoionisation-equilibrium modelling \cite{1992ApJ...395..275D} of the X-ray afterglow spectra of \grb s have generally resulted in poor fits, either returning results consistent with a neutral medium (see Methods) or leaving the ionisation parameter totally unconstrained \citep[e.g.][]{2002ApJ...577..680P}. This contrast is further compounded by the fact that moderate signal-to-noise soft X-ray spectra of GRBs are reasonably fitted by neutral absorption \cite{2000ApJ...542..914W} models.

This logical conflict arises because photoionised absorber models generally assume photoionisation equilibrium in the medium. This, however, cannot be the case for gas photoionised by a \grb\ because i) the ionising flux decreases by several orders of magnitude over the combined prompt and afterglow durations and ii) recombination timescales at densities consistent with a star forming environment (as we find, i.e. $log(n) [cm^{-3}]$ between 1 and 5) are longer than the total duration of the prompt and afterglow phases. The combination of non-equilibrium photoionisation affecting a region extending over several tens of pc and geometric dilution of the flux over such a region results in a highly stratified medium (See Fig \ref{fig:schematic}). The structure inside this medium can be precisely described only by solving proper time-evolving photoionisation and radiative transfer equations. Past attempts to implement time-evolving ionisation codes have focused on the early time-variable properties of the absorption \cite[e.g.][]{2002ApJ...580..261P} or to a qualitative comparison of the optical/X-ray absorbing columns \cite{2013ApJ...774..115K}. 

Crucially there have been no attempts as of yet to fit the late-time X-ray spectrum with a detailed time-evolving photoionisation model. At such late times, the bulk of the ionisation has already taken place in the medium and the gas loses practically no photoionising energy given the long recombination timescale. The ionisation structure thus depends basically only on the total ionising photon budget and on the number density as a function of the distance from the GRB, i.e. $n(r)$. This allows to break the $n \cdot r^2$ degeneracy intrinsic into the definition of the (equilibrium) ionisation parameter $\xi$ ($= L/nr2$, \cite{1992ApJ...395..275D}). This can be understood because for a same $N_H= n\times r$, larger densities require smaller size, $r\approx n^{-1}$, in turn implying that the region is subjected to a stronger ionisation, because the flux at a given distance varies with $r^{-2}$. In this study, we make use of a state-of-the-art model \cite[TEPID, hereinafter,][]{L22} that self-consistently solves the ionic balance and the radiative transfer as a function of both time and distance from the ionising source. We apply TEPID to a sample of \grb s observed with \xmm. For each burst we use the observed ionising source luminosity and light curve to produce spectral table-models to fit the EPIC-PN X-ray spectra.

\section*{Results}
\subsection*{Photoionisation in a sample of bright X-ray afterglow spectra}

\begin{figure}
    \centering
    \includegraphics[width=0.8\textwidth]{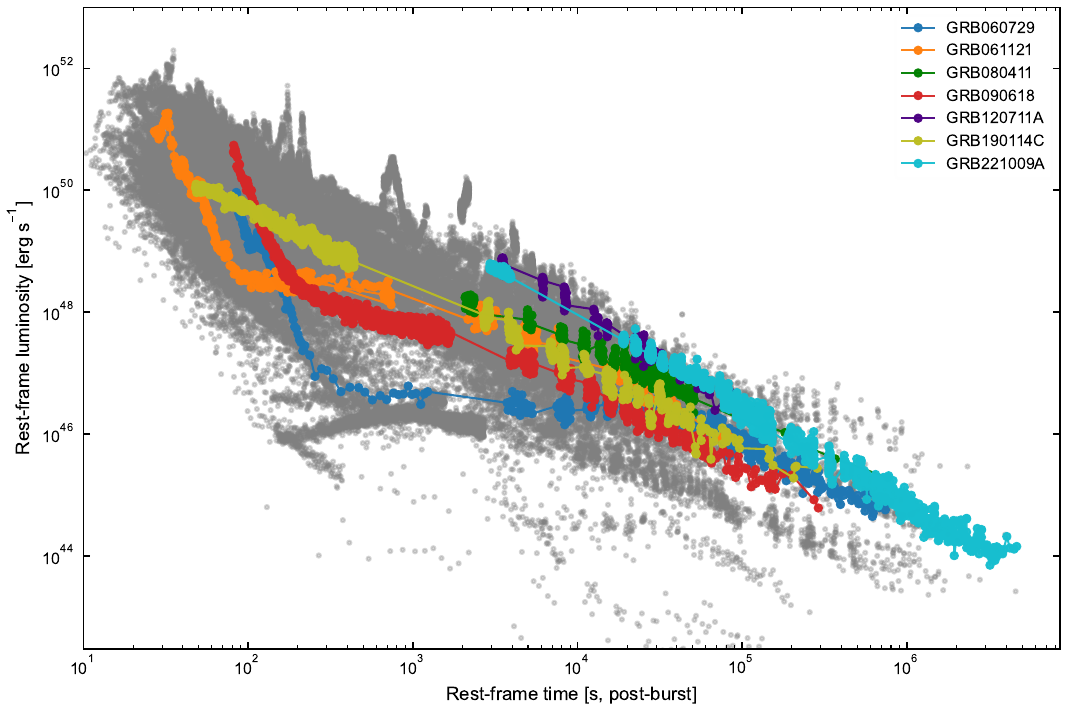}
    \caption{\textbf{Luminosity distribution of the \textit{Swift-XRT} sample of lGRBs with known redshift}: The luminosity light curves for the full sample are plotted in the rest-frame band of 0.3 - 10 keV as grey points. Superimposed on this distribution are the light curves of our sample GRBs, colour-coded per GRB. Our sample is distributed towards the centre of the scatter. This clearly shows no bias toward high luminosity GRBs is introduced by our selection criteria.}
    \label{fig:lumin_select}
\end{figure}

We adopted the following criteria to select a sample of bright \grb s observed by \xmm: 1) availability of spectroscopic redshift and 2) EPIC-PN spectra with $\ge 10^5$ counts over the 0.3–10 keV band. This yields seven GRBs with afterglow observations starting from 5 hrs to 4 days after the GRB, which are listed in row 1 and 2 of Tab. \ref{tab:master} (A full observational log is included as Supplementary Table 1). As shown in Fig \ref{fig:lumin_select}, our criteria do not introduce biases toward luminous \grb s (energetics and obscuration biases checks are shown in Extended Data, Fig.~\ref{fig:bias_check}), making our sample, while small, representative of the broader \grb\ population. 

EPIC-PN data have been reduced with standard procedures, producing spectra that have been fitted with {\sc Xspec} \cite{1996ASPC..101...17A}. We fit the data using $\chi^2$ statistics and we estimate the errors on the TEPID best fit parameters through an MCMC chain. Finally, we estimate the significance of the statistical improvement brought by TEPID over the alternative model using the Bayes Factor (BF) and the Akaike Information Criterion (AIC). We model the time-evolving photoionisation in the medium with TEPID by implementing for each burst the observed prompt-to-afterglow light curve (with the exception of 221009A where we use a parametric light curve, see Methods).

We model the X-ray spectrum as a power law with an intrinsic absorption at the redshift of the burst, in addition to the absorption due to our Galaxy. While we have used a neutral model for Galactic absorption and there is a warm component in the Galactic ISM, this has no effect on the results of our analysis (discussed further in Methods). For each \grb, we adopt three absorption models in increasing order of both complexity and physical motivation: i) a neutral absorber, ii) a TEPID absorber (TEPID-only) and iii) as the previous but with the addition of a neutral screen, whose column has been derived for each individual burst from metal absorption lines in the optical spectra (TEPID+optical, except for GRB 221009A). For the fit with a neutral absorber we fixed the metallicity to the solar value. In fact, it is customary to recast the X-ray column as its hydrogen-equivalent, derived assuming solar metallicity. For consistency, we first fixed the metallicity to solar also for the TEPID fits. 

We find that for six GRBs of our sample, neutral absorber fits leave systematic residuals across the energy range of the spectral fit (see Extended Data Fig 2 for the case of GRB 060729, and further details in Methods). For all of these, we find that our non-equilibrium photoionised absorber model fits improve upon the standard neutral absorber fit ($7 \lesssim \Delta\chi^{2} \lesssim 130$), with no systematic residuals compared to the neutral fits. For 4 GRBs of our sample: GRB 060729, GRB 080411, GRB 090618 and GRB 120711A, the improvement in the fit ($40 < \Delta\chi^{2} < 130$) is conclusively significant for both TEPID-only and TEPID+optical fits (BF $>$ 100 and $\Delta$AIC $>$ 10). The improvement ($\Delta\chi^{2}=10$) is significant, but not conclusive ($BF > 100$, but $\Delta \mathrm{AIC} \sim 9$) for the TEPID-only fit of GRB 221009A. Finally, the improvement ($\Delta\chi^{2} = 7$) is moderately significant for the TEPID+optical fit of GRB 061121 ($BF > 100$, but $\Delta \mathrm{AIC} \sim 7$, see Methods for discussion of statistical significance thresholds.) For the GRBs with metal absorption lines in their optical spectra, we find that because the X-ray has opacity to higher ionisation material, the column is systematically larger than the optically derived one, further pointing to the medium being ionised by the GRB. Within the TEPID formulation, we can also fit for the size ($r$) along with the column density (N$_{H,X,TEPID}$) and density ($n_{0}$) of the absorbing medium. The best-fit parameters and associated 90\% confidence intervals for the neutral model and the \tepi\ model with solar metallicity are presented in Tab \ref{tab:master}. 

The capability of the TEPID absorber model to improve the fit lies in the unique ionic composition and stratification of the medium as a function of the distance from the GRB. This leads to a unique spectral profile, due to a unique combination of photoelectric opacities arising from several ionisation stages of multiple elements within the photoionised medium that are absent in a neutral model. (see Extended Data Fig. 3). At rest-frame energies below 1 keV, the absorption is largely dominated by He and H, while at energies above 1 keV, metal contributions start to become dominant. Across our sample, we find that on average helium contributes more than half of the observed continuum opacity, with contributions also from H and metals between a few percent to tens of percent. 

Taking into account that He \citep{2013ApJ...768...23W} and H play a role in the absorption of the X-ray spectrum, we then verified the metallicity dependence of our results. GRB hosts are known to have sub-solar values, although with a large scatter that encompasses solar-like abundances \cite{Levesque_2010, 2010AJ....139..694L, 2015A&A...581A.125K}. Therefore, we set the host metallicity as a lower limit. On the other hand, lGRBs are known to be associated with star-formation in their hosts, and star-forming regions are subjected to localised metal enrichment prior to the explosion of the GRB progenitor compared to the ISM of the host galaxy, with the enrichment higher ($\sim 0.2$ dex above metallicity of the host ISM, see \cite{2021A&A...650A.103D}) in denser and more actively star-forming regions. \cite{2021A&A...650A.103D, 2019ApJ...887...80K}. Conservatively, we therefore set solar metallicity as an upper limit and add the metallicity as a fitting parameter free to vary in this range. The best-fit parameters and associated 90\% confidence intervals for \tepi\ with metallicity free are also presented in Tab \ref{tab:master}. (for clarity of presentation, the best-fit parameters for all \tepiO\ fits are included in Supplementary Table 3. The main conclusions on the size and density of the absorber are in any case unchanged)

The TEPID best-fits with metallicity free are consistent with those in which the metallicity is fixed to solar, as is their statistical significance (see also Methods) with the metallicity being essentially unconstrained. However, there is an increase in the 90\% confidence region of $r$ for the TEPID fits with metallicity free. This happens because the X-ray absorption results from an ionisation pattern that has a set fractional contribution of H, He and metals. This means that when the metallicity is lowered, the fit increases the size of the region (and thus decreasing both N$_H$ and log($n_0$)) so as to accumulate the same amount of metals. As a result, we find that even after including the effects of metallicity in our analysis, our conclusions on the size and density of the absorbing medium are unaffected. 

An important point is that our inferred number densities are two orders of magnitude higher than those inferred from multiwavelength modelling of the GRB jet \citep{2022MNRAS.511.2848A}. This is because the densities probe two different regions of the circumburst medium. The afterglow-inferred density probes the region extending up to a few parsecs, which has been cleared out by strong mass loss from the Wolf-Rayet progenitor, creating a wind-blown bubble in which the jet expands. On the other hand, TEPID probes the absorption from the photoionised medium outside this low number density bubble and, thus, corresponds to the outer medium, within the star-forming region. 

Finally, absorption by the Inter-Galactic Medium/Warm Hot Intergalactic Medium (IGM/WHIM) has been invoked as a possible explanation for the X-ray absorption of the GRB afterglow \citep{2023ApJ...953..158G}. To check the contribution of IGM/WHIM absorption in our sample, we plot the opacities of our sample against the IGM opacities from \citep{2023ApJ...953..158G} and \citep{2006ApJ...650..560C} in Extended Data Fig \ref{fig:IGM_opacity}. As can be clearly seen, the expected IGM/WHIM opacity is nearly an order of magnitude lower in the redshift range of our sample, thereby showing that the contribution to the observed X-ray spectrum is negligible. We can therefore rule out IGM absorption in the case of our sample (More details on both points in Methods.)

\section*{Discussion}
\subsection*{Photoionised gas in the GRB vicinity and the link to star-forming regions}

The results of our analysis represent the first systematic, rigorous fitting of the late-time X-ray afterglow spectrum with an accurate time-evolving photoionisation code specifically tuned to model the GRB's effects on the environment. From the TEPID best-fit we derive the size and density of the absorbing medium and find that these regions have physical sizes $r=5-50$ pc and number densities $n=10^{2.5-4}$ cm$^{-3}$. Such sizes and densities for the absorbing medium have been inferred in the past from the observed distributions of the X-ray column \citep{2001ApJ...549L.209G, 2002ApJ...565..174R}: by comparing the optical and X-ray columns \citep{2013ApJ...774..115K} or extinction and X-ray columns \citep{2013MNRAS.432.1231C}, and, by theoretical considerations on the stripping of electrons \citep{2013ApJ...768...23W}, but which we now estimate directly from the X-ray spectrum of the GRB. Additionally, from this direct fitting, we also find that the GRB X-ray absorption is found to come from a combination of He \citep{2013ApJ...768...23W} and highly-ionised metals within the GRB vicinity \citep{2011A&A...525A.113S}. Thus, by directly fitting the GRB X-ray spectrum with a physically accurate model that correctly takes into account the time-dependent photoionisation in the medium due to the \grb\ allows us to show that the X-ray absorber in GRBs is a combination of all these: the \grb\ X-ray spectrum is primarily absorbed by photoionised gas situated within the star-forming region that embeds the GRB.

\begin{figure}
    \centering
    \includegraphics[width=0.9\textwidth]{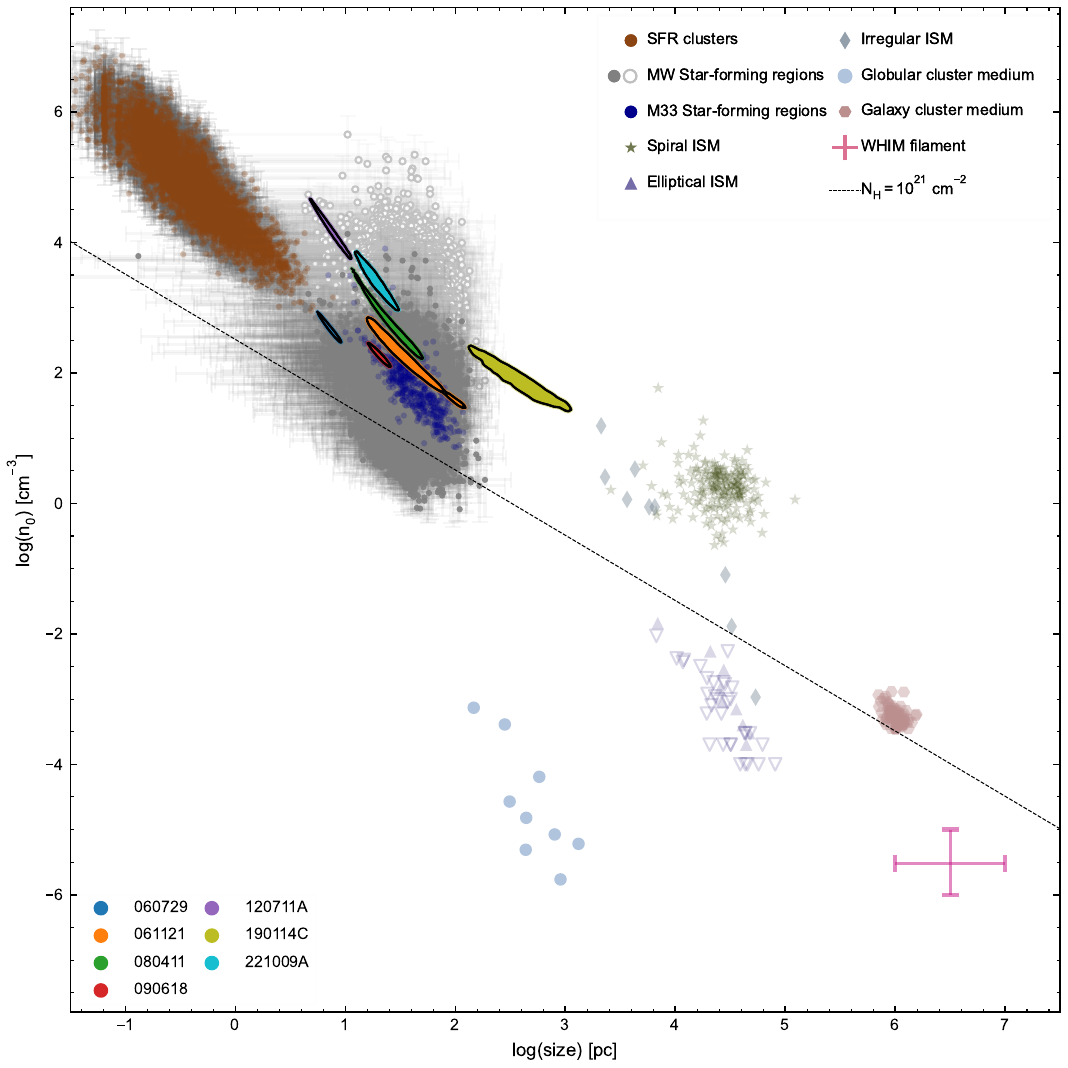}
    \caption{\textbf{Number density versus size distribution of gravitationally bound objects in the Universe}: The coloured, filled contours are the 90\% confidence regions of the best fits we obtain by fitting the \xmm\ spectrum with the TEPID-only photoionised model, leaving the metallicity free to vary in the range per GRB. In order of increasing physical size, we plot clumps within star-forming regions (brown circles, \cite{2019MNRAS.484.4444U}), Galactic (grey filled circles) and extragalactic (dark blue circles) star forming regions (\cite{2017ApJ...834...57M} and \cite{2017A&A...601A.146C}, respectively), globular clusters (lightblue circles,\cite{2024A&A...686A.283P}), galaxies (spiral as green filled stars, irregular as blue filled diamonds and ellipticals as purple triangles (downward open triangles mark upper limits) \cite{2007A&A...465...71T, 2008AJ....136.2563W, 1980ApJ...242..931S}), galaxy clusters (magenta hexagons, \cite{2018MNRAS.478.3072C}) and a WHIM filament (red cross, \cite{2008Sci...319...55N}). The black dashed line plots a column density of $10^{21}$ cm$^{-2}$. The open, light grey circles show the SFR number density assuming a volume filling fraction of 3\% for dense gas. Where plotted, the errors on the size of the objects are taken as reported in the respective reference, while the errors on density are either taken as reported or propagated from the mass and size when required.} 
    \label{fig:master}
\end{figure}

We plot the TEPID-only best-fit contours in the density versus size plane in Fig. \ref{fig:master}. The best-fit contours for the six GRBs of our sample are tightly clustered around the parameter space occupied by star-forming regions (SFR), both in our own Galaxy (grey circles)\cite{2017ApJ...834...57M}, and in external galaxies (dark blue circles)\cite{2017A&A...601A.146C}. To further emphasise this remarkable association we include in Fig. \ref{fig:master} other gravitationally-bound objects in the Universe for increasing size: clumps within SFRs \cite{2019MNRAS.484.4444U}, globular clusters \cite{2009MNRAS.396.1096V, 2015MNRAS.446.2226M}, galaxies\cite{2007A&A...465...71T, 2008AJ....136.2563W, 1980ApJ...242..931S}, galaxy clusters \cite{2018MNRAS.478.3072C} and a filament of the WHIM. Considering the differences in the individual GRB ionisation, in the \xmm\ observing times and in the redshift distribution of our GRBs, the observed clustering of the TEPID best-fit regions (now estimated directly from the X-ray spectrum) within the distance vs number density plane in Fig. \ref{fig:master} is indeed remarkable, particularly considering the observed consistency with SFR environments.

An obvious point of intrigue is the TEPID best-fit of 190114C, which seems to be differing from the other six. This GRB has occurred in a merging galaxy at 170 pc from the core \cite{2020A&A...633A..68D} and the host has a higher molecular gas fraction than isolated field galaxies, suggesting a cold and dusty ISM. Additionally, the EPIC-PN spectrum has very low statistics below $E_{rest} = 2$ keV (see Extended Data Fig. 4). This likely suggests that the X-ray absorption for this GRB is dominated by this cold ISM. Furthermore, we cannot reliably disentangle the contribution of the ISM to the observed X-ray absorption using the neutral ISM column density from the optical spectrum. (See Methods for a discussion of the optical spectrum column of this burst). To test if we can still constrain the presence of any photoionised region, we simulated a spectrum assuming a neutral+TEPID model, with the neutral column set to the best-fit X-ray neutral model, finding that when refitting this spectrum we cannot distinguish between the TEPID and neutral models. As a result of all these factors, we conclude that this lGRB is likely an outlier for our sample. 

The distribution of the \tepi\ best-fit contours seem to be slightly skewed to higher densities compared to the average densities of SFRs in Fig. \ref{fig:master}. We have derived the number density assuming a uniform distribution of the mass within SFRs. However, SFRs are known to have complex sub-structure of dense gas. Star-formation (especially that of young, massive stars) inside these environments is associated with the densest regions that occupy only a small fraction of the total volume, $\sim 3$\% from observations \cite{1985A&A...152..371P}. Using this reduced volume fraction in the density estimation would give us a higher number density than assuming a uniform distribution of mass within the SFR. We have shown this in Fig. \ref{fig:master} with the open light grey points. This shows that the connection with SFRs is enhanced after considering effects of clumping and suggests that the lGRB in our sample may be preferentially harboured in the denser parts of SFRs, where most massive stars are located.

The good agreement between the best fit size and density of the absorbing medium with SF regions makes a strong case for the progenitors being massive stars, i.e. collapsars. Collapsars have long been understood to be the dominant progenitor channel for long duration GRBs, however, this picture has been challenged in recent times by the discovery of merger-driven long GRBs. To check the robustness of our results we then investigated contamination from merger-driven long GRBs \cite{2024Natur.626..737L}. We have considered the key properties related to collapsars and present them in Tab. \ref{tab:collapsar}. We consider both intrinsic properties of the emission (spectral lag) and extrinsic properties of the \grb\ environment (star-formation in host galaxy, association with supernova and physical offset). All bursts in our sample exhibit spectral lag in their prompt light curves, as compared to no spectral lag in the light curves of merger-driven lGRB 211211A and lGRB 230307A \citep{Peng_2024}. A supernova has been observed for four out of the seven, and four have a host galaxy that exhibits active star-formation. Lastly, the physical offset for six out of seven GRBs are consistent with the observed distribution of offsets of collapsar GRBs \cite{2016ApJ...817..144B} (no host detection to date for GRB 080411). The offsets of the merger-driven long GRBs, instead, are more similar to the observed offset distribution of canonical sGRBs. This leads us to conclude that there is no contamination from merger-driven long GRBs in our sample and lends further support to collapsars as the predominant channel of the long GRB population. Our results open the perspective to extend, with upcoming sensitive and high spectral-resolution X-ray facilities like \textit{NewAthena}, the study of star forming regions embedding GRBs throughout the cosmic ages and up to the high-redshift era \cite{2022ExA....54...23P}. 

\setcounter{figure}{0}
\renewcommand{\figurename}{Extended Data Figure} 

\section{Methods}

\subsection{Sample bias checks}

Given that we have imposed a flux selection criteria for our sample, we may have a significant bias toward higher luminosity or less obscured GRBs. We performed a check for luminosity biases by plotting our GRBs in luminosity space in Fig. \ref{fig:lumin_select} (A plot of the same sample in flux space is shown in the Supplementary Data,  Fig. 1). All the GRBs in our sample are concentrated at the centre of the spread, thereby showing that there is no strong bias in terms of luminosity. We further check for bias toward higher energy output GRBs by plotting the cumulative distribution of the isotropic energy release (E$_{iso}$) for a combined sample of \grb s \cite{2012MNRAS.421.1256N, 2016ApJ...831...28T, 2017ApJ...837..119A} in the left panel of Extended Data Fig. 1. The GRBs of our sample are distributed between the median and the third quartile, suggesting a marginal bias for higher energy outputs. Finally, we check for a low obscuration bias by plotting the cumulative distribution of column densities from \cite{2010MNRAS.402.2429C} in the right panel. The column densities of the \grb\ in our sample are those derived from the neutral absorber fits to the \xmm spectra. The \grb s in our sample are distributed between the first quartile and the median, suggesting a marginal bias towards lower column densities. Both of these marginal biases can be accounted for by our flux-selection criteria, and we thus conclude that our sample is not overly biased and, while small, is representative of the \grb\ population. 

\subsection{Modelling the lGRB circumburst medium with TEPID}\label{modelling}

\subsubsection{Photoionising output of the GRB}

We base our modelling on the GRB setting in \cite{L22} and we update TEPID in order to allow to take in as an input the GRB-specific light curve, to better track the evolution of the ionising luminosity. When available, we use the combined BAT-XRT 0.3-10 keV unabsorbed lightcurve as derived by \cite{BusrtAn10} (retrieved from: \url{https://www.swift.ac.uk/burst\_analyser/}). For GRB 120711A, which was discovered by \textit{INTEGRAL} and thus has only \textit{XRT} data from \textit{Swift}, we use the time-resolved \textit{Fermi} fits from \cite{2016A&A...588A.135Y} and extrapolate each time-resolved fit to the 0.3-10 keV range and then use this extrapolated lightcurve with available \textit{XRT} data to create a combined light curve. For the extrapolated light curve, we apply an absorption correction by assuming the column density of the \textit{XRT} best-fit. For a detailed discussion on the differences between the ionising light curve settings used in this work with that in \cite{L22}, we refer the reader to Supplementary Section 1.2. 

\subsubsection{Physical settings of the circumburst medium}

We assume a constant density profile medium that is pre-ionised by a Wolf-Rayet progenitor at $t = 0$ s. As discussed in \cite{L22}, even though the circumburst medium is ionised by the Wolf-Rayet progenitor, this pre-ionisation is primarily due to UV photons and is thus irrelevant to the observed absorption at X-ray energies. Furthermore, the GRB ionising output is several orders of magnitude more powerful and can thus completely wash out the pre-ionisation from the progenitor. 

Following on from \cite{L22}, we simulate the GRB circumburst medium starting at a given distance ($r_{inner}$) from the GRB and spanning a range of ambient number density ($n_0$). The number density and the distance from the GRB are connected to the column density (\nh), as:

\begin{equation*}
    N_H = n_0 \times (r_{out} - r_{inner})
\end{equation*}

Where $r_{inner}$ and $r_{outer}$ are the inner and outer radii of the GRB circumburst medium. As a result, while we fit the X-ray spectrum using N$_H$ and $log(n_0)$ as free parameters, we can derive the distance of the medium from the GRB for a given value of \logn. We start our model computations at an inner radius of 1 pc ($\sim 10^{18}$ cm) to ensure that the X-ray emission from the forward shock (arising at $\sim 10^{16-17}$ cm) is fully enclosed. The computation is then performed up to an $r_{out}$ such that, for the given value of \logn, log$N_H$ = 24 \cms. For each burst, the grid of TEPID X-ray absorption spectra covers the following parameter range:

\begin{itemize}
    \item log\nh\ \cms\ $\in [18,24]$ in 300 steps
    \item \logn\ $($cm$^{-3}) \in [1,5]$ in 50 steps
\end{itemize}

This stepping is chosen to have a higher resolution of our fitting grid than the typical errors on these parameters from our best-fits. From the grid of tabulated ionic populations, we can produce the spectrum by folding the tables through a custom version of the PHASE spectral engine \cite{2003ApJ...597..832K, L22} and then fit it to the observed EPIC-PN spectrum.

\subsubsection{Host metallicity}

We have also included the host metallicity in our TEPID analysis. For three GRBs of our sample we have direct measurements of the host metallicity (expressed typically as 12 + log(O/H), which for solar metallicity is 8.69) from emission lines in the host galaxy spectrum, which we report inside the parenthesis for each GRB; GRB 061121 (8.5, subsolar by 0.2 dex), \cite{2017A&A...599A.120V, 2019A&A...623A..26P} GRB 190114C (8.3, subsolar by 0.4 dex) \cite{2020A&A...633A..68D} and GRB 221009A (7.7, subsolar by 1.0 dex) \cite{2024NatAs...8..774B}. For the remaining four, we estimate the metallicity based on any available information: stellar mass and redshift for GRB 060729 (8.4, subsolar by 0.29 dex) \cite{2011MNRAS.413..669C} and redshift only for GRB 080411 (8.5, subsolar by 0.2 dex), GRB 090618 (8.4, subsolar by 0.3 dex) and GRB 120711A (8.5, subsolar by 0.2 dex) \cite{2023ApJ...954...13G}. 
We therefore bracket the metallicity between the host ISM as a lower limit and solar as an upper limit, leaving it free to vary when fitting the X-ray spectrum. Our assumption of solar metallicity as the upper limit is based on the fact that GRB hosts are known to have a large scatter that encompasses sub-solar values to solar-like abundances \cite{Levesque_2010, 2010AJ....139..694L, 2015A&A...581A.125K} and that local overabundances are known to exist in star-forming regions \citep{2021A&A...650A.103D}. Our derived values are fully consistent with the observed average for, and distribution in, the redshift range of our sample \citep{Graham_2023}.

\subsection{X-ray spectral analysis}

\subsubsection{Data reduction}

We perform standard data extraction after downloading the data from the \xmm\ archive (\url{http://nxsa.esac.esa.int/nxsa-web/}). Data were reduced using standard procedures for point-source extraction using the \xmm\ Science Analysis Software (\textit{XMM SAS}) v19.1.0. Event lists were generated using the \textit{epchain} task from which the EPIC-PN spectrum was extracted after filtering the data for high background and other standard filtering criteria (\url{https://www.cosmos.esa.int/web/xmm-newton/sas-thread-pn-xmmselect-spectrum}). We focus on the EPIC-PN dataset because the PN has the best efficiency in the energy range (0.2-7.5 keV) that we fit the spectrum over and unlike the MOS, is not significantly affected by pile-up effects at the high flux for some lGRBs of our sample. The PN spectra were rebinned to have 20 counts in the background plus source spectrum per energy bin. We note that for GRB 221009A, the Galactic absorption is very high due to its low latitude ($b$=+4 degrees). This has significantly affected the number of counts below 1 keV in the extracted spectrum, and thus, we performed our fit starting from 0.9 keV. 

\paragraph{Pile-up checks}

We have performed the recommended pile-up checks following the official data reduction cookbook (\url{https://www.cosmos.esa.int/web/xmm-newton/sas-thread-epatplot}) by producing the pattern plots for the EPIC-PN spectra. None of the PN spectra of the GRBs in our sample suffer from pile-up effects. 

For GRB 221009A, which is the brightest in our sample (and can thus be treated as an extreme case), we performed a further check with the procedure of \citep{pileup}. By computing the counts per frame, (0.3 counts frame$^{-1}$) and using their Fig 6, we find that spectral distortion is less than 2\% and thus comfortably within the recommended limits. 

\subsubsection{Fitting the EPIC-PN spectra}

We fit the spectra within {\sc xspec} \citep{1996ASPC..101...17A} adopting three model fits for each GRB X-ray spectrum, in increasing order of complexity:

\begin{enumerate}
    \item A neutral absorber at the redshift of the source, via \ztba\ \cite{2000ApJ...542..914W}
    \item The GRB-specific TEPID absorber (TEPID-only, hereafter)
    \item A combined TEPID and neutral absorber. The $N_H$ of the neutral absorber is restricted to the computed range of \nho. (\tepiO\ hereafter). This TEPID+optical model is most representative of the physical stratification of the medium that we expect along the LOS to the GRB, with a dense region (modelled by TEPID) within 100 pc and a tenuous ISM (modelled by \ztba) outside this region.
\end{enumerate}

In all fits (neutral, \tepi\ and \tepiO), we correct for Galactic absorption as estimated from w3nh (https://heasarc.gsfc.nasa.gov/cgi-bin/Tools/w3nh/w3nh.pl) according to the GRB sky position and model the intrinsic GRB emission as a power law. Apart from log\nh\ and \logn, which are left free to vary, we fixed the following parameters of TEPID: the time of the observation from the onset of the burst, the turbulent broadening of the absorption lines (in $km/s$) and the redshift. 

We fix the time at the midpoint of the \xmm\ observation in the rest-frame of the GRB since the spectrum for the midpoint is quite the same as the spectrum made with the average abundances over the observing time. As can be seen in modelled photon spectra as plotted in Extended Data Fig. 3, there are a wealth of absorption lines due to ionic species at various ionisation stages. These lines are too weak and narrow to be detected in the EPIC-PN data, requiring a larger area and higher spectroscopic resolution. Therefore, we fixed the velocity to a fiducial value of 200 $km\,s^{-1}$, i.e. below the resolution of the EPIC-PN instrument, and since we do not expect any broadening. In any case, at the resolution of the EPIC-PN, the absorption is dominated by the continuum, and the effects from the narrow lines are negligible so freezing the velocity has no systematic effect in our analysis. Finally, we run a $10^6$ step MCMC chain with a $10^5$-step burn-in to sample the best-fit model parameter space robustly. 

For the \tepiO\ fit, we have used the optical spectrum to infer the amount of neutral material to disentangle the contribution of the host galaxy ISM as described in the next section. 

\paragraph{Effects of warm Galactic ISM}

The Galactic ISM consists of both warm and cold components. To test the effects of our choice of a neutral Galactic medium, we included the contribution of this warm component in our fit by using the model of \citep{FN16}. To do this, we assumed that the warm component has a temperature of 5000 K. At such a temperature, \citep{FN16} find that the total column is composed of 50\% neutral material. Keeping this in mind, we then used their model by freezing the total column density to double the value derived from H1 density maps (w3Nh) and then reran the TEPID fits to our sample of GRBs, finding that there is no difference in the best-fits using either Galactic model, thus showing that our choice of neutral Galactic ISM does not affect our results in any way. 

\paragraph{Comparison with IGM opacity}

Absorption by the IGM/WHIM along the LOS has been invoked for explaining the excess absorption seen in the GRB X-ray spectrum \citep{Behar_2011, 10.1093/mnras/stt400, Rahin_2019, 10.1093/mnras/stab335, 2023ApJ...953..158G}. For computing the IGM/WHIM opacity, we use the latest results of \citep{2023ApJ...953..158G}. We plot the IGM opacity at 0.5 keV as a function of the redshift in Extended Data Fig. 4
(indigo line) with the associated 90\% uncertainty (shaded indigo region), along with the measurements and upper limits on opacity derived from GRB afterglow spectra in the analysis of \citep{Rahin_2019}. 

Superimposed on this plot as coloured diamonds are the opacities at 0.5 keV for five GRBs of our sample (GRB 060729 (blue), GRB 061121 (orange), GRB 080411 (green), GRB 090618 (red) and GRB 120711A (purple). We do not include GRB 190114C ($z=0.425$) which is strongly absorbed by the host galaxy below 0.9 keV (but has negligible Galactic absorption) and GRB 221009A ($z=0.151$) which is strongly absorbed by our Galaxy below 0.9 keV (and at the redshift of which the IGM contribution is even lower). As can be clearly seen, the opacity of the GRBs in our sample is at least an order of magnitude higher than the IGM opacity at the corresponding redshift, thereby showing that the IGM absorption is negligible.  

To further verify this, we also performed our own independent computation of the predicted IGM opacity by averaging 30 lines of sight using the hydrodynamical simulations of \citep{2006ApJ...650..560C} up to a redshift of 0.8. We plot this with the black square in the above plot, which is consistent within uncertainties to the values from \citep{2023ApJ...953..158G}. This leads us to conclude that due to the IGM absorption being clearly subdominant in the redshift range covered by our sample ($0.151 < z < 1.405$), the main contribution to the X-ray opacity is coming from gas in the vicinity of the GRB. 

\subsection{Analysis of metal absorption lines from the optical spectrum}\label{optical}

For the TEPID+optical fit, we use the optical spectrum to infer the amount of neutral material attributable to the host galaxy ISM in order to disentangle the absorption contributions from the host ISM compared to the medium within 100 pc of the GRB location, as can be seen schematically in Fig. \ref{fig:schematic} \cite[see also][]{{2011A&A...525A.113S}}. \grb\ optical afterglow spectra have absorption lines from singly-ionised metals, that are known to arise in the ISM of the host galaxy, far away from the GRB site. We use these lines to estimate the column of neutral material along the LOS. The standard procedure for estimating the column density of material producing an optical absorption feature is using the Curve-of-Growth \citep[COG,][]{1978ppim.book.....S} that uses:

\begin{itemize}
    \item The rest-frame equivalent width (EW)
    \item The oscillator strength (\textit{f}-value) of the atomic transition producing the line
\end{itemize}

While all lGRBs in our sample have optical spectra taken within a few hours of the trigger \cite{2006GCN..5373....1T, 2006GCN..5826....1B, 2008GCN..7587....1T, 2009GCN..9518....1C, 2012GCN.13441....1T, 2019GCN.23695....1S, 2022GCN.32648....1D}, the redshift of GRB221009A is too low to have singly ionised metal features redshifted in the optical spectrum, we therefore cannot estimate the neutral host column with the same procedure as the rest of the GRBs in our sample. For all GRBs, we fit the COG to EW measurements either collected from the literature \cite[GRB 060729, GRB 061121, GRB 080411 from][]{2009ApJS..185..526F, 2012A&A...548A..11D} or measured by us directly from the optical spectrum (Spectra for GRB 090618 and GRB 120711A were obtained via private communication, the spectrum for GRB 190114C is publicly available on the ESO archive \citep{2019GCN.23710....1K}). Where we have performed our own EW measurements, we verify our values against the averages reported by \cite{2012A&A...548A..11D}. 

For estimating the Doppler parameter for the COG (\textit{b}), we use doublets or multiple lines of the same ion, whichever available, and then assume the same COG for other ion lines in the spectrum. We use the same COG because not all spectra in our sample have multiple lines from each ion consistently available.

Accordingly, we estimate the neutral column assuming the same COG for the optical spectra of GRB 060729, GRB 080411, GRB 090618, GRB 120711A and GRB 190114C. For GRB 061121, the spectrum has multiple lines available for several ions including two different ions from \ion{Mg}{}: \ion{Mg}{I} and \ion{Mg}{II}. When fitting the COG for these ions, we found that the solutions converged on two distinct values for \textit{b}. We therefore decomposed the absorption into two systems, one corresponding to the \ion{Mg}{I} best-fit ($b = 63\ km/s$) and the other to the \ion{Mg}{II} best-fit ($b = 72\ km/s$). For the estimation of the neutral column, we assume the \ion{Mg}{II} best-fit. Since we assume the same COG for all elements in the same spectrum, we end up underestimating the uncertainty on the corresponding derived columns. We therefore assume an uncertainty corresponding to 2\% on the ionic column density (corresponding to $\sim 0.3$ dex), which is propagated to the uncertainty on each ion's hydrogen equivalent column density. The ionic columns per absorption feature are reported in column 3 of Extended Data Tab. \ref{tab:ew_analysis}

In order to compare the optical-inferred column density to the X-ray-inferred column density, we must account for the contributions of material that are missed by the optical but contribute to the X-ray. These consist of two corrections for ionisation fraction and dust depletion, before the column can be finally expressed as hydrogen-equivalent consistent with the X-ray analysis. To derive a dust correction, it is recommended to estimate and use the relative abundance of Zn to Fe, under the assumption that little-to-no Zn is depleted \citep{2013A&A...560A..88D, 2018A&A...613L...2D}. However, we unfortunately do not have Zn features available consistently in all the optical spectra of our sample of GRBs. While they are present in the spectra of GRB 061121 and 080411, they are blended with features from other ions. Therefore we cannot apply this method directly and systematically across all GRBs in our sample. Instead, we estimated an average Zn/Fe from the values computed for GRB hosts by \citep{2013A&A...560A..88D}. Using data from their Table 1, we find an average Zn/Fe = 0.7 with a standard deviation of $\sigma_{Zn/Fe}$ = 0.5. Using the average value, we used the element-wise linear depletion sequences computed in \citep{2016A&A...596A..97D} for estimating the corresponding depletion correction. For estimating the uncertainty on this correction, we used the extreme values of Zn/Fe bracketed by the standard deviation to derive two extreme values for the element-wise depletion, and then used the semi-scatter of the two extreme values as the uncertainty. Then, this uncertainty is propagated to the final estimate of the optical column used in the TEPID+optical fits.

To correct for the ionisation fraction, we use the Galactic ISM model from \cite{FN16} which includes internal contributions of stars in the disk and external meta-galactic contribution. Finally, we assume solar metallicity using abundances of \cite{2009ARA&A..47..481A} to convert the optical metal columns to hydrogen equivalent. In the following, we refer to this column as \nho. 

The final estimate of \nho\ is computed as a weighted average of all the individual hydrogen equivalent columns estimated from each element in the spectrum, and the uncertainty on the final value is the propagated uncertainty on each ion summed in quadrature. The COG Doppler factor, the complete list of lines and EWs per GRB and the correction factors assumed in our analysis, are reported in Supplementary Tab 3, final \nho\ values for each ion in each optical spectrum and the range used in the \tepiO\ fit. For a comparison of the columns derived from the optical analysis with those from the X-ray analysis, we refer the reader to Supplementary Data Section 1.3. 

\paragraph{GRB 190114C: effects of a higher dust correction}

As we have discussed in the main text, the X-ray absorption is likely dominated by the ISM of the host galaxy for GRB 190114C. Considering this, the discrepancy between \nho\ ($\sim 2.1\times10^{21}\,cm^{-2}$) and \nhc\ ($4\times10^{22}\,cm^{-2}$) for this GRB is puzzling. One possible explanation for this could be the assumed dust correction in our calculation of \nho, since the ISM is cold and dusty, as is reported in from observations from \cite{2020A&A...633A..68D}. If the medium is colder than the templates we assume, then the fraction of \ion{Mg}{} that is depleted into dust is much larger, $\sim 1.5$ dex instead of $\sim 0.8$ dex \citep{1996ARA&A..34..279S}. Assuming a higher dust correction increases the \nho\ to $\sim 6.3\times10^{21}\,cm^{-2}$. Considering the 1-$\sigma$ uncertainty on this \nho\ measurement, the upper limit of the confidence interval would therefore become $\sim 10^{22}\,cm^{-2}$. This is within the order of magnitude of the neutral and \tepi\ best-fit column densities, which lends further support to our conclusion that the cold ISM of the host is indeed dominating the observed absorption. 

\paragraph{Moderately ionised gas: TEPID predictions vs optical measurements}

The optical spectra of our sample were taken by different instruments and generally start from 3000 \AA. Therefore, rest-frame UV lines from moderately ionised gas (such as \ion{C}{IV}) are reported only in the spectrum of GRB 061121 \citep{2009ApJS..185..526F}. Taking the blended EWs reported in \citep{2009ApJS..185..526F} we fit the COG, finding a solution in the highly-saturated scenario, deriving $b \in [66-104]$ km/s, which gives log N$_{\ion{C}{IV}}\in [16.3,18.0]$. From TEPID, the predicted column has a similar range log N$_{\ion{C}{IV}}\in [16.0,18.0]$. Thus, the column of C IV as predicted by TEPID based on fitting the X-ray spectrum to the bulk continuum is in reasonable agreement with the column determined from the blended \ion{C}{IV} feature as reported in the low-resolution optical spectrum.

To compare the predicted column densities of \ion{C}{IV}, \ion{N}{V} and \ion{O}{VI} for the TEPID best-fits of our sample with measured columns from GRB observations, we used the combined samples of \cite{2008A&A...491..189F}, \cite{2008ApJ...685..344P} and \cite{2018MNRAS.479.3456H}, and plotted our predictions for the corresponding ion against the measured values from high-resolution optical spectroscopy in Supplementary Fig. 4. When multiple components in one ion are present, we performed a first check by summing the two components (in order to ensure that we dont violate the maximum limit), and then also by plotting each individual component. These are plotted as grey circles in the plot above. The TEPID predictions are plotted as vertical, coloured bars per GRB. We find that that for all three ions, the two distributions are entirely consistent. 

\subsection{Estimation of statistical significance for model selection}

We verified the {\sc Xspec} $\Delta\chi^{2}$ results with a Bayesian X-ray fit performed using the Bayesian X-ray Analysis package \cite{2014A&A...564A.125B}. From the maximum likelihood estimate (MLE) and the Bayesian evidence ($\mathcal{Z}$) obtained as part of the results of the Bayesian fit, we then estimate and use the Akaike Information Criterion (AIC) and the Bayes Factors ($BF$) respectively, as two separate tests for the significance of the improvement by the more complex model (TEPID) with respect to the simpler model (neutral). 

\subsubsection*{Akaike Information Criterion}

For a given model with number of parameters $\theta$ and maximum likelihood estimate $\mathcal{L}(d|\theta)_{MLE}$, the AIC can be estimated as:

\begin{equation}
    \mathrm{AIC} = -2\ ln(\mathcal{L}(d|\theta)_{MLE}) + 2\theta
\end{equation}

When comparing two model fits to a given dataset, the model that gives the lower value of AIC is preferred over the other. The model with lower AIC is called best model (in our case TEPID) whereas the other model is called the candidate model (in our case the neutral model). The significance of the improvement in the fit can be numerically quantified by the difference $\Delta$AIC = $AIC_{neutral} - AIC_{TEPID}$. From \citep{burnham2004multimodel, fabozzi2014basics}, a $\Delta$AIC $> 10$, shows conclusive support for TEPID over the neutral model, a $\Delta$AIC between 4 and 7 suggests moderate support for TEPID over the neutral model and a $\Delta$AIC $< 2$ suggests no support for TEPID over the neutral model.

For four GRBs of our sample: GRB 060729, GRB 080411, GRB 090618 and GRB 120711A; $\Delta$AIC is higher than 10 for both \tepi\ and \tepiO\ fits, showing conclusively strong support for the TEPID absorber over the neutral absorber. This holds for both TEPID fits with metallicity fixed to solar and for metallicity free to vary.

For GRB 221009A $\Delta$AIC = 9 for the \tepi\ fit. Given that this is higher than the threshold of 7, but less than 10, we interpret only strong support for TEPID over the neutral absorber, again with the significance unchanged whether fixing metallicity to solar or leaving it free to vary.

Finally, for GRB 061121, for the \tepiO\ fit with metallicity fixed to solar the $\Delta$AIC = 7. Since this is exactly at the threshold of 7, we conservatively interpret moderate support for the TEPID+optical absorber based on a combination of the $\Delta$AIC and $BF$. This significance is unaffected for the \tepiO\ fit with metallicity free. For the \tepi\ fit with metallicity free, the $\Delta$AIC = 3. Given the additional free parameter, we therefore dont consider the improvement to be significant for the \tepi\ fit with metallicity free.

\subsubsection*{Bayes Factor}

For a given model with number of parameters $\theta$ that has likelihood $\mathcal{L(D,\theta})$, and a prior distribution $\pi(\theta)$, the Bayesian evidence is estimated as:

\begin{equation}
\mathcal{Z} = \int \mathcal{L}(d,\theta) \pi(\theta) d\theta
\end{equation}

For comparing two competing models, the Bayes factor is estimated as the ratio of the two evidences. For our analysis, we estimate this as 

\begin{equation}
    BF = \frac{\mathcal{Z}_{TEPID}}{\mathcal{Z}_{neutral}}
\end{equation}

Within this formulation, $BF > 100$ \cite{Lee_Wagenmakers_2014} shows decisively strong support for TEPID over the neutral model. 

For four GRBs of our sample: GRB 060729, GRB 080411, GRB 090618 and GRB 120711A, we find that the $BF$ is much larger than this threshold, which we interpret as decisively strong support for the TEPID absorber compared to the neutral absorber for all TEPID best-fits, both with metallicity fixed to solar and left free to vary.

For GRB 221009A $BF = 4000$ and is thus considerably higher than the threshold, but the support for the TEPID-only fit as determined by the $\Delta$AIC (9) is not as strong as for the other four. We therefore, conservatively conclude that there is only strong but not conclusive, support for the \tepi\ best-fit compared to the neutral model. This conclusion is unchanged for the fit with metallicity free. 

For GRB 061121, the \tepiO\ best-fit $BF = 160$ which while larger than the threshold for strong evidence, is not as high as that for the other five. Combined with the $\Delta$AIC = 7, we therefore conservatively conclude that there is only moderate support for the \tepiO\ absorber over the neutral absorber. For the \tepi\ fit with the metallicity free to vary, the best-fit $BF=170$, with a $\Delta$AIC = 3. However, since this includes an additional parameter in the fit, we dont consider the improvement to be significant. The significance for the \tepiO\ fit with metallicity free is unaffected.

\subsection{Results of spectral fitting}

We discuss as an example of our results the case of GRB 060729. Accordingly, we show the count spectrum and best-fit model photon spectrum plots only for this lGRB as part of the Extended Data Figures. For all other fit-specific plots (count and model spectra, ionic column distributions and MCMC chain contours) for the \tepi\ and \tepiO\ best-fits of the rest of the sample, we refer the reader to Supplementary section 1.4. 

We note here that the best-fit contours presented in Fig. \ref{fig:master} are produced from MCMC chains that were run in a distance-wise configuration. Instead, the errors on \nhx\ reported in Tab. \ref{tab:master} were computed from a separate set of MCMC chains run in the column-density wise configuration. While the best-fits obtained from either approach is exactly the same, in this way, we could accurately determine the errors on $N_{X,TEPID}$, $log(n_0)$ and the size without propagation, as error propagation is not straightforward when the parameters are correlated. 

For GRB 060729, we show in the top panel of Extended Data Fig. 2 the count spectra of 060729 with the model predictions of the best-fitting neutral model (green) and the TEPID-only model (orange), with the corresponding residuals in the middle and bottom panels respectively. The neutral best-fit model leaves systematic residuals in the soft X-ray band (E$_{obs}$ from $0.2 - 1.0$ keV), and thus the fit while reasonable ($\chi^{2}/DOF = 1249/1191$) cannot adequately describe the observed spectral profile. We have fitted all the spectra of our sample with a photoionisation-equilibrium model \cite{1992ApJ...395..275D}, finding that for all of them the best-fit ionisation parameter ($\xi$) converges on anomalously low values, being essentially consistent with a neutral medium. This shows that a photoionisation-equilibrium model is unable to correctly model the GRB X-ray absorption, and would lead to the erroneous conclusion that the medium is likely neutral. The fit with our time-evolving photoionisation model instead is clearly much better ($\chi^{2}/DOF = 1120/1190$, $\Delta\chi^{2}=129$) being able to reproduce the spectral profile and remove most of the residuals that can be seen in the neutral best-fit. 

More generally, the \tepi\ fits improve upon the neutral fits for five bursts: GRB 060729, GRB 080411, GRB 090618, GRB 120711A, GRB 221009A ($10 \lesssim \Delta\chi^{2} \lesssim 129$), while the fit for GRB 061121 is identical to the neutral fit ($\Delta\chi^{2} = 1$). As discussed before, this improvement is conclusively significant for GRB 060729, GRB 080411, GRB 090618 and GRB 120711A based on the \textit{BF} ($> 100$) and $\Delta$AIC ($>10$) and significant, but not conclusive, for GRB 221009A ($BF > 100$ but $\Delta$AIC = 9). 

The resulting spectral profile can be seen in the top panel of Extended Data Fig. 3, where we plot the TEPID best-fit model spectrum with the neutral best-fit model. The TEPID spectrum can describe the X-ray spectrum much more accurately. To show how this comes about, we plot the element-wise breakdown of the ionic population to the opacity in the bottom panel of Extended Data Fig. 3. The improvements in the TEPID spectral profile are arising from a unique ionic population that combines absorption due to the combination of multiple ionisation stages of H, He and metals existing in a physical region of size 6 pc with number density 10$^{2.8}$ cm$^{-3}$.

Out of the six bursts where we have \nho\ measurements, we find that for five: GRB 060729, GRB 061121, GRB 080411, GRB 090618 and GRB 120711A, the \tepiO\ fit improves upon the neutral fit ($7 \lesssim \Delta\chi^{2} \lesssim 131$). With the exception of GRB 061121, the best-fit column and number densities for the remaining GRBs are relatively unchanged from the \tepi\ best-fits resulting in essentially the same ionic distribution and photon spectrum as the \tepi\ fits. Furthermore, the \tepiO\ fits are statistically identical to the \tepi\ best-fits as shown by the similar $BF$ and $\Delta$AIC values. 

For GRB 061121, the best-fit column ($3\times10^{22}\ \mathrm{cm}^{-2}$) and number density (10$^{3.4}$ cm$^{-3}$) both increase (by a factor of $\sim 10$) with an associated improvement over the neutral fit ($\Delta\chi^{2} = 7$) that is moderately significant ($\Delta$AIC$ \sim 7$). This suggests that if the neutral absorption from the ISM is separately accounted for, the presence of a photoionised environment could potentially be recovered. 

After bracketing the metallicity for each GRB as described before, we left it free to vary between the limits, finding that it is essentially unconstrained (as is expected in the case of low-resolution spectra), and as a result the best-fit parameters are not significantly affected. However, the 90\% confidence region for the size of the TEPID best-fit regions with the metallicity free does slightly increase. This happens because the fit tries to recover a given fractional contribution of H, He and metals, thus for lower metallicity necessitating a larger size of the absorbing medium to accumulate sufficient metals. Our conclusions, drawn from directly fitting the X-ray spectrum for the first time with a physically accurate model for the medium ionised by the GRB thus show that GRB X-ray absorption comes from dense, ionised gas within star-forming regions embedding the GRB inside of the host galaxy \citep{2007ApJ...660L.101W, 2011A&A...525A.113S, 2012MNRAS.421.1697C, 2013ApJ...768...23W, 2013ApJ...774..115K, 2002ApJ...565..174R, 2001ApJ...549L.209G, 2013MNRAS.432.1231C, 2012ApJ...754...89W, 10.1093/mnras/stw2423}.

\subsection{Differing afterglow- and TEPID-inferred densities}

The circumburst density profile and the number density of a GRB medium can be inferred from the multiwavelength afterglow data by either fitting the broadband light curves \citep[e.g.][]{2022MNRAS.511.2848A} or using synchrotron closure relations based on characteristic frequencies \citep[e.g.][]{2018ApJ...866..162G}. The circumburst density profile is found to be either ISM-like ($n \propto r^{0}$) or wind-like ($n \propto r^{-2}$), typically on the order of $1-10\,\mathrm{cm}^{-3}$ \citep[see][]{2022MNRAS.511.2848A}. From the TEPID best-fits, we derive number densities of $\sim 10^3 \mathrm{cm}^{-3}$. Our best-fitting number densities are thus two orders of magnitude higher than the afterglow-modelled number densities. This is because the densities we infer are due to absorption from a dense star-forming region than the low-density, diffuse medium inferred from the afterglow jet modelling, we explain this further below.

The progenitor of the GRB is expected to undergo mass loss throughout its lifetime. In particular, as the progenitor is a Wolf-Rayet (WR) star, the average mass-loss rate in the WR phase is $\sim 10^{-5} M_{\odot}\ yr^{-1}$ with average wind velocities of the order of $10^{3}\,km/s$ \citep{2007ARA&A..45..177C}. Several studies have modelled the mass loss history of the progenitor and documented its effects on the physical nature of the circumburst medium have found that the WR wind clears out the surrounding medium up to a few parsec at the expected densities of star-forming regions, thereby forming a wind-blown bubble. \citep[e.g.]{2022MNRAS.515.2591C}  This suggests that the afterglow inferred density is probing the wind bubble region cleared by the WR wind through which the GRB jet expands and which is thus at the deceleration radius length scale ($\lesssim 10^{17}\,cm$).  

The TEPID densities, on the other hand, arise from X-ray absorption by the medium inside the dense and high ionisation region starting at a distance of $\sim 1$ pc from the progenitor and which is thus clearly distinct from the inner region within 1 pc of the burst. At the densities of TEPID best-fits ($\sim 10^{3}$ cm$^{-3}$), the expansion of the wind bubble is found to be strongly suppressed due to the denseness of the surrounding medium. The clearing out of the medium is therefore not expected to exceed a few pc at most \citep{2022MNRAS.515.2591C}, thereby being consistent with the inner radius of our TEPID simulations.

We further note that afterglow modelling for five of the GRBs in our sample suggests that the jet is indeed expanding into a wind-like medium, which lends support to our explanation for the different physical origins of the afterglow- and TEPID-inferred densities. \citep[see][]{1207t90, 2020ApJ...890....9A, 2022MNRAS.515.2591C, 2025ApJ...978...29T}.

\subsection{Placing TEPID results on the density vs size plane }

We contextualise the size and number density of the TEPID absorber best-fits by comparing them with the density and size of other gravitationally bound objects in the Universe. For this, we have considered the medium within clumps inside star-forming regions; star-forming regions both in our Galaxy and an external galaxy; globular clusters; spiral, elliptical and irregular galaxies; galaxy clusters and a filament of the WHIM. We estimate the number densities of the medium inside these objects in a consistent manner with our TEPID model by assuming that the total mass of the structure is evenly distributed in a given geometrical shape.

We assume a spherical geometry for star-forming regions, clumps inside them, globular clusters and galaxy clusters and we estimate the density as:

\begin{equation}
    n_0 = \frac{M}{\frac{4}{3} \pi r^{3} m_p}
\end{equation}

Where $M$ is the total mass of the object in g, $r$ is the radius in cm and $m_p$ is the mass of the proton. We take the virial mass and radius as reported by the corresponding catalogues for the clumps and extragalactic star-forming regions \cite{2019MNRAS.484.4444U, 2017A&A...601A.146C}. For Galactic star-forming regions we use the luminosity mass. Since the size estimation for some Galactic SFRs is affected by their Galactocentric radius, we plot the associated uncertainty in size and also propagate it to the estimation of the number density. \cite{2017ApJ...834...57M} For globular clusters, we use the ICM gas mass estimates and the tidal radius in physical units from \cite{2024A&A...686A.283P}. For galaxy clusters, we use the ICM mass and cluster radius reported by \cite{2018MNRAS.478.3072C}

For the galaxies we assume a cylindrical geometry, from which we can derive the number density as:

\begin{equation}
    n_0 = \frac{M}{2 \pi r^{2} h m_p}
\end{equation}

Where $M$, $r$ and $m_p$ are the same as above and h is the representative scale height, which we assume to be 300 pc. For all galaxies, we use the isophotal radius converted to physical units and luminous mass as reported by \cite{2007A&A...465...71T, 2008AJ....136.2563W, 1980ApJ...242..931S}. Finally, for the WHIM filament, we directly plot the range of number density and length scale reported by \cite{2008Sci...319...55N}. 

\section*{Data Availability}
A minimal set of primary data (and Python codes) used to produce the plots in this work will be shared upon reasonable request addressed to the corresponding author. The referenced datasets used for various comparisons are included in the bibliography and are publicly available.

\section*{Code Availability}
The TEPID code is proprietary software and is not publicly available. A minimal set of data can be made available on reasonable request to the corresponding author.

\section*{Acknowledgements}
A.L.T, L.P, A.L., F.N. acknowledge support from the European Union Horizon 2020 programme under the AHEAD2020 project (grant agreement number 871158). A.L.T and L.P acknowledge support from ASI (Italian Space Agency) through Contract No. 2019-27-HH.0. A.L. and F.P acknowledge support from the PRIN-MUR-2022 “Advanced X-ray modeling of black hole winds” (DRAGON), ID: 2022K9N5B4. Y.K. acknowledges support from grant PAPIIT-UNAM IN102023. We thank Nial Tanvir for sharing the Gemini/GMOS-S optical spectrum of GRB 120711A and S Bradley Cenko for the P200/DBSP optical spectrum of GRB 090618. Based on observations obtained with XMM-Newton, an ESA science mission with instruments and contributions directly funded by ESA Member States and NASA. 

\section*{Author contributions}
A.L.T, L.P, A.L and F.N initiated the work for this study, A.L.T was in-charge of overall activities, analysis (with contributions from L.P, A.L and F.N) and writing of the manuscript. A.L.T and L.P set the conditions and identified the sample of GRBs. A.L, F.N, L.P, Y.K and A.L.T developed the first version of the TEPID model. A.L.T reduced the XMM spectra (with contributions from B.G) and compiled and analysed the optical spectra (with contributions from S.S). Additionally, S.S and F.N contributed in the estimation of the hydrogen-equivalent optical columns. All authors provided feedback on the text in the manuscript.

\section*{Competing Interests} 
The authors declare no competing financial interests
\clearpage
\section*{Tables}
\setlength{\tabcolsep}{4pt}
\begin{table}[h]
\caption{\textbf{Summary of the X-ray best-fits for our sample of GRBs}. We present the results of the neutral, \tepi\ with solar and \tepi\ with free metallicity fits along with the fit statistics corresponding to the best fit in each case. The computed significance estimators for each fit are also reported. Errors are quoted at 90\%.}
\centering
\begin{tabular}{c|c c c c c c c}
\toprule
Parameters & \multicolumn{6}{c}{Value}
\\
GRB & 060729 & 061121 & 080411 & 090618 & 120711A & 190114C & 221009A\\
$z$ & 0.54 & 1.314 & 1.031 & 0.54 & 1.405 & 0.425 & 0.151 \\
\midrule
& \multicolumn{7}{c}{\textbf{{\sc TBabs} fit}} \\
\midrule
\nhc\ $(10^{21} {cm}^{-2})$ & $0.91\pm0.03$ & $5.7\pm0.2$ & $2.5\pm0.1$ & $1.7\pm0.04$ & $9.8\pm0.1$ & $46\pm1$ & $4.9\pm$0.2 \\
$\alpha_{neutral}$ & $2.1\pm0.01$ & $2.02\pm0.02$ & $2.05\pm0.03$ & $1.98\pm0.02$ & $1.95\pm0.02$ & $1.98\pm0.04$ & $1.74\pm0.01$ \\
$\chi^2/DOF$ & 1249/1191 & 767/783 & 580/528 & 807/741 & 937/881 & 967/950 & 1249/1237 \\
AIC & 1260 & 784 & 598 & 820 & 950 & 963 & 1257  \\
\midrule
& \multicolumn{7}{c}{\textbf{\tepi\ fit}, solar metallicity} \\
\midrule
\nhx\ $(10^{21} cm^{-2})$ & 9.1$^{+13.7}_{-0.8}$ & 22.6$^{+13.8}_{-14.5}$ & 70.1$^{+61.4}_{-19.1}$ & 11.5$^{+3.8}_{-1.2}$ & 323.6$^{+252.4}_{-130.1}$ & 57.5$^{+8.5}_{-2.6}$ & 199.0$^{+125.6}_{-57.2}$ \\
\logn\ $(cm^{-3})$ & 2.8$^{+0.4}_{-0.1}$ & 2.6$^{+0.3}_{-1.2}$ & 3.2$^{+0.3}_{-0.8}$ & 2.4$^{+0.2}_{-0.1}$ & 4.4$^{+0.4}_{-0.5}$ & 1.7$^{+0.2}_{-0.7}$ & 3.7$^{+0.3}_{-0.3}$ \\
size (pc) & 6$^{+0.7}_{-1.8}$ & 18$^{+77}_{-5}$ & 15$^{+22}_{-3}$ & 16$^{+2}_{-4}$ & 6$^{+3}_{-2}$ & 325$^{+1665}_{-88}$ & 14$^{+5}_{-3}$ \\
$\alpha_{ionised}$ & 2.05$^{+0.01}_{-0.02}$& $2.03^{+0.02}_{-0.03}$ & $1.99^{+0.02}_{-0.03}$ & $1.93^{+0.02}_{-0.03}$ & 1.86$^{+0.02}_{-0.02}$ & $2.01^{+0.03}_{-0.03}$ & 1.66$^{+0.01}_{-0.02}$ \\
$\chi^2/DOF$ & 1120/1190 & 766/782 & 544/527 & 758/740 & 894/880 & 978/949 & 1239/1236 \\
$\Delta \chi^2$ ($\Delta DOF = 1$) & 129 & 1 & 36 & 49 & 43 & -11 & 10 \\
AIC & 1121 & 785 & 567 & 773 & 905 & 973 & 1248  \\
BF & $10^{30}$ & 9 & $10^{8}$ & $10^{11}$ & $10^{10}$ & 0.01 & $4652$ \\
\midrule
& \multicolumn{7}{c}{\textbf{\tepi\ fit}, free metallicity} \\
\midrule
\nhx\ $(10^{21} cm^{-2})$ & 8$^{+13.2}_{-0.7}$ & 21.4$^{+8.8}_{-11.6}$ & 67.1$^{+51.7}_{-32.3}$ & 11.5$^{+3.0}_{-1.9}$ & 323.6$^{+238.7}_{-149.7}$ & 112$^{+6.2}_{-47}$ & 158$^{+224.3}_{-80.9}$ \\
\logn\ $(cm^{-3})$ & 2.7$^{+0.4}_{-0.1}$ & 2.5$^{+0.2}_{-1.1}$ & 3.2$^{+0.3}_{-0.9}$ & 2.4$^{+0.2}_{-0.2}$ & 4.2$^{+0.5}_{-0.4}$ & 2.1$^{+0.2}_{-0.7}$ & 3.6$^{+0.3}_{-0.6}$ \\
size (pc) & 7$^{+1}_{-2.4}$ & 23$^{+90}_{-8}$ & 14$^{+24}_{-3}$ & 16$^{+5}_{-3}$ & 6$^{+3}_{-2}$ & 250$^{+636}_{-122}$ & 14$^{+11}_{-3}$ \\
$\alpha_{ionised}$ & 2.05$^{+0.01}_{-0.02}$ & $2.02^{+0.04}_{-0.02}$ & $1.99^{+0.03}_{-0.03}$ & $1.93^{+0.02}_{-0.03}$ & $1.86^{+0.02}_{-0.01}$ & $2.01^{+0.03}_{-0.04}$ & 1.67$^{+0.02}_{-0.02}$ \\
$\chi^2/DOF$ & 1117/1189 & 760/781 & 544/526 & 758/739 & 893/879 & 974/948 & 1238/1235 \\
$\Delta \chi^2$ ($\Delta DOF = 2$) & 133 & 7 & 36 & 49 & 43 & -7 & 11 \\
AIC & 1122 & 781 & 569 & 776 & 907 & 969 & 1250 \\
BF & $10^{31}$ & 171 & $10^{8}$ & $10^{11}$ & $10^{8}$ & 0.01 & $9290$ \\
    \botrule
    \end{tabular}
    \label{tab:master}
\end{table}
\clearpage
\begin{table}[h]
\caption{\textbf{Properties supporting a collapsar progenitor for our sample}. We compiled from the literature the T$_{90}$ (as measured by trigger instrument), the $E_{peak}$, the $E_{iso}$, Amati correlation (\cite{2012MNRAS.421.1256N}) and spectral lag as measured from the prompt phase, the detection of an associated supernova, presence of star-formation in the host and physical and normalised offsets of each lGRB as measured from the afterglow observations. All references for the properties reported in this table are available in the Supplementary section 1.5. If a property is not available for the corresponding GRB, it is marked as N/A. }
\label{tab:collapsar}%
\begin{tabular}{@{}lcccccccc@{}}
\toprule
GRB & T$_{90}$ & $E_{peak}$ & $E_{iso, 52}$ & Spectral Lag & Supernova & Star-formation & Amati Correlation & Offset (normalised) \\
 & $(s)$ & (keV) & (erg) &  &  &  &  & (kpc) \\
\midrule
060729 & 115  & 60* & 1.6 & Y & Y & Y & Y & 2.15 (1.1)  \\
061121 & 22  & 607 & 28 & Y & N/A & Y & Y & $\sim$2.5 (0.35)  \\
080411 & 56 & 266 & 24 & Y & N/A & N/A & Y & N/A  \\
090618 & 113  & 187 & 25.3 & Y & Y & N/A & Y & 4.4 (0.3) \\
120711A & 119  & 1061 & 100 & Y & N/A & N/A & Y & 0.137 (0.1)  \\
190114C & 362  & 856 & 25 & Y & Y & Y & Y & 0.17 (0.09) \\
221009A & 290 & 3700 & 1000 & Y & Y & Y & Y & 0.65 (0.53)  \\
\botrule
\end{tabular}
\end{table}

\clearpage

\section*{Extended Data Figures}
\begin{figure}
    \centering
    \includegraphics[width=0.95\textwidth]{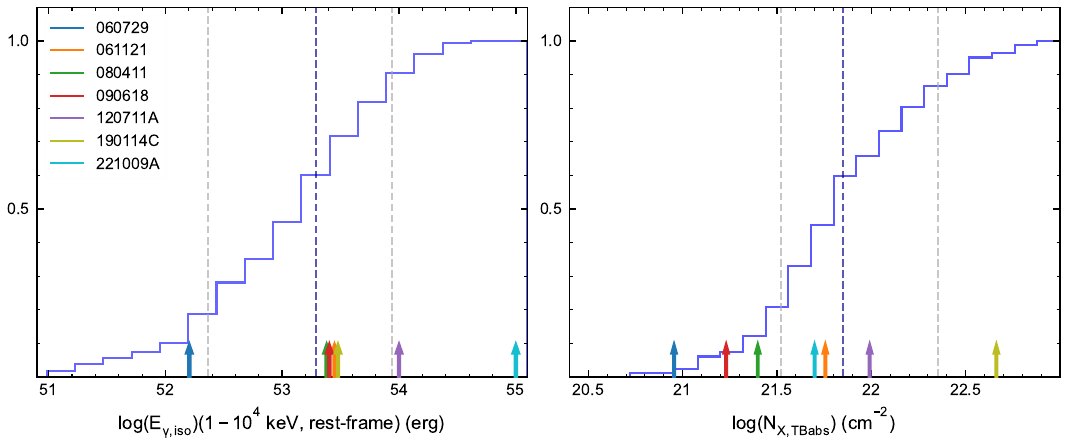}
    \caption{\textbf{Cumulative distribution functions of GRB properties for bias checks} \textit{Left:} Rest-frame $E_{iso}$ (measured in 1-10$^{4}$ keV) of a sample of GRBs from \cite{2012MNRAS.421.1256N, 2016ApJ...831...28T, 2017ApJ...837..119A}. \textit{Right:} Intrinsic X-ray absorption column density of the sample of GRBs from \citep{2010MNRAS.402.2429C}. In both panels, the vertical arrows mark the value for the respective burst in our sample, the blue dashed line shows the sample median and the grey lines show the first and third quartiles.}
    \label{fig:bias_check}
\end{figure}
\clearpage
\begin{figure}
    \centering
    \includegraphics[width=0.49\textwidth]{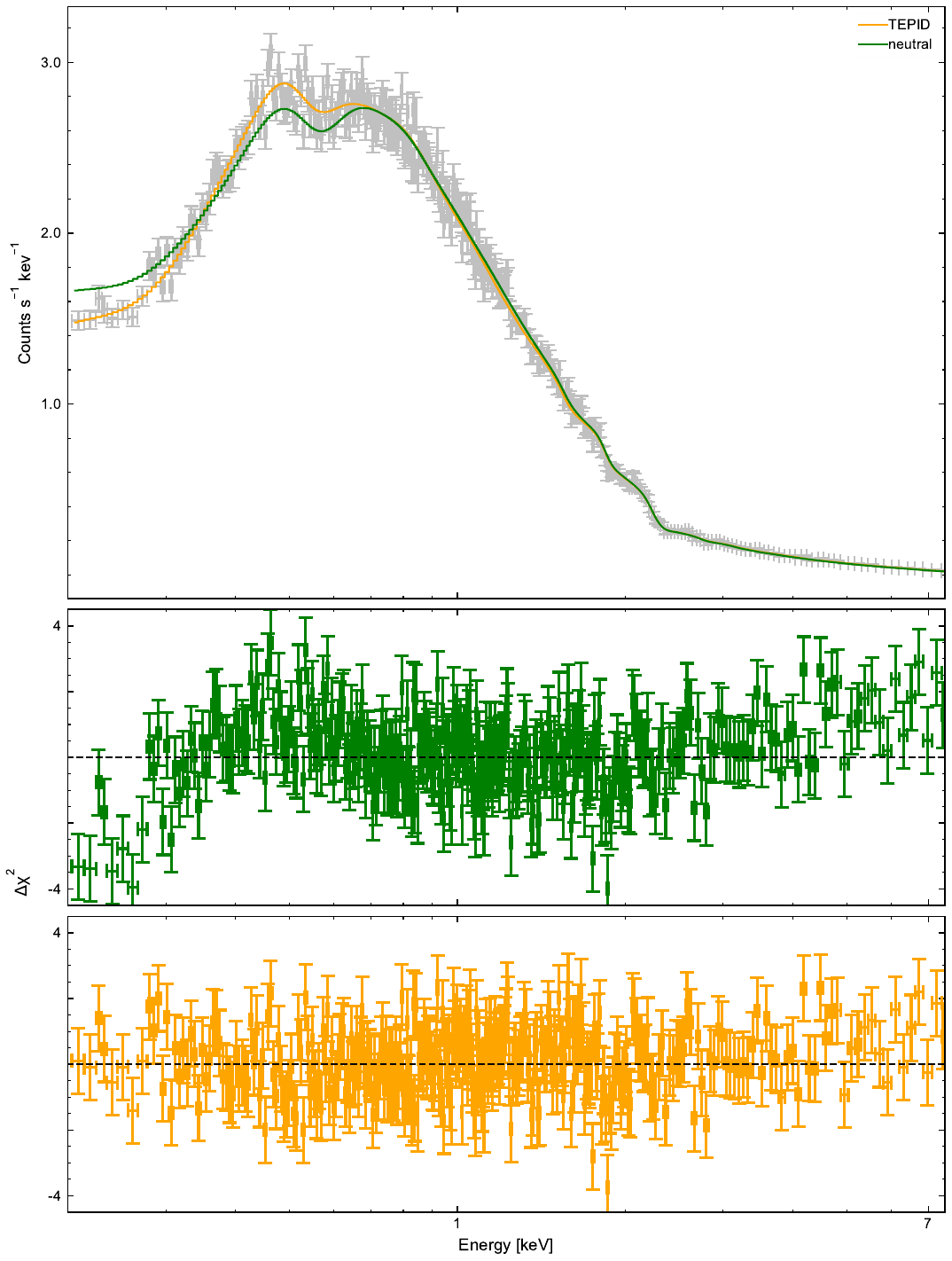}
    \caption{\textbf{EPIC-PN count spectrum of GRB 060729 with folded model predictions}: \textit{(Top panel)}: The EPIC-pn spectrum is plotted in grey points and has been rebinned for plotting purposes. The y-errors represent the 1-sigma uncertainty on the count-rate per bin, and the x-errors represent the bin size. The best-fit model predictions of the neutral model and TEPID model are plotted as green and orange lines respectively. \textit{(Middle and bottom panels)}: Corresponding residuals for the neutral and TEPID model, respectively, colour-coded as above. The data have been heavily rebinned for plotting purposes. The residuals below 1 keV seen from the neutral model-fit are not present in the TEPID model-fit, showing the improvement in the fit.}
    \label{fig:count_spec}
\end{figure}
\clearpage
\begin{figure}
    \centering
    \includegraphics[width=0.6\textwidth]{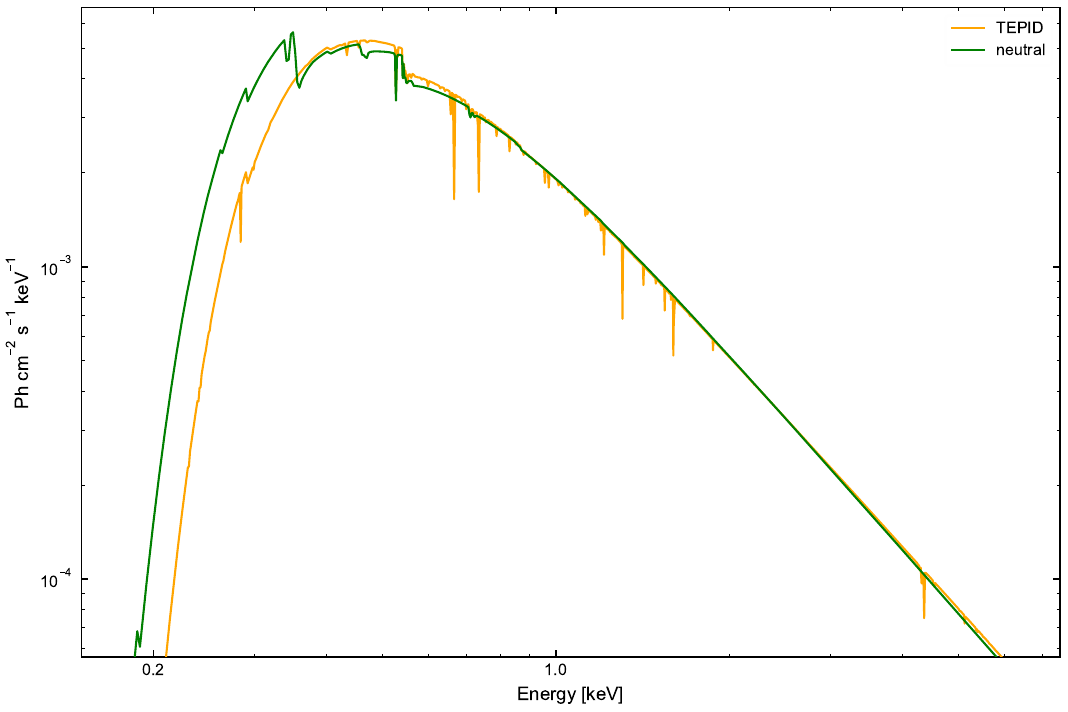}
    \includegraphics[width=0.6\textwidth]{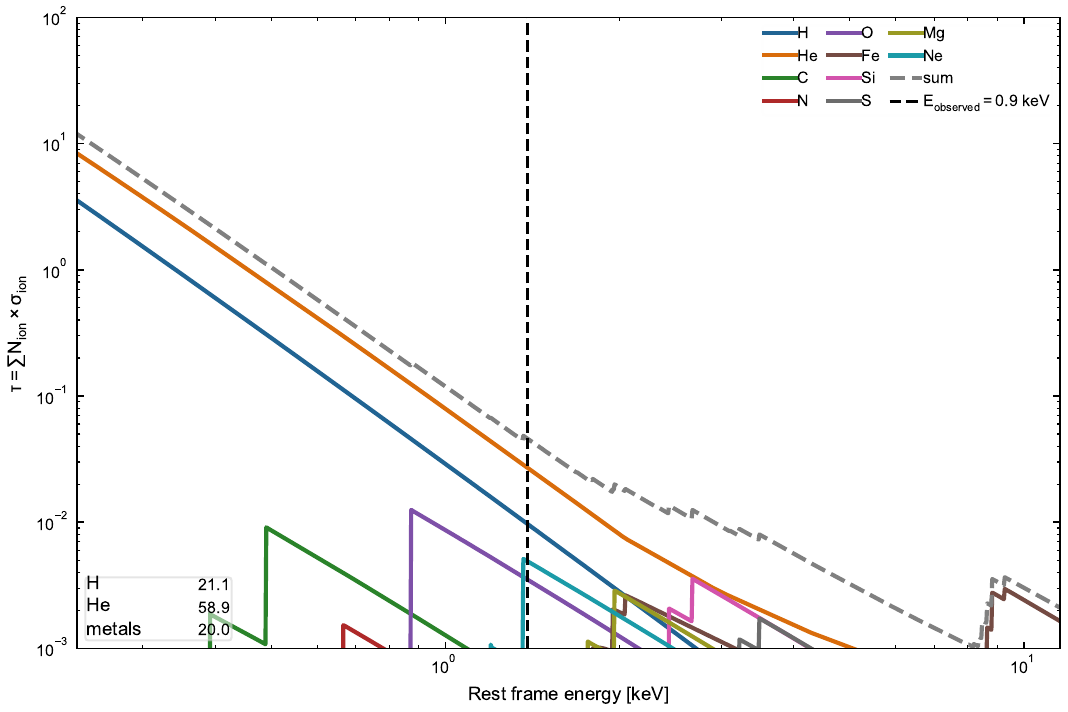}
    \caption{\textbf{Photon spectrum and element-wise opacity breakdown for the GRB 060729 TEPID-only best-fit} \textit{Top panel}: The TEPID best-fit model photon spectrum is plotted in orange as a function of observed energy. For comparison, the best-fit neutral model is plotted in green. \textit{Bottom panel}: Opacity contributions per element (summed over all ions) as a function of rest-frame energy. At $E_{rest}< 1$ keV, H and He are contributing to the absorption, instead at $E_{rest} >$ 1 keV metal contributions start to become significant. The vertical dashed line marks the E$_{obs} = 0.9$ keV, with percent contributions of H, He and metals reported in the bottom left.}
    \label{fig:phot_spec1}
\end{figure}

\begin{figure}[!ht]
\centering
  \includegraphics[width=0.7\textwidth]{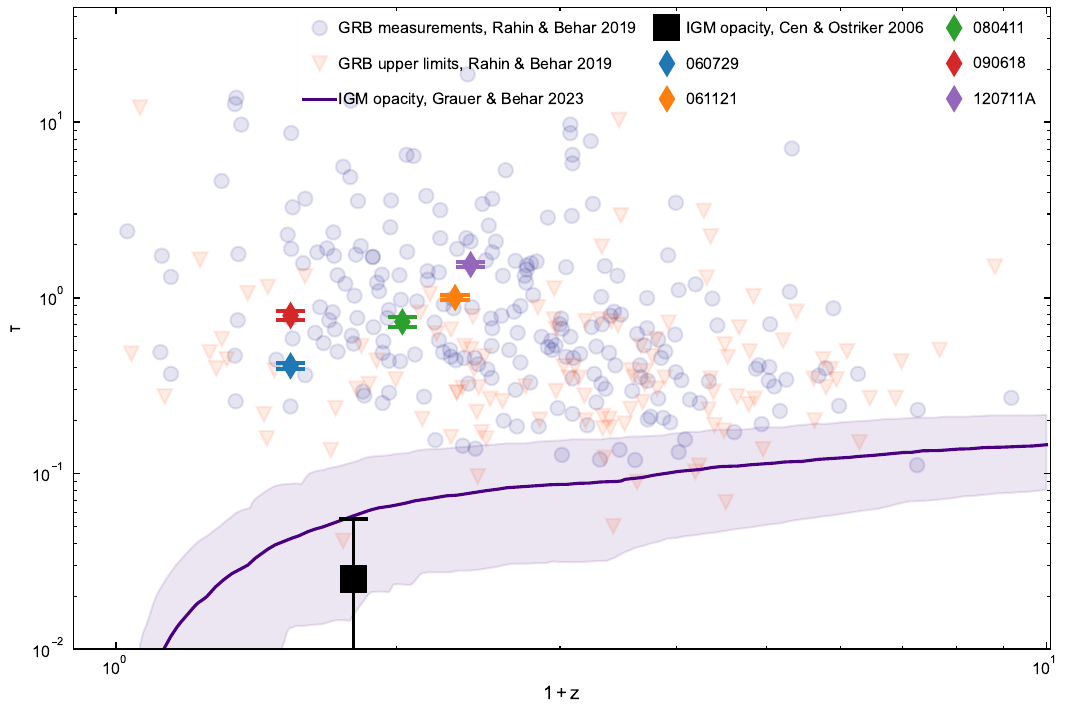}
  \caption{\textbf{Distribution of 0.5 keV opacity as a function of redshift.} The indigo line plots the IGM opacity from \citep{2023ApJ...953..158G}, along with the 90\% uncertainty (shaded, transparent indigo region). The coloured diamonds plot the opacity of GRBs from our sample, with the uncertainty taken from the best-fit X-ray neutral column. The blue circles and orange downward triangles mark the GRB opacity measurements from \citep{Rahin_2019}. The black square plots the opacity determined from the hydrodynamic simulations of \citep{2006ApJ...650..560C}. }
  \label{fig:IGM_opacity}
\end{figure}
\clearpage
\begin{figure}[!ht]
    \centering
    \includegraphics[width=0.7\linewidth]{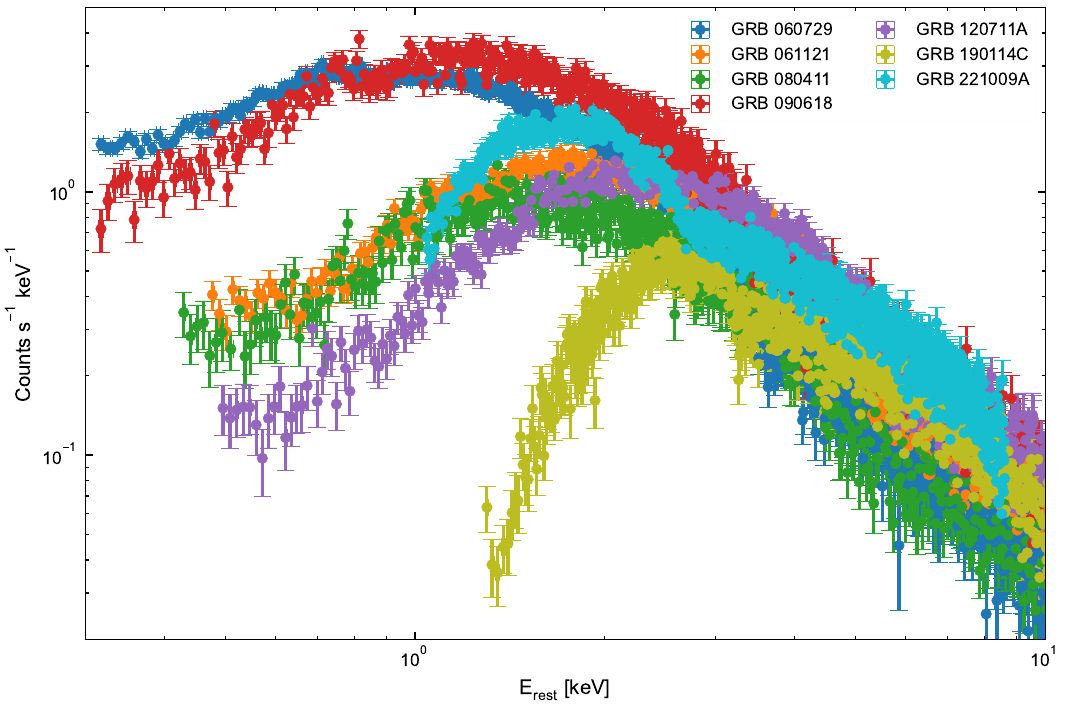}
    \caption{\textbf{Distribution of the EPIC-pn X-ray spectra of our sample of GRBs in the rest-frame.} The spectra have been cut to the fitted energy range and then redshifted to the rest-frame . As can be seen, the spectrum of GRB 190114C does not have counts below 2 keV, likely suggesting that the X-ray absorption is dominated by a cold and dusty component.}
    \label{fig:rest_spec}
\end{figure}

\clearpage
\setcounter{figure}{0}
\setcounter{table}{0}
\renewcommand{\figurename}{Supplementary Figure} 
\renewcommand{\tablename}{Supplementary Table} 

\section{Supplementary Information}

\subsection{Sample selection and observation log}

\begin{figure}[b]
    \centering
    \includegraphics[width=0.7\textwidth]{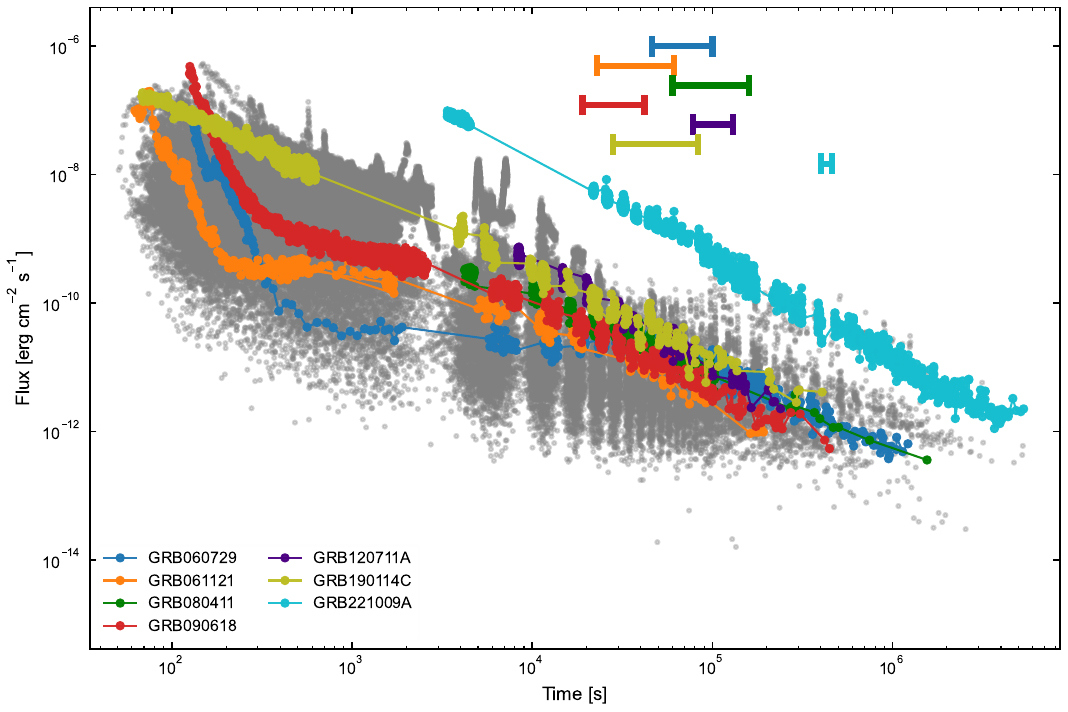}
    \caption{\textbf{Lightcurves of XRT sample of GRBs with known redshift in flux space (0.3-10 keV)}. The horizontal bars at the top show the duration of the \textit{XMM} observations. The light curves of our sample are clearly towards the top of the scatter at the starting time of \textit{XMM} observation.}
    \label{fig:flux_select}
\end{figure}

\begin{table}[!ht]
    \centering
    \caption{\textbf{Observational log of \xmm\ observed bursts we identify for our analysis.} Column 2 lists the redshift, Column 3 lists the trigger instrument, Column 4 lists the trigger time and Column 5 shows the start time (shown as elapsed time post-trigger, $\Delta$t) and duration of the \xmm\ observation in ks, Column 6 lists the start time of the optical observation (shown again as elapsed time post trigger $\Delta$t$_{optical}$) in hour, and Column 7 lists the spectrograph that performed the observation.}
    \begin{tabular}{c c | c c | c | c c }
        \toprule 
        & & \multicolumn{2}{c|}{Prompt emission} & \multicolumn{1}{c|}{X-ray afterglow} & \multicolumn{2}{c}{Optical afterglow} \\
        \cmidrule{3-7}
        GRB & \textit{z} & Trigger instrument & Trigger time & $\Delta$t (+duration) & $\Delta$t$_{optical}$ & Spectrograph \\
        & & & (UTC) & (post-trigger, ks) & (post-trigger, h) & \\
        \midrule
        060729 & 0.54 & \textit{Swift-BAT} & 29-07-2006 19:12:29 & 45(+60) & 13.5 & \textit{VLT/FORS2}  \\
        061121 & 1.314 & \textit{Swift-BAT} & 21-11-2006 15:22:29 & 22(+40) & 0.2 & \textit{Keck/LRIS} \\ 
        080411 & 1.031 & \textit{Swift-BAT} & 11-04-2008 21:15:32 & 60(+100) & 2.4 & \textit{VLT/FORS1} \\
        090618 & 0.54 & \textit{Swift-BAT} & 18-06-2009 08:08:29 & 19(+23) & 0.3 & \textit{LICK/KAST} \\
        120711A & 1.405 & \textit{INTEGRAL-IBIS} & 11-07-2012 02:44:48 & 78(+52) & 7.5 & \textit{Gemini/GMOS-S} \\
        190114C & 0.425 & \textit{Swift-BAT} & 14-01-2019 20:57:03 & 28(+55) & 4.8 & \textit{VLT/X-Shooter} \\
        \midrule
        221009A & 0.151 & \textit{Fermi-GBM} & 09-10-2022 13:16:59 & 400(+60)\footnotemark[1] & 11.6 & \textit{VLT/X-Shooter} \\
        \botrule
    \end{tabular}
    \footnotetext[1]{This is the second \textit{XMM} observation, given the extreme luminosity, the first observation was performed with the thick filter.}
    \label{tab:sample}
\end{table}

The full observational log of the sample of GRBs that we identify for our analysis are presented in Supplementary Tab \ref{tab:sample}. We checked the homogeneity of our sample selection by plotting and comparing the the lightcurves of all \textit{Swift} observed GRBs with known redshift, in flux units in Supplementary Fig. \ref{fig:flux_select} in grey. The lightcurves of our sample GRBs are plotted in colour, and the horizontal bars indicate the duration of the \xmm\ observation. We have clearly selected the GRBs that are bright \textit{at the starting time} of \xmm\ observations.

\subsection{Modelling the GRB photoionising output}

The original TEPID GRB model \cite{L22} used a parametric light curve built upon the compilation of 650 \textit{Swift} GRBs presented in \cite{2013MNRAS.428..729M}. The lightcurve consists of an initial constant luminosity phase (the prompt) lasting 85 sec, followed by a $t^{-2}$ luminosity decrease until 100 s and, after that, a $t^{-1}$ decrease (the afterglow).
\begin{table}[h]
    \caption{\textbf{0.3-10 keV energetics per GRB} Energies are integrated up to 100s for the prompt phase and from 100s to 1 Ms for the afterglow (in rest-frame time). For the actual lightcurve, we integrate the real luminosity light curve and normalise the parametric curve so to have the same prompt energy. We list the total irradiated energy (as E(prompt) +E(afterglow)), the fractions irradiated in the two phases, and the fraction up to midpoint of the XMM observation.}
    \label{tab:spec_analysis}      
    \begin{tabular}{@{}c|cccc|cccc@{}}        
    \toprule 
    GRB & \multicolumn{4}{|c}{Actual lightcurve} & \multicolumn{4}{|c}{Parametric LC of \cite{L22}} \\
    \midrule
     & Total & $\frac{Prompt}{Total}$ & $\frac{Afterglow}{Total}$ & E$_{t_{obs}}$ (frac) & Total & $\frac{Prompt}{Total}$ & $\frac{Afterglow}{Total}$ & E$_{t_{obs}}$ (frac) \\
    \midrule 
    060729 & 8.0 $\times\ 10^{52}$ & 0.96 & 0.04 & 7.9 $\times\ 10^{52} $ (0.99) & 6.2 $\times\ 10^{52}$ & 0.6 & 0.4 & 5.4 $\times\ 10^{52}$ (0.9)\\ 
    061121 & 3.0 $\times\ 10^{53}$ & 0.99 & 0.01 & 3.0 $\times\ 10^{53} $ (0.99) & 2.5 $\times\ 10^{53}$ & 0.6 & 0.4 & 1.9 $\times\ 10^{53}$ (0.8) \\
    080411 & 8.7 $\times\ 10^{54}$ & 0.98 & 0.02 & 8.7 $\times\ 10^{54} $ (0.99) & 6.8 $\times\ 10^{54}$ & 0.6 & 0.4 & 6.0 $\times\ 10^{54}$ (0.9) \\
    090618 & 9.6 $\times\ 10^{53}$ & 0.64 & 0.36 & 8.7 $\times\ 10^{53}$ (0.9) & 7.2 $\times\ 10^{53}$ & 0.6 & 0.4 & 6.9 $\times\ 10^{53}$ (0.9) \\
    120711A & 1.9 $\times\ 10^{54}$ & 0.2 & 0.8 & 1.6 $\times\ 10^{54}$ (0.8) & 1.6 $\times\ 10^{53}$ & 0.6 & 0.4 & 1.4 $\times\ 10^{53}$ (0.9) \\
    190114C & 4.6 $\times\ 10^{53}$ & 0.97 & 0.03 & 4.6 $\times\ 10^{53}$ (0.9) & 7.3 $\times\ 10^{53}$ & 0.6 & 0.4 & 6.4 $\times\ 10^{53}$ (0.9) \\
    221009A & - & - & - & - & 2.0 $\times\ 10^{54}$ & 0.63 & 0.4 & 1.7 $\times\ 10^{54}$ (0.9) \\
    \botrule
    \end{tabular}
\end{table}

We summarise the differences obtained in the total energy budgets by using the parametric versus the actual light curves in Supplementary Tab. \ref{tab:spec_analysis} for the bursts in our sample. For comparing the two approaches we normalised the parametric curve so to have the same prompt energy. We report the fractions of the total ionising energies injected into the medium in the prompt phase (integrating up to 100 s), the fraction in the afterglow phase (integrating from 100s to the 1 Ms) and the fraction injected in the medium up to the midpoint of the \xmm\ observation.

In the case of computations that use the actual light curve, the fraction of energy in the prompt phase is consistently higher than the corresponding value for the parametric light curve. This difference arises because the assumed temporal evolution is flat in the parametric light curve, compared to the highly variable, even increasing, evolution observed in the actual lightcurves. Conversely, the parametric lightcurve appears to consistently overestimate the fraction of the energy present in the afterglow. This is because in the actual lightcurve, the flux decays much steeper than the fiducial index of -1 in the parametric lightcurve, and thus, the resulting luminosity integral up to longer times is smaller. 

However, crucially, if the parametric lightcurve is correctly normalised to match the integrated energy from the prompt phase light curve, it can be seen that the total energy (prompt+afterglow) is consistent within a factor of a few from either approach. Additionally, it can also be seen that the fraction of energy injected into the medium up to the point of a typical \xmm\ observation ($\gtrsim$ 10 ks, rest-frame) is equal between either approach. Thus, the impact on the TEPID results obtained by using either light curve model should not be significant. However, using the actual GRB lightcurve completely eliminates potential errors introduced through normalisation matching the prompt phase lightcurve and allows to track the observed temporal behaviour of the ionising luminosity much more accurately than using the parametric lightcurve approach. Therefore, throughout this paper, we have used the actual GRB lightcurve, when possible. 

We briefly outline the effects of increasing time and ionising energy on the modelled absorption spectrum. We show in Supplementary Fig. \ref{fig:pop_sampl} the distribution of the modelled spectra at a fiducial time of 10 ks (rest-frame) for increasing total ionising energy (left panel) and the distribution for increasing time using the parametric light curve.

\begin{figure}[!ht]
    \centering
    \includegraphics[width=\linewidth]{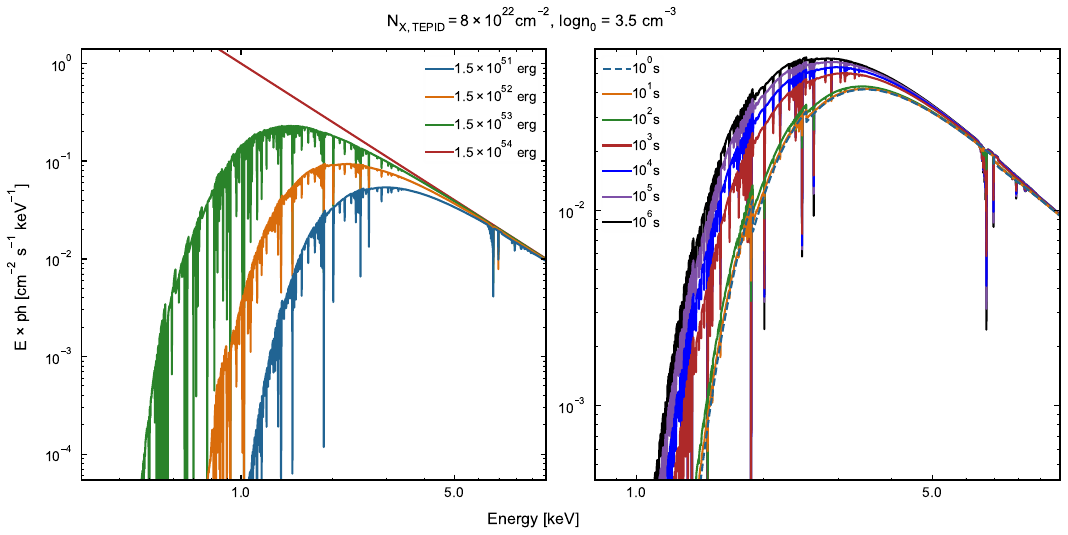}
    \caption{\textbf{Evolution of the TEPID modelled soft X-ray spectrum}. \textit{Left panel}: as a function of the total irradiated $0.3-10$ keV energy, at a time of 10 ks post-GRB. \textit{Right panel}: as a function of time for a fixed $0.3-10$ keV energy of $10^{51}$ erg. These spectra were produced assuming the parametric light curve of \cite{L22}. Note the different scales of the y-axis in either panel.}
    \label{fig:pop_sampl}
\end{figure}

In both cases, we set \nh=$8\times10^{22}$ cm$^{-2}$ and \logn = 3.5 cm$^{-3}$ as representative of the best-fit results of our sample. For the spectra in the right panel and the blue spectrum in the left panel, we assume a total $0.3-10$ keV ionising energy of $1.5\times10^{51}$ erg. Therefore, the photon spectra shown in the left panel of Fig. \ref{fig:pop_sampl} are essentially showing the effect of increasing total ionising energy at the time of the \xmm\ observation over four orders of magnitude. With increasing energy, the medium becomes steadily more transparent and becomes totally transparent for an energy of $10^{54}$ erg. 

The time evolution, on the other hand, is less drastic. While the overall transparency of the medium increases with increasing time, the most dramatic jump is observed from 10$^{0}$ s to 10$^{3}$ s. This dramatic increase is driven by the flat evolution of the prompt emission up to 100 sec, as the medium becomes steadily and rapidly ionised with a constant flux of ionising photons. As a result of this, the innermost region of the cloud becomes totally ionised. After this time, the increase in ionisation is much less drastic, although the higher ionisation states keep becoming populated. This is because after the first 100 s, we enter the afterglow monotonic decay in which the number of ionising photons also reduces. For what concerns this paper, the time evolution expected at the time of \textit{XMM} observations (between 30 and 110 ksec, observer frame) and on longer time scales and later is small (a factor $\lesssim 0.1$, except for GRB 060729 which has some re-brightening in the afterglow) from 10 ksec to 1 Msec. Indeed, the energy in the prompt+afterglow up to the observation time in the GRB rest-frame is typically $\gtrsim 75-90\%$ of the total energy up to 1 Msec.

\subsection{Analysis of the metal absorption lines from the optical spectrum}

\begin{sidewaystable}[h]
    \caption{\textbf{Analysis of metal absorption lines in the optical spectra}. For each GRB, we list instrument that secured the spectrum, derived Doppler parameter $b$, ions and their equivalent widths, wavelengths, and oscillator strengths, the ionic column, the ionic fraction correction, the dust correction factor and the final estimate of \nho. For the first three GRBs, EWs are taken from \citep{2009ApJS..185..526F}. Spectra for 090618 and 120711A are from private comm. and the spectrum of 190114C is from \citep{2019GCN.23710....1K}.}
    \label{tab:ew_analysis}      
    \addtolength{\tabcolsep}{-0.4em}
    \begin{tabular}{lccccccccccc}   
    \toprule
    GRB & Spectrograph & $b$ & Ion & $EW_{rest}$ & $\lambda_{rest}$ & $f_{oscillator}$ & logN$_{ion}$ & Ionic fraction & Dust fraction & logN$_{H, optical}$ & logN$_{H, optical}$ \\
     & & ($km/s$) &  & (\AA) & (\AA) &  & (cm$^{-2}$) & & (dex) & (per ion, cm$^{-2}$) & (cm$^{-2}$) \\
    \midrule
    060729 & VLT/FORS2\footnotemark[1] & 22.4 & \ion{Fe}{II} & $0.45 \pm 0.13$ & 2600.2 & 0.239 & $13.7\pm0.3$ & 1.0 & $0.9\pm0.6$ & $19.13\pm0.7$ & 19.13 $\pm 0.5$\\
    & & & \ion{Mg}{II} & $1.09 \pm 0.04$ & 2796.3 & 0.608 & $14.5\pm0.3$ & 0.75 & $0.5\pm0.3$ & $19.12\pm0.4$ \\
    & & & & $0.73 \pm 0.04$ & 2803.5 & 0.303 &  &  &  &  \\ 
    & & & \ion{Mg}{I} & $0.59 \pm 0.04$ & 2853.0 & 1.8 & $13.1\pm0.3$ & & - \\ 
    & & & \ion{Ca}{II} (K) & $0.75 \pm 0.03$ & 3934.8 & 0.65 & $13.3\pm0.3$ &  & - \\
    & & & \ion{Ca}{II} (H) & $0.55 \pm 0.03$ & 3969.6 & 0.322 & - & - & - \\
    & & & & & & & & & & \\
    061121 & Keck/LRIS\footnotemark[1] & 63.0 & \ion{Mg}{I} & $0.21 \pm 0.02$ & 1827.9 & 0.024 & $14.5\pm0.3$ & 1.0 & - \\
    & & & & $2.46 \pm 0.04$ & 2853 & 1.8 &  &  & - \\
    & & & \ion{Mg}{II} & $1.5 \pm 0.04$ & 2796.4 & 0.608 & $14.0\pm0.3$ & - & $0.5\pm0.3$ & $19.8 \pm 0.4$ \\
    & & & & $1.2 \pm 0.03$ & 2803.5 & 0.303 & & & & \\
    & & 71.2 & \ion{Mg}{II} & $2.93 \pm 0.04$ & 2796.4 & 0.608 & $15.6\pm0.3$ & 0.75 & $0.5\pm0.3$ & $20.6 \pm 0.4$ & 21.4$\pm0.5$ \\
    & & & & $3.00 \pm 0.03$ & 2803.5 & 0.303 & & & & \\
    & & & \ion{Cr}{II} & $0.73 \pm 0.03$ & 2056.3 & 0.105 & $14.5\pm0.3$ & 0.9 & $0.6\pm0.3$ & $21.5\pm0.4$ \\
    & & & & $0.47 \pm 0.03$ & 2066.2 & 0.051 &  & & & \\
    & & & \ion{Fe}{II} & $1.84 \pm 0.03$ & 1608.5 & 0.059 & $16.7\pm0.3$ & 1.0 & $0.9\pm0.6$ & $22.1\pm0.7$ \\
    & & & & $2.60 \pm 0.03$ & 2374.5 & 0.031 & & & & \\
    & & & & $3.20 \pm 0.09$ & 2586.7 & 0.069 & & & & \\
    & & & \ion{Ni}{II} & $0.33 \pm 0.03$ & 1709.6 & 0.032 & $14.7\pm0.3$ & 1.0 & $1.0\pm0.6$ & $21.5\pm0.7$ \\
    & & & & $0.53 \pm 0.03$ & 1741.6 & 0.043 & & & & \\
    & & & & $0.31 \pm 0.03$ & 1751.9 & 0.029 & & & & \\
    & & & \ion{Mn}{II} & $1.33 \pm 0.04$ & 2576.9 & 0.351 & $14.17\pm0.3$ & 1.0 & $0.6\pm0.5$ & $21.3\pm0.6$ \\
    & & & & $1.30 \pm 0.03$ & 2594.5 & 0.271 & & & & \\
    & & & & & & & & & \\
    080411 & VLT/FORS1\footnotemark[1] & 18.6 & \ion{Fe}{II} & $0.34 \pm 0.03$ & 2249.9 & 0.0074 & $15.7\pm0.3$ & 1.0 & $0.9\pm0.4$ & $21.1\pm0.7$ & 20.5$\pm0.5$ \\
    & & & & $0.33 \pm 0.03$ & 2260.8 & 0.0024 & & & & \\
    & & & \ion{Mg}{II} & $1.54 \pm 0.07$ & 2796.3 & 0.6080 & $15.4 \pm 0.3$ & 0.75 & $0.5\pm0.3$ & $20.4\pm0.4$ \\
    & & & & $1.58 \pm 0.07$ & 2803.5 & 0.303 & & & & \\
    & & & \ion{Mn}{II} & $0.22\pm0.02$ & 2576.9 & 0.35 & $13.2 \pm 0.3 $ & 1.0 & $0.6\pm0.5$ & $20.4\pm0.6$ \\
    & & & \ion{Ca}{II} (K) & $1.1 \pm 0.15$ & 3934.8 & 0.65 & $13.6\pm0.3$ & - & - \\
    & & & \ion{Ca}{II} (H) & $1.1 \pm 0.15$ & 3969.6 & 0.322 &  & - & - &  \\
    & & & \ion{Mg}{I} & $0.6\pm0.05$ & 2853.0 & 1.8 & $13.7\pm0.3$ & - & - \\
    & & & & & & & & & & \\
    090618 & P200/DBSP\footnotemark[2] & 40.6 & \ion{Fe}{II} & $1.87 \pm 0.5$ & 2600.2 & 0.239 & $15.6\pm0.3$ & 1.0 & $0.9\pm0.6$ & $21.0\pm0.7$ & 20.7$\pm0.6$ \\
    & & & \ion{Mg}{II} & $1.70 \pm 0.2$ & 2796.3 & 0.608 & $15.5\pm0.3$ & 0.75 & $0.5\pm0.3$ & $20.4\pm0.5$ \\
    & & & & $1.92 \pm 0.3$ & 2803.5 & 0.303 & & & & \\
    & & &&  & & & & & & \\
    120711A & Gemini/GMOS-S\footnotemark[2] & 58.7 & \ion{Fe}{II} & $1.51 \pm 0.12$ & 2374.5 & 0.031 & $15.6\pm0.3$ & 1.0 & $0.9\pm0.6$ & $21.0\pm0.7$ & $21.0\pm0.7$ \\
    & & & & $2.6 \pm 0.20$ & 2387.7 & 0.32 &  &  &  & \\
    & & & & $1.63 \pm 0.13$ & 2586.5 & 0.0691 &  &  &  & \\
    & & & & $2.01 \pm 0.09$ & 2600.2 & 0.239 &  &  &  & \\
    & & & & & & & & & & \\
    190114C & VLT/X-Shooter \footnotemark[3] & 121.9 & \ion{Ca}{II} (K) & $3.94 \pm 0.1$ & 3934.8 & 0.65 & $14.83\pm0.3$ & - & - \\
    & & & \ion{Ca}{II} (H) & $3.61 \pm 0.07$ & 3969.6 & 0.322 &  & - & - &  \\
    & & & \ion{Mg}{II} & $5.22 \pm 0.32$ & 2796.3 & 0.608 & $15.8\pm0.3$ & 0.75 & $0.7\pm0.2$ & $20.8\pm0.4$ & 20.8$\pm0.4$ \\
    & & & & $5.0 \pm 0.20$ & 2803.5 & 0.303 &  &  &  & \\
    & & & \ion{Mg}{I} & $4.01\pm0.29$ & 2853.0 & 1.8 & $14.3\pm0.3$ & - & - \\
    \botrule
    \end{tabular}
\end{sidewaystable}

\begin{figure}[h]
\centering
  \includegraphics[width=0.49\textwidth]{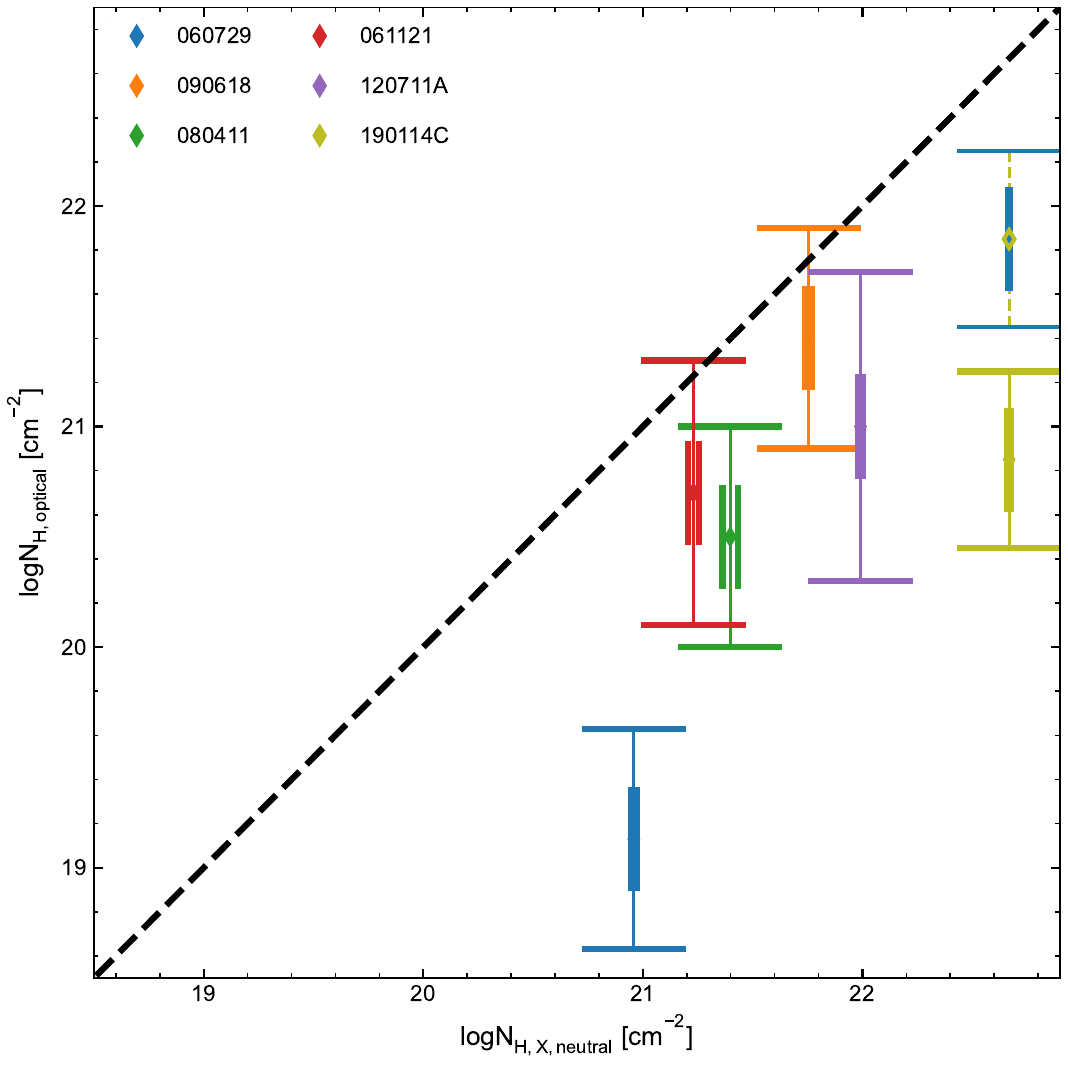}
  \includegraphics[width=0.49\textwidth]{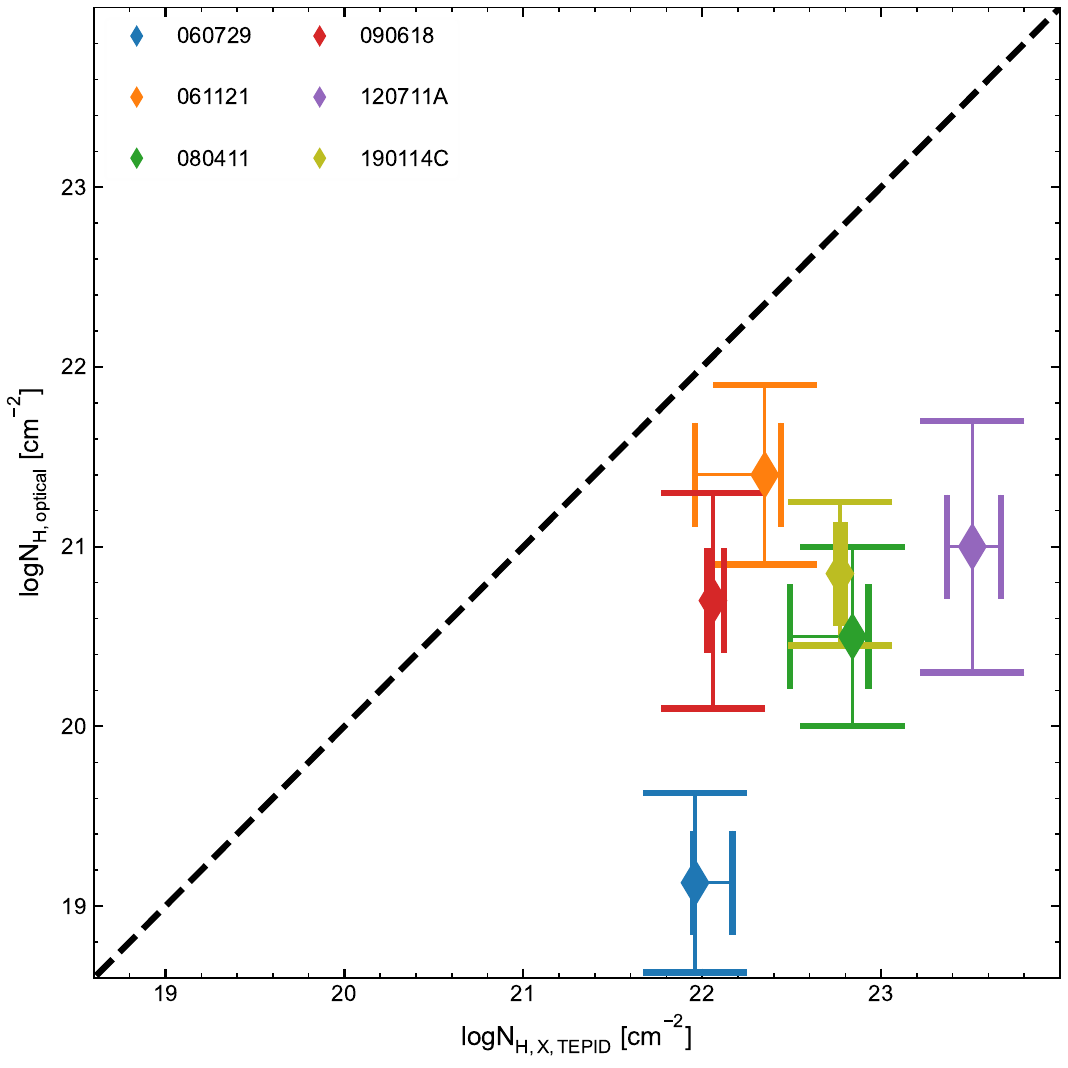}
  \caption{\textbf{Comparison of column densities from X-ray and optical spectra} \textit{Left panel}: X-ray versus optical neutral hydrogen equivalent column densities. X-ray columns are from the neutral best-fit, optical columns are from metal absorption lines. The dashed line for GRB 190114C shows the \nho\ range assuming a higher dust correction (details in Methods).
  \textit{(Right panel)} TEPID versus optical column densities for the GRBs of our sample. TEPID columns are from the \tepi\ best-fit assuming solar metallicity. In both panels, the dashed line marks where the two columns are equal. For both panels, the errors are at 1-$\sigma$.}
  \label{fig:optical_columns}
\end{figure}

We compare the distributions of \nho\ and the best-fit \nhc\ in the left panel of Supplementary Fig \ref{fig:optical_columns}, finding that for all six bursts with metal absorption lines in their spectra (except 221009A), the optical columns fall short by a factor of $\sim 3-100$ compare to the neutral X-ray column. This factor is in fact higher ($\sim 10-1000 $) when comparing the best-fit \nhx\ with the \nho, as shown in the right panel of Fig. \ref{fig:optical_columns}. Such an excess is fully consistent with the previous results of \citep{2007ApJ...660L.101W} and \citep{2011A&A...525A.113S}.

\begin{figure}[h]
    \centering
    \includegraphics[width=\linewidth]{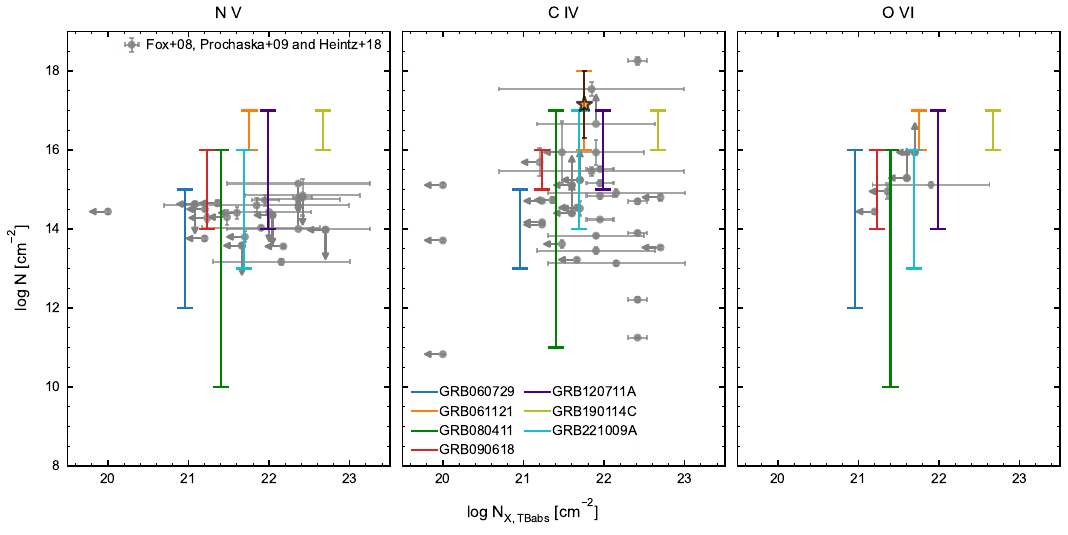}
    \caption{\textbf{TEPID predicted moderate ionisation columns versus high resolution optical measurements.} From left to right, the panels plot \ion{N}{V}, \ion{C}{IV} and \ion{O}{VI} from the combined sample of \cite{2008A&A...491..189F}, \cite{2008ApJ...685..344P} and \cite{2018MNRAS.479.3456H}, with the 1-$\sigma$ errors when there is a measurement. The TEPID predictions for our sample are color-coded per GRB, and are derived for the 90\% TEPID confidence region. We also overplot the actual measurement of C IV for GRB 061121 as derived from a blended, low-resolution detection reported in \cite{2009ApJS..185..526F} as an orange star with black 1-$\sigma$ error bars. The distribution of the TEPID predicted columns for our sample are fully consistent with the observed distribution as derived from high-resolution spectroscopy.}
    \label{fig:cols_optical}
\end{figure}
\clearpage

\subsection{Results of spectral fitting}

This section includes a table summarising the results of the \tepiO\ fits and fit-specific figures for each burst in our sample. These include the count spectra with the best-fit model predictions for neutral and TEPID models (Supplementary Fig. 10-16), the photon spectra of the best-fit neutral and TEPID absorber (Supplementary Fig. 17-23), the ionic column distributions for the TEPID best-fit parameters (Supplementary Fig. 24-31, and the individual contour plots of the MCMC chains (Supplementary Fig. 32-28). We include separate plots for the TEPID-only and TEPID+optical best-fits of GRB 061121 since the results for these fits are different. 

\begin{figure}[b]
    \centering
    \includegraphics[width=0.49\textwidth]{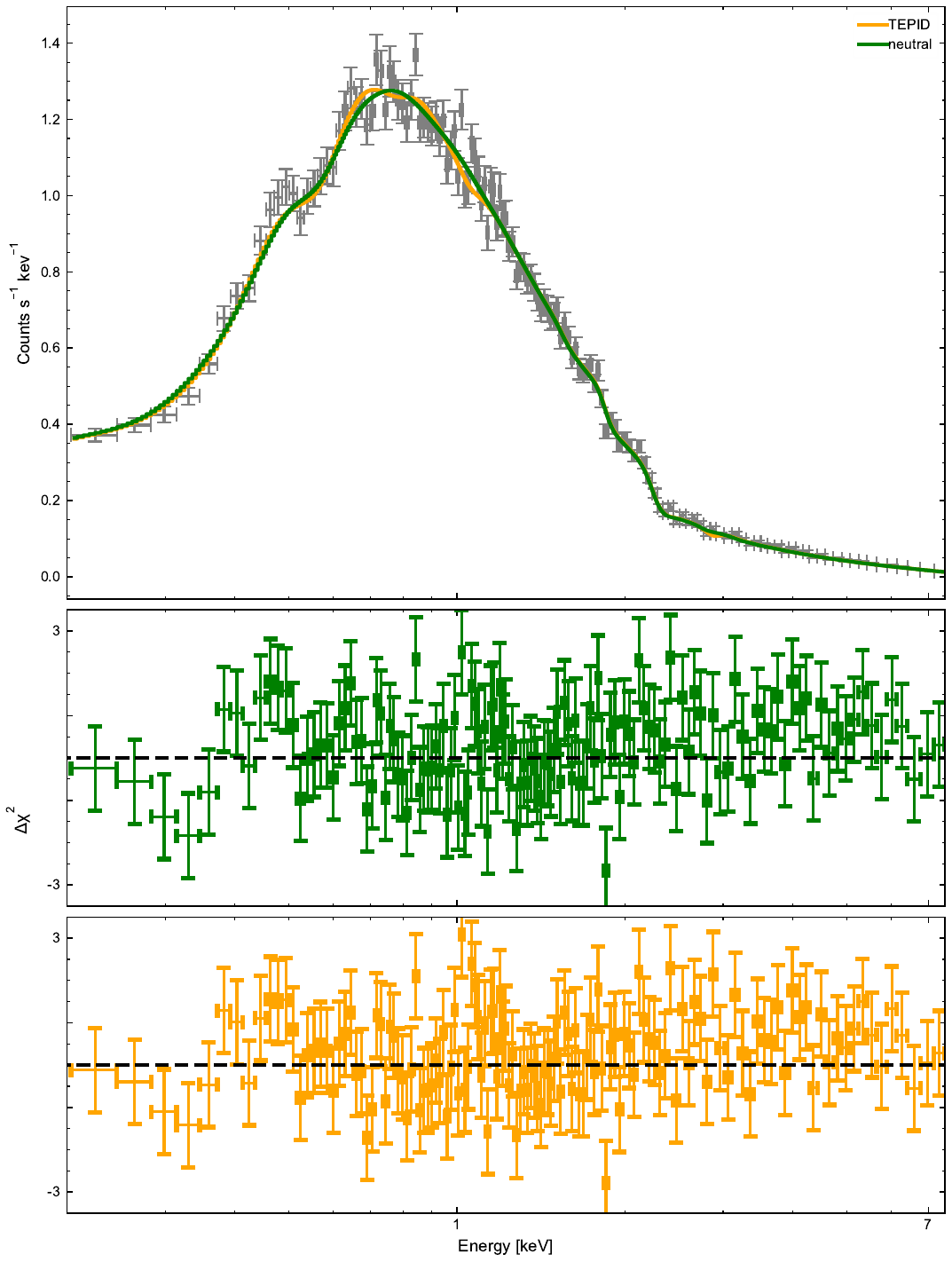}
    \caption{\textbf{EPIC-PN count spectrum of GRB 061121 with folded model predictions}: \textit{(Top panel)}: The EPIC-pn spectrum is plotted in grey points and has been rebinned for plotting purposes. The y-errors represent the 1-sigma uncertainty on the count-rate per bin, and the x-errors represent the bin size. The best-fit model predictions of the neutral model and TEPID model are plotted as green and orange lines respectively. \textit{(Middle and bottom panels)}: Corresponding residuals for the neutral and TEPID model, respectively, colour-coded as above. The residuals below 1 keV seen from the neutral model-fit are not present in the TEPID model-fit, showing the improvement in the fit.}
    \label{fig:count_spec2}
\end{figure}

\begin{figure}
    \centering
    \includegraphics[width=0.49\textwidth]{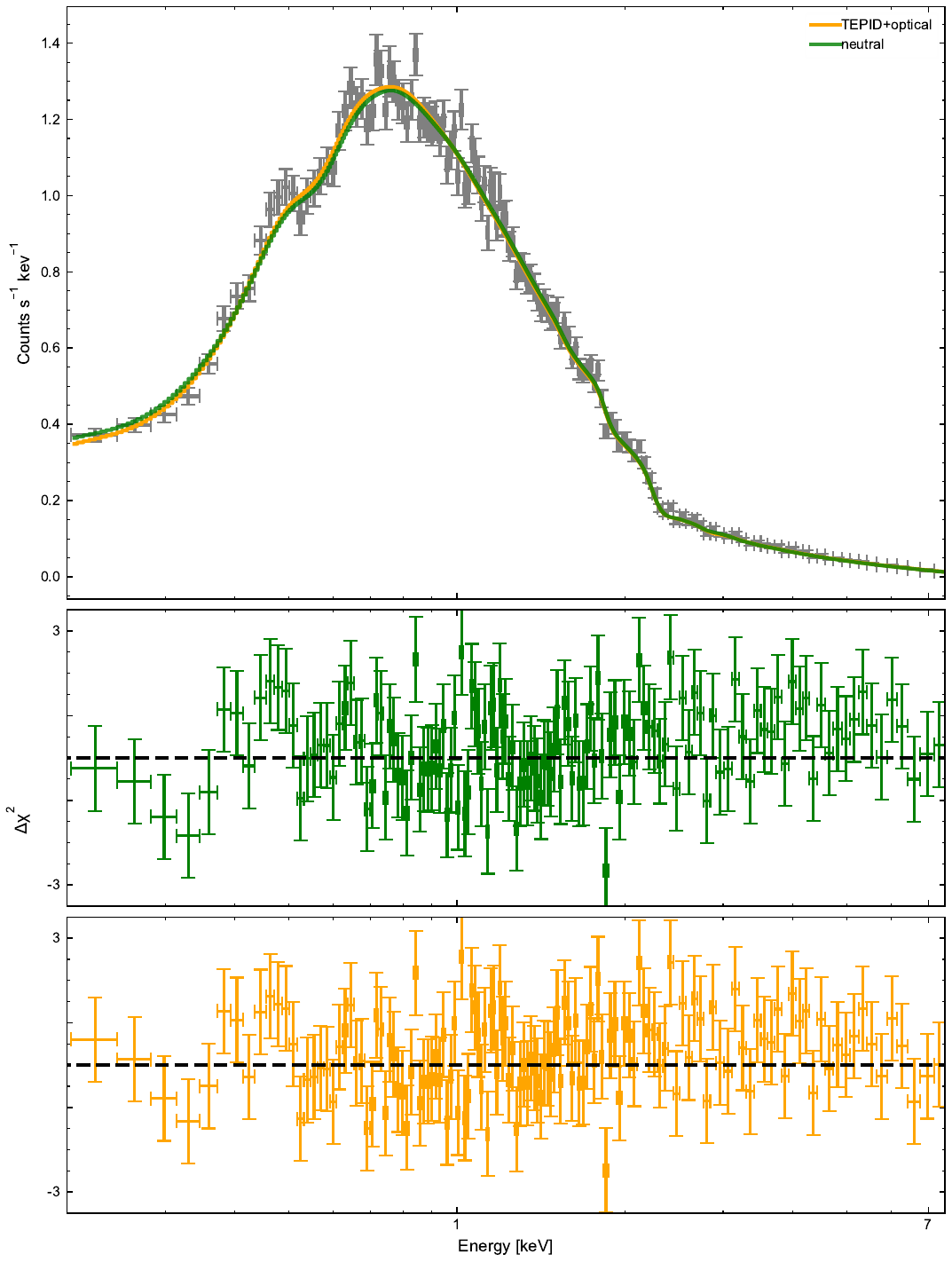}
    \caption{\textbf{EPIC-PN count spectrum of GRB 061121 with folded model predictions}: \textit{(Top panel)}: The EPIC-pn spectrum is plotted in grey points and has been rebinned for plotting purposes. The y-errors represent the 1-sigma uncertainty on the count-rate per bin, and the x-errors represent the bin size. The best-fit model predictions of the neutral model and TEPID+optical model are plotted as green and orange lines respectively. \textit{(Middle and bottom panels)}: Corresponding residuals for the neutral and TEPID model, respectively, colour-coded as above. The residuals below 1 keV seen from the neutral model-fit are not present in the TEPID model-fit, showing the improvement in the fit.}
    \label{fig:count_spec2opt}
\end{figure}

\begin{figure}
    \centering
    \includegraphics[width=0.49\textwidth]{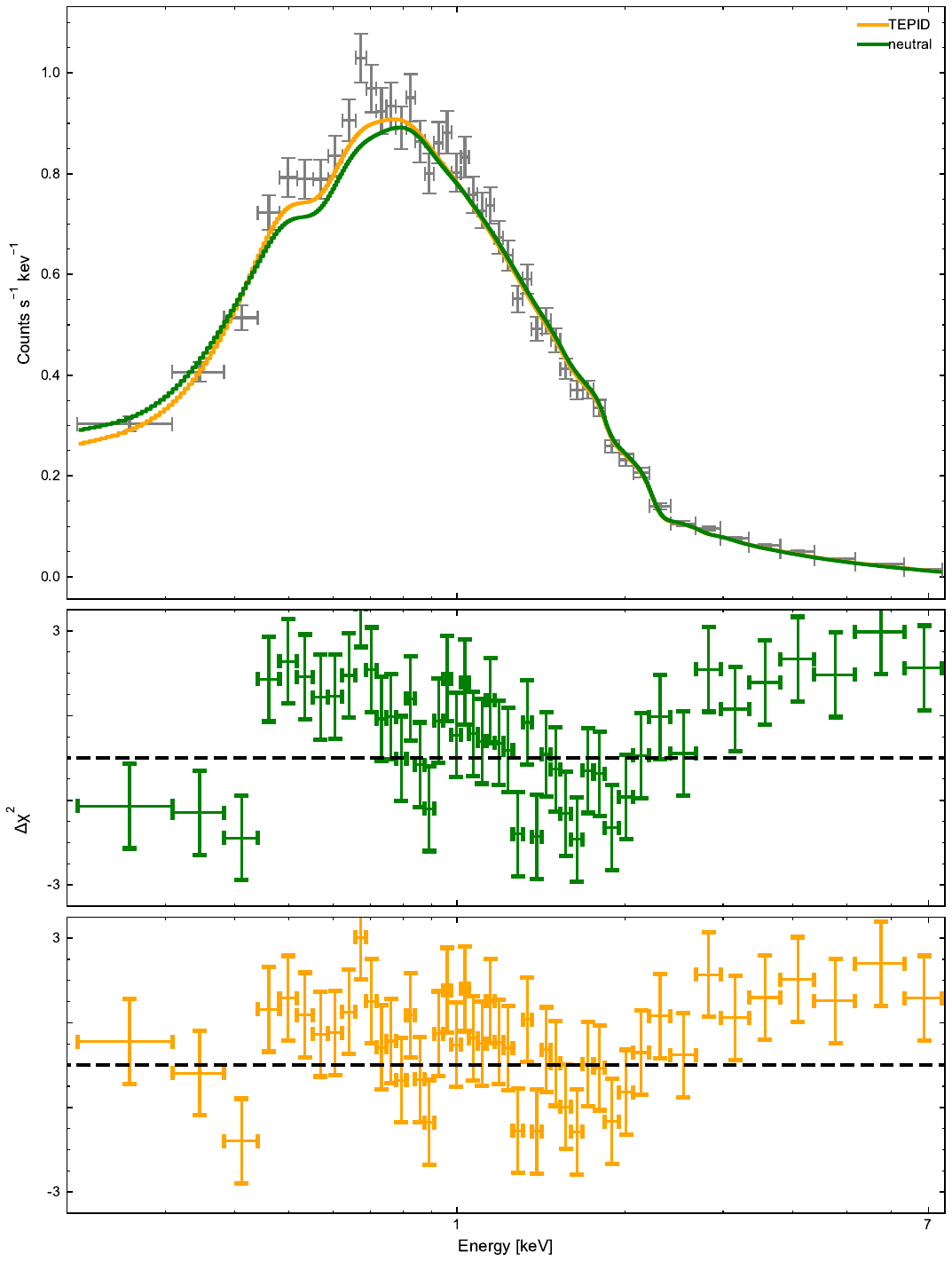}
    \caption{\textbf{EPIC-PN count spectrum of GRB 080411 with folded model predictions}: \textit{(Top panel)}: The EPIC-pn spectrum is plotted in grey points and has been rebinned for plotting purposes. The y-errors represent the 1-sigma uncertainty on the count-rate per bin, and the x-errors represent the bin size. The best-fit model predictions of the neutral model and TEPID model are plotted as green and orange lines respectively. \textit{(Middle and bottom panels)}: Corresponding residuals for the neutral and TEPID model, respectively, colour-coded as above. The residuals below 1 keV seen from the neutral model-fit are not present in the TEPID model-fit, showing the improvement in the fit.}
    \label{fig:count_spec3}
\end{figure}

\begin{figure}
    \centering
    \includegraphics[width=0.49\textwidth]{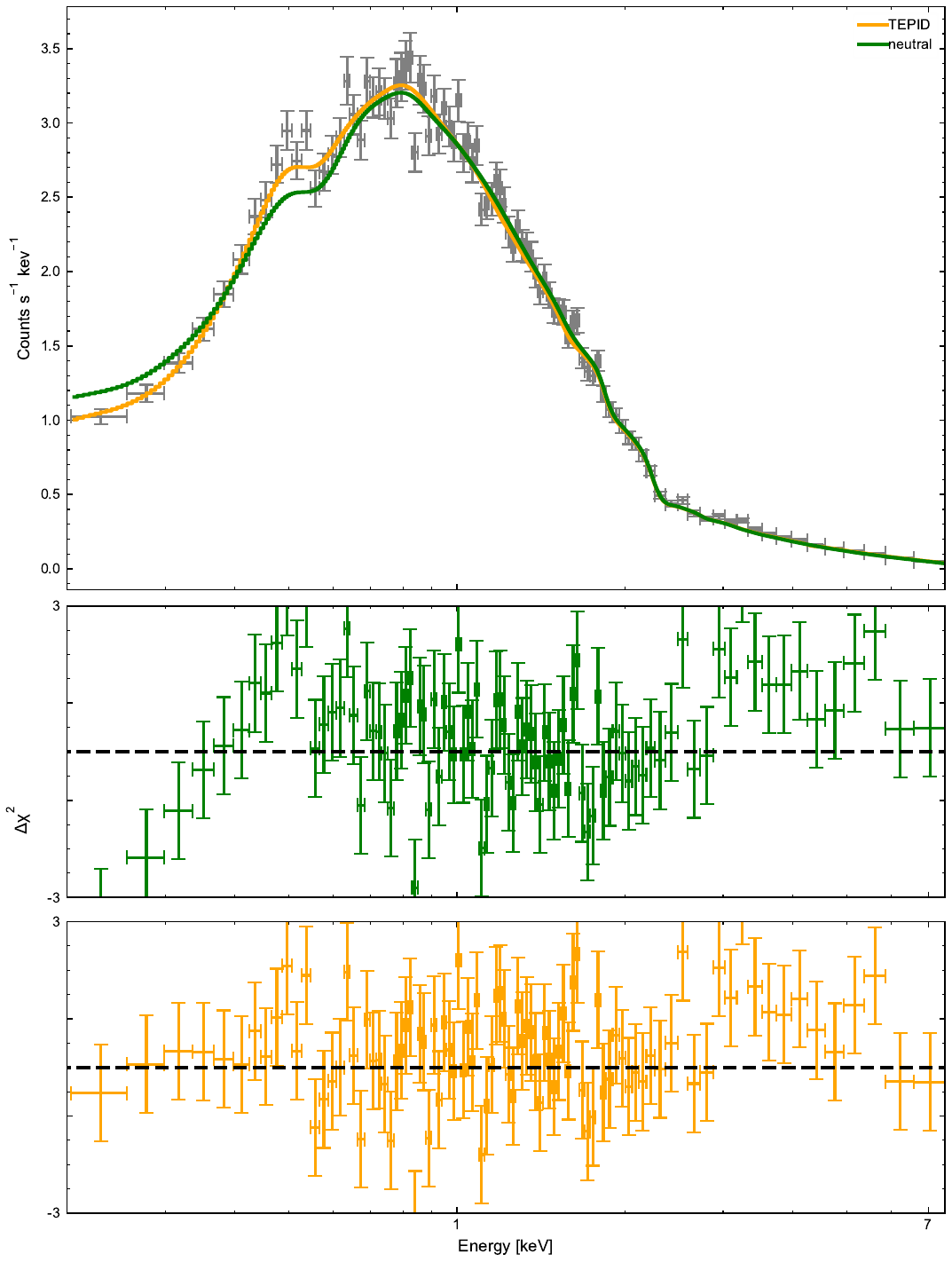}
    \caption{\textbf{EPIC-PN count spectrum of GRB 090618 with folded model predictions}: \textit{(Top panel)}: The EPIC-pn spectrum is plotted in grey points and has been rebinned for plotting purposes. The y-errors represent the 1-sigma uncertainty on the count-rate per bin, and the x-errors represent the bin size. The best-fit model predictions of the neutral model and TEPID model are plotted as green and orange lines respectively. \textit{(Middle and bottom panels)}: Corresponding residuals for the neutral and TEPID model, respectively, colour-coded as above. The residuals below 1 keV seen from the neutral model-fit are not present in the TEPID model-fit, showing the improvement in the fit.}
    \label{fig:count_spec4}
\end{figure}

\begin{figure}
    \centering
    \includegraphics[width=0.49\textwidth]{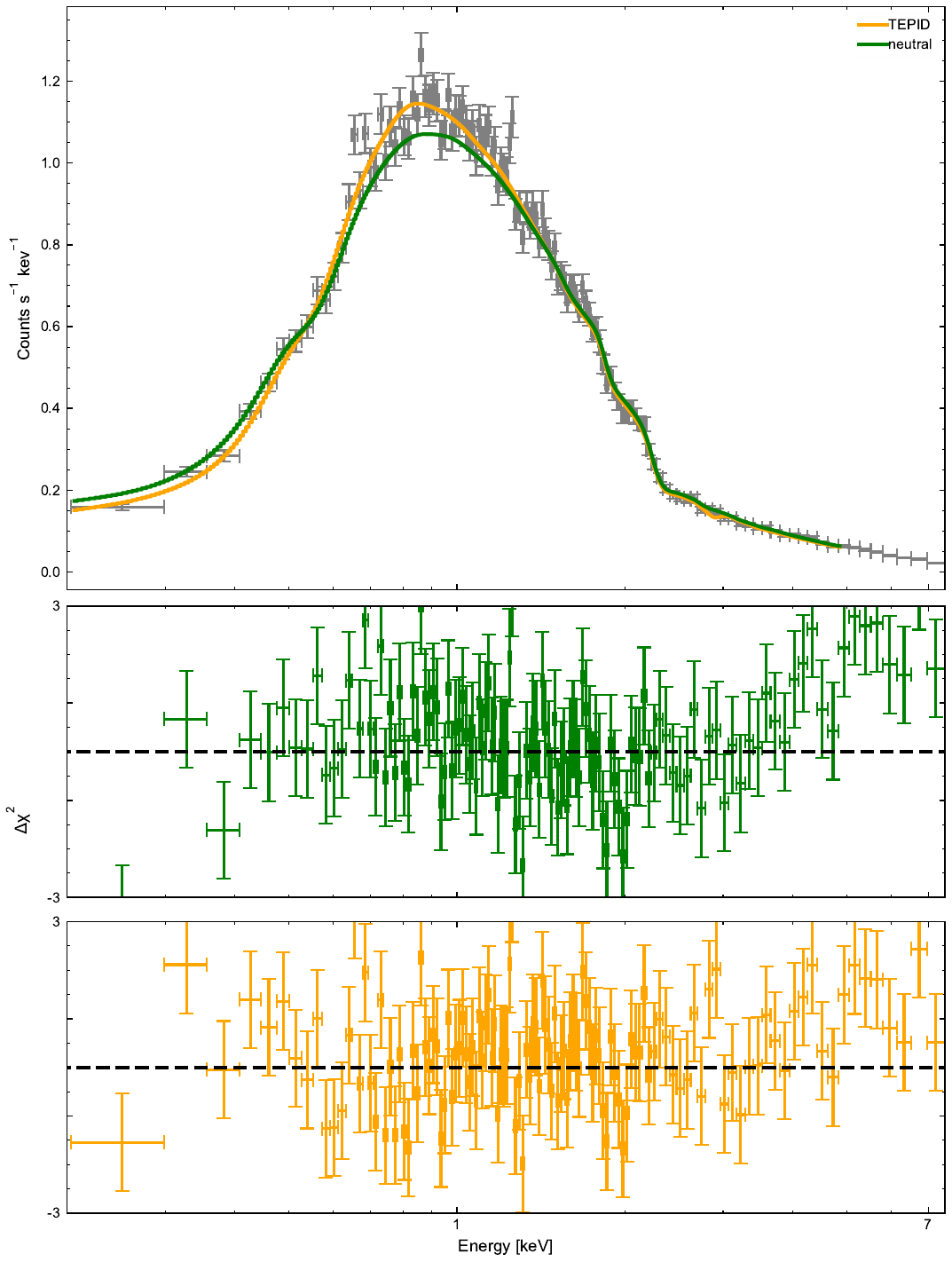}
    \caption{\textbf{EPIC-PN count spectrum of GRB 120711A with folded model predictions}: \textit{(Top panel)}: The EPIC-pn spectrum is plotted in grey points and has been rebinned for plotting purposes. The y-errors represent the 1-sigma uncertainty on the count-rate per bin, and the x-errors represent the bin size. The best-fit model predictions of the neutral model and TEPID model are plotted as green and orange lines respectively. \textit{(Middle and bottom panels)}: Corresponding residuals for the neutral and TEPID model, respectively, colour-coded as above. The residuals below 1 keV seen from the neutral model-fit are not present in the TEPID model-fit, showing the improvement in the fit.}
    \label{fig:count_spec5}
\end{figure}

\begin{figure}
    \centering
    \includegraphics[width=0.49\textwidth]{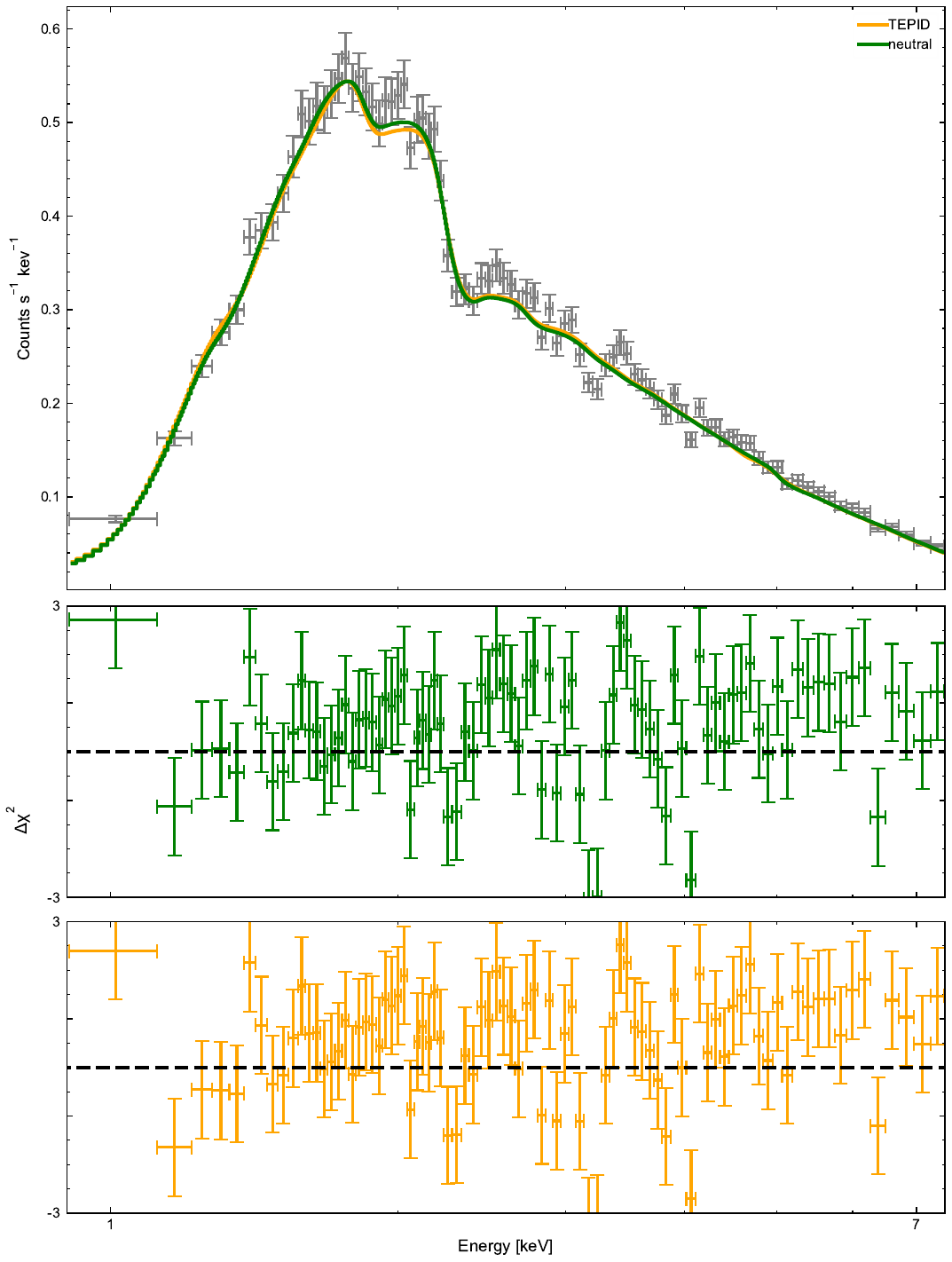}
    \caption{\textbf{EPIC-PN count spectrum of GRB 190114C with folded model predictions}: \textit{(Top panel)}: The EPIC-pn spectrum is plotted in grey points and has been rebinned for plotting purposes. The y-errors represent the 1-sigma uncertainty on the count-rate per bin, and the x-errors represent the bin size. The best-fit model predictions of the neutral model and TEPID model are plotted as green and orange lines respectively. \textit{(Middle and bottom panels)}: Corresponding residuals for the neutral and TEPID model, respectively, colour-coded as above. The residuals below 1 keV seen from the neutral model-fit are not present in the TEPID model-fit, showing the improvement in the fit.}
    \label{fig:count_spec7}
\end{figure}

\begin{figure}
    \centering
    \includegraphics[width=0.49\textwidth]{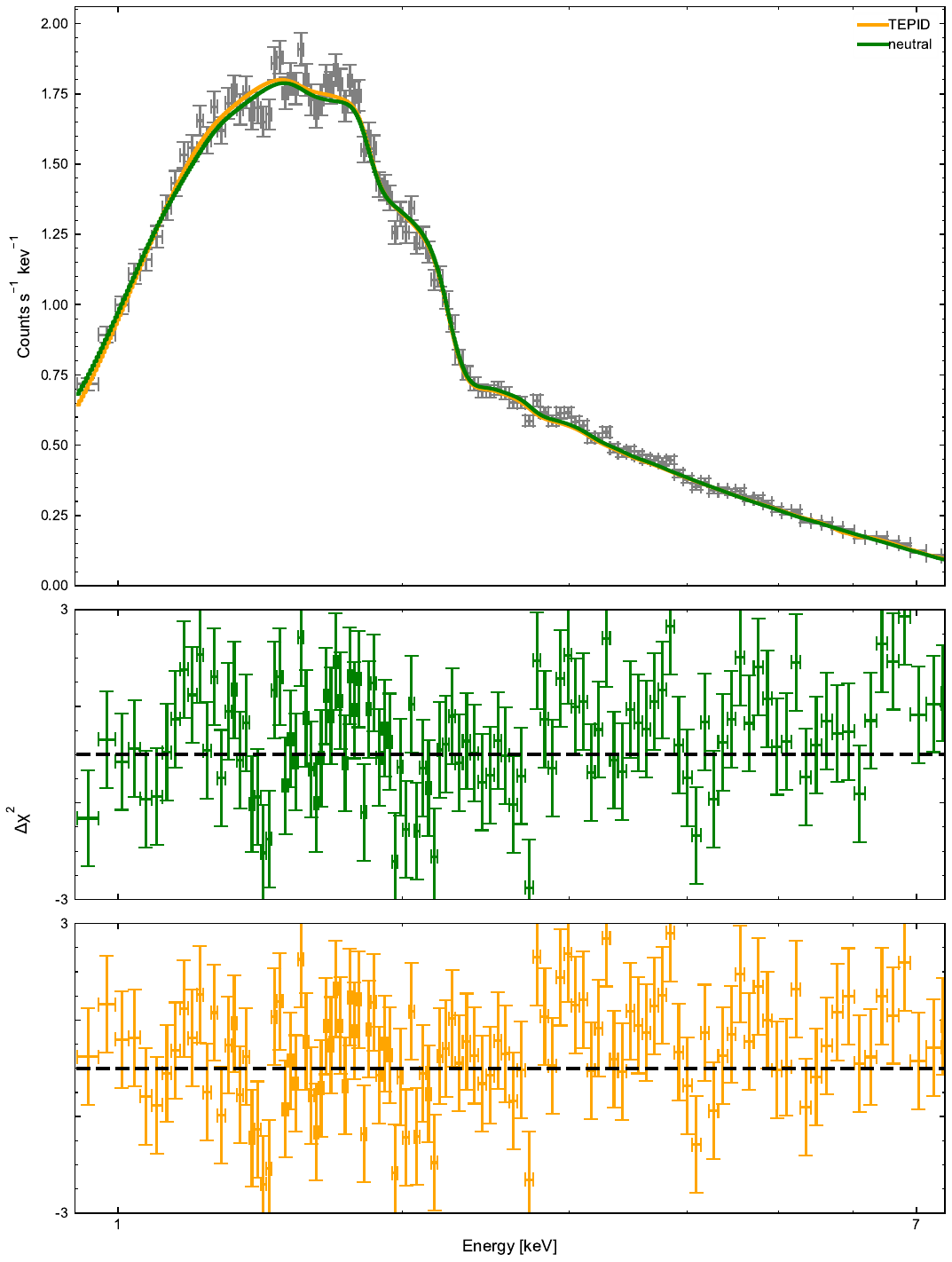}
    \caption{\textbf{EPIC-PN count spectrum of GRB 221009A with folded model predictions}: \textit{(Top panel)}: The EPIC-pn spectrum is plotted in grey points and has been rebinned for plotting purposes. The y-errors represent the 1-sigma uncertainty on the count-rate per bin, and the x-errors represent the bin size. The best-fit model predictions of the neutral model and TEPID model are plotted as green and orange lines respectively. \textit{(Middle and bottom panels)}: Corresponding residuals for the neutral and TEPID model, respectively, colour-coded as above. The residuals below 1 keV seen from the neutral model-fit are not present in the TEPID model-fit, showing the improvement in the fit.}
    \label{fig:count_spec6}
\end{figure}
\clearpage

\begin{table}[h]
\caption{Summary of the best-fit results of the \tepiO\ fit with solar metallicity and and \tepiO\ fit with free metallicity including the fit statistics corresponding to the best fit in each case and significance estimators. The summary of neutral model fits is also included for reference. All errors are quoted at 90\%. GRB 221009A is not included in this table as due to the lack of suitable redshifted metal lines, we cannot estimate an \nho.}
\centering
\begin{tabular}{c|c c c c c c}
\toprule
Parameters & \multicolumn{5}{c}{Value}
\\
GRB & 060729 & 061121 & 080411 & 090618 & 120711A & 190114C\\
$z$ & 0.54 & 1.314 & 1.031 & 0.54 & 1.405 & 0.425\\
\midrule
& \multicolumn{6}{c}{\textbf{{\sc TBabs} fit}} \\
\midrule
\nhc\ $(10^{21} {cm}^{-2})$ & $0.91\pm0.03$ & $5.7\pm0.2$ & $2.5\pm0.1$ & $1.7\pm0.04$ & $9.8\pm0.1$ & $46\pm1$ \\
$\alpha_{neutral}$ & $2.1\pm0.01$ & $2.02\pm0.02$ & $2.05\pm0.03$ & $1.98\pm0.02$ & $1.95\pm0.02$ & $1.98\pm0.04$ \\
$\chi^2/DOF$ & 1249/1191 & 767/783 & 580/528 & 807/741 & 937/881 & 967/950  \\
AIC & 1260 & 784 & 598 & 820 & 950 & 963 \\
\midrule
& \multicolumn{6}{c}{\textbf{\tepiO\ fit}, solar metallicity} \\
\midrule
\nhx\ $(10^{21} cm^{-2})$ & 9.5$^{+13.4}_{-1.2}$ & 44.2$^{+114.2}_{-39.8}$ & 64.2$^{+54.6}_{-17.0}$ & 11.7$^{+71.4}_{-3.8}$ & 323.6$^{+238.7}_{-141.6}$ & 56.5$^{+8.3}_{-2.3}$ \\
\logn\ $(cm^{-3})$ & 2.8$^{+0.3}_{-0.2}$ & 3.5$^{+0.4}_{-2.4}$ & 3.2$^{+0.6}_{-0.8}$ & 2.4$^{+0.5}_{-0.2}$ & 4.5$^{+0.2}_{-0.7}$ & 1.7$^{+0.5}_{-0.6}$ \\
size (pc) & 6$^{+1}_{-1.3}$ & 6$^{+74}_{-2}$ & 15$^{+23}_{-7}$ & 15$^{+3}_{-7}$ & 5$^{+5}_{-1}$ & 340$^{+1013}_{-205}$ \\
$\alpha_{ionised}$ & 2.05$^{+0.02}_{-0.02}$ & $1.99^{+0.05}_{-0.01}$ & $1.99^{+0.03}_{-0.03}$ & $1.93^{+0.03}_{-0.02}$ & $1.86^{+0.03}_{-0.01}$ & $2.01^{+0.04}_{-0.04}$ \\
$\chi^2/DOF$ & 1118/1189 & 760/781 & 545/526 & 758/739 & 894/879 & 977/948 \\
$\Delta \chi^2$ ($\Delta DOF = 2$) & 131 & 7 & 35 & 49 & 43 & -10 \\
AIC & 1123 & 776 & 564 & 772 & 903 & 969 \\
BF & $10^{31}$ & 166 & $10^{7}$ & $10^{10}$ & $10^{8}$ & 0.01 \\
\midrule
& \multicolumn{6}{c}{\textbf{\tepiO\ fit}, free metallicity} \\
\midrule
\nhx\ $(10^{21} cm^{-2})$ & 8.1$^{+14.5}_{-0.9}$ & 41.6$^{+109.6}_{-37.8}$ & 71.9$^{+49.8}_{-22.8}$ & 11.5$^{+50.2}_{-3.0}$ & 323.6$^{+232.9}_{-151.9}$ & 100.0$^{+16.1}_{-36.9}$ \\
\logn\ $(cm^{-3})$ & 2.7$^{+0.3}_{-0.1}$ & 3.3$^{+0.5}_{-2.2}$ & 3.2$^{+0.7}_{-0.9}$ & 2.4$^{+0.4}_{-0.2}$ & 4.4$^{+0.3}_{-0.6}$ & 2.1$^{+0.1}_{-0.7}$ \\
size (pc) & 7$^{+1}_{-2}$ & 7$^{+71}_{-3}$ & 14$^{+26}_{-6}$ & 16$^{+4}_{-7}$ & 5$^{+4}_{-1}$ & 231$^{+788}_{-42}$ \\
$\alpha_{ionised}$ & 2.05$^{+0.02}_{-0.02}$& $1.99^{+0.05}_{-0.02}$ & $1.99^{+0.03}_{-0.03}$ & $1.93^{+0.02}_{-0.03}$ & 1.86$^{+0.03}_{-0.01}$ & $1.97^{+0.05}_{-0.03}$ \\
$\chi^2/DOF$ & 1117/1188 & 760/780 & 545/525 & 759/738 & 893/878 & 974/947 \\
$\Delta \chi^2$ ($\Delta DOF = 3$) & 130 & 7 & 36 & 49 & 43 & -11 \\
AIC & 1118 & 777 & 565 & 772 & 903 & 973 \\
BF & $10^{32}$ & 318 & $10^{8}$ & $10^{11}$ & $10^{10}$ & 0.01 \\
    \botrule
    \end{tabular}
    \label{tab:master}
\end{table}

\begin{figure}
    \centering
    \includegraphics[width=0.7\textwidth]{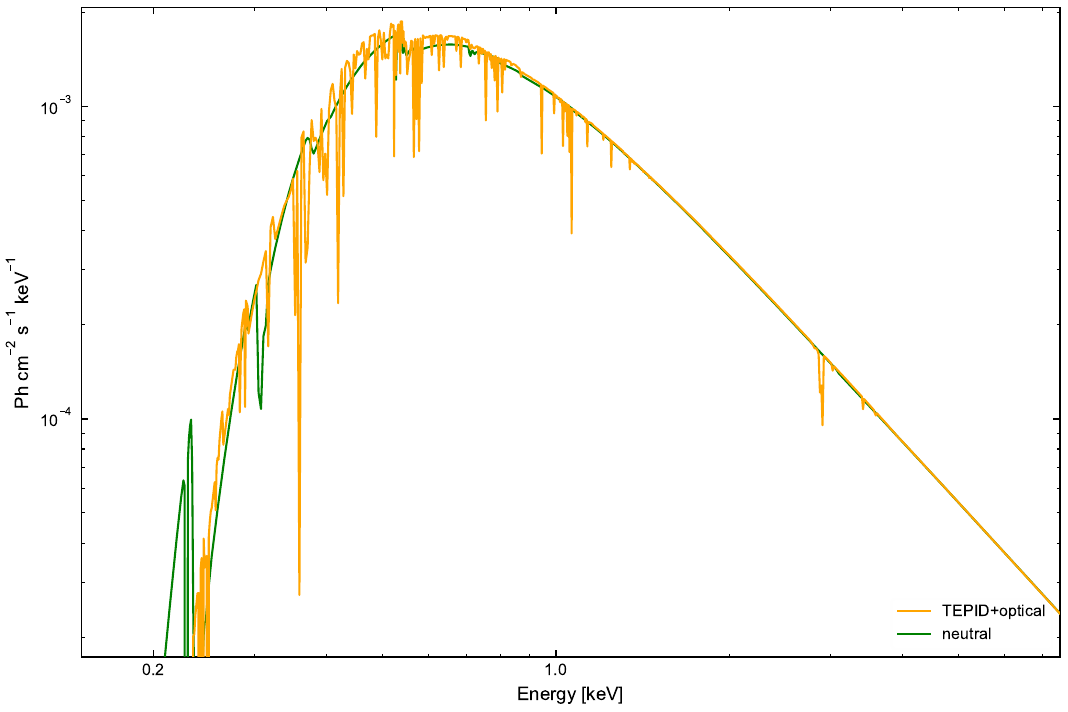}
    \includegraphics[width=0.7\textwidth]{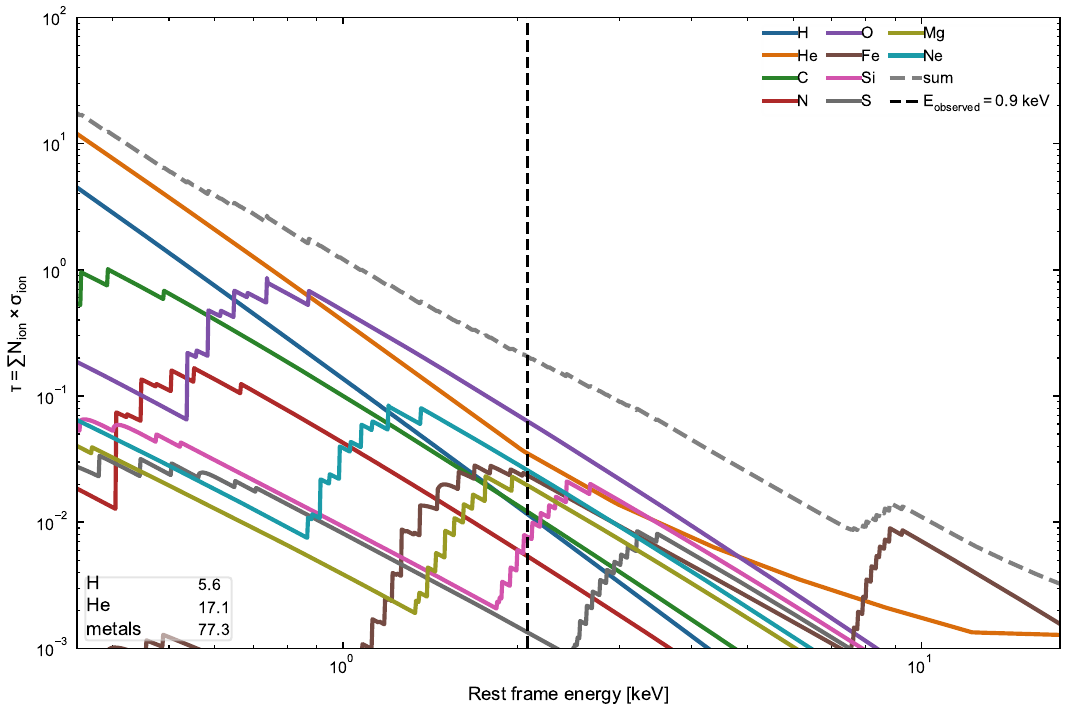}
    \caption{\textbf{Photon spectrum and element-wise opacity breakdown for the GRB 061121 TEPID-only best-fit} \textit{Top panel}: The TEPID best-fit model photon spectrum is plotted in orange as a function of observed energy. For comparison, the best-fit neutral model is plotted in green. \textit{Bottom panel}: Opacity contributions per element (summed over all ions) as a function of rest-frame energy. At $E_{rest}< 1$ keV, H and He are contributing to the absorption, instead at $E_{rest} >$ 1 keV metal contributions start to become significant. The vertical dashed line marks the E$_{obs} = 0.9$ keV, with percent contributions of H, He and metals reported in the bottom left.}
    \label{fig:phot_spec2}
\end{figure}
\clearpage
\begin{figure}
    \centering
    \includegraphics[width=0.7\textwidth]{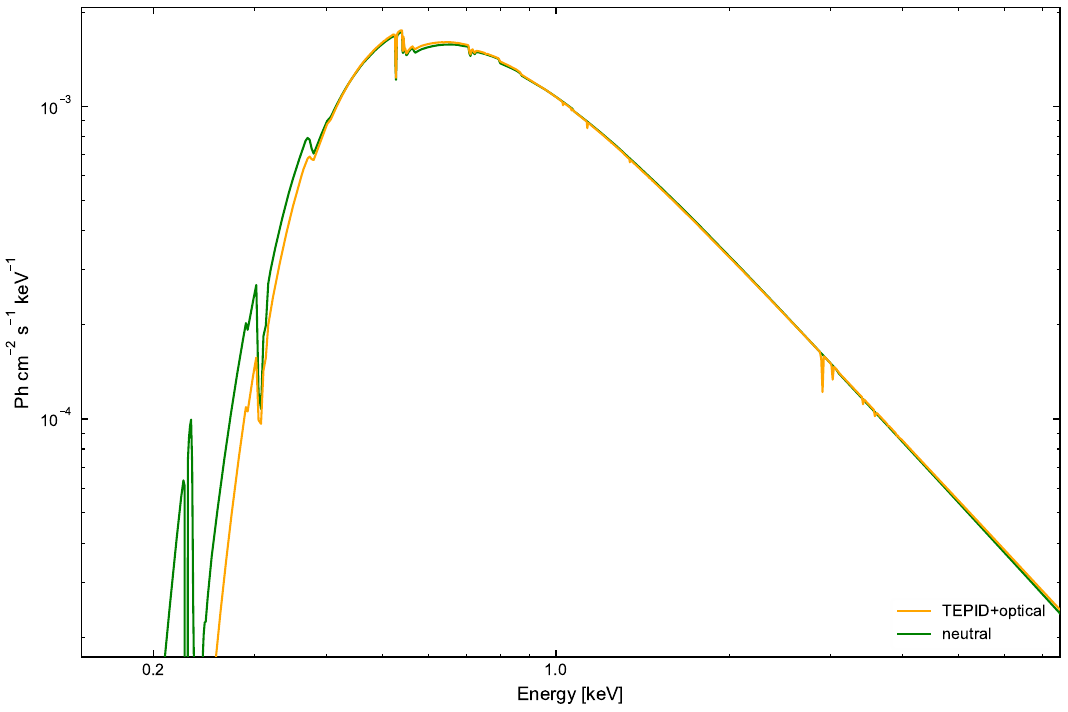}
    \includegraphics[width=0.7\textwidth]{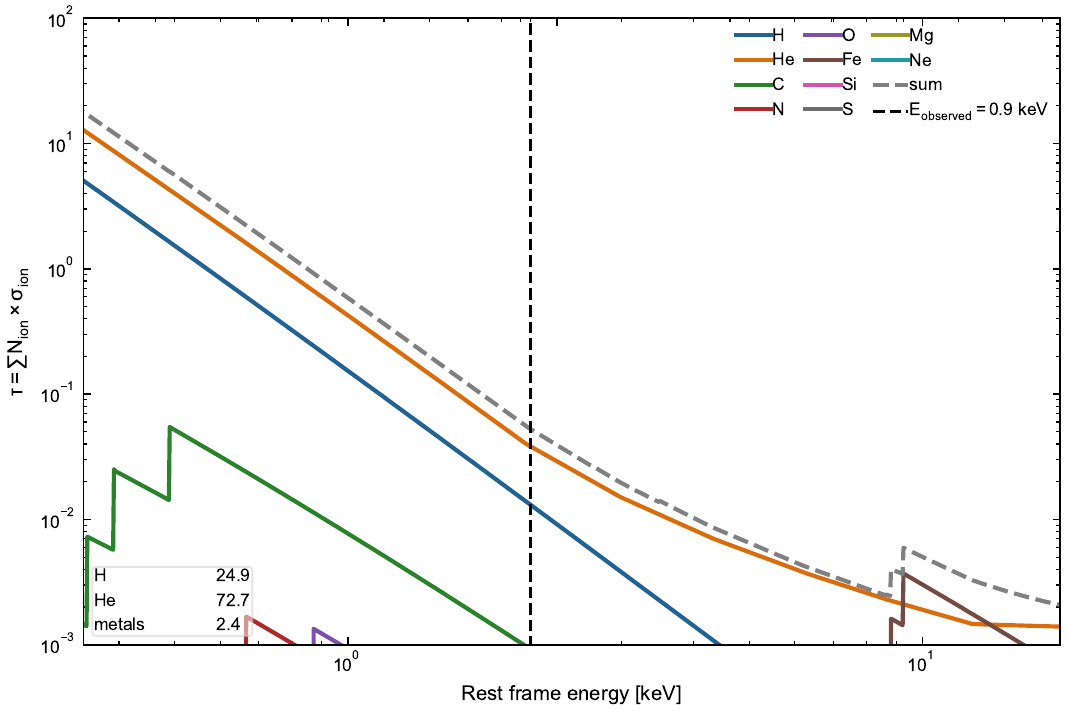}
    \caption{\textbf{Photon spectrum and element-wise opacity breakdown for the GRB 061121 TEPID+optical best-fit} \textit{Top panel}: The TEPID best-fit model photon spectrum is plotted in orange as a function of observed energy. For comparison, the best-fit neutral model is plotted in green. \textit{Bottom panel}: Opacity contributions per element (summed over all ions) as a function of rest-frame energy. At $E_{rest}< 1$ keV, H and He are contributing to the absorption, instead at $E_{rest} >$ 1 keV metal contributions start to become significant. The vertical dashed line marks the E$_{obs} = 0.9$ keV, with percent contributions of H, He and metals reported in the bottom left.}
    \label{fig:phot_spec3}
\end{figure}
\clearpage
\begin{figure}
    \centering
    \includegraphics[width=0.7\textwidth]{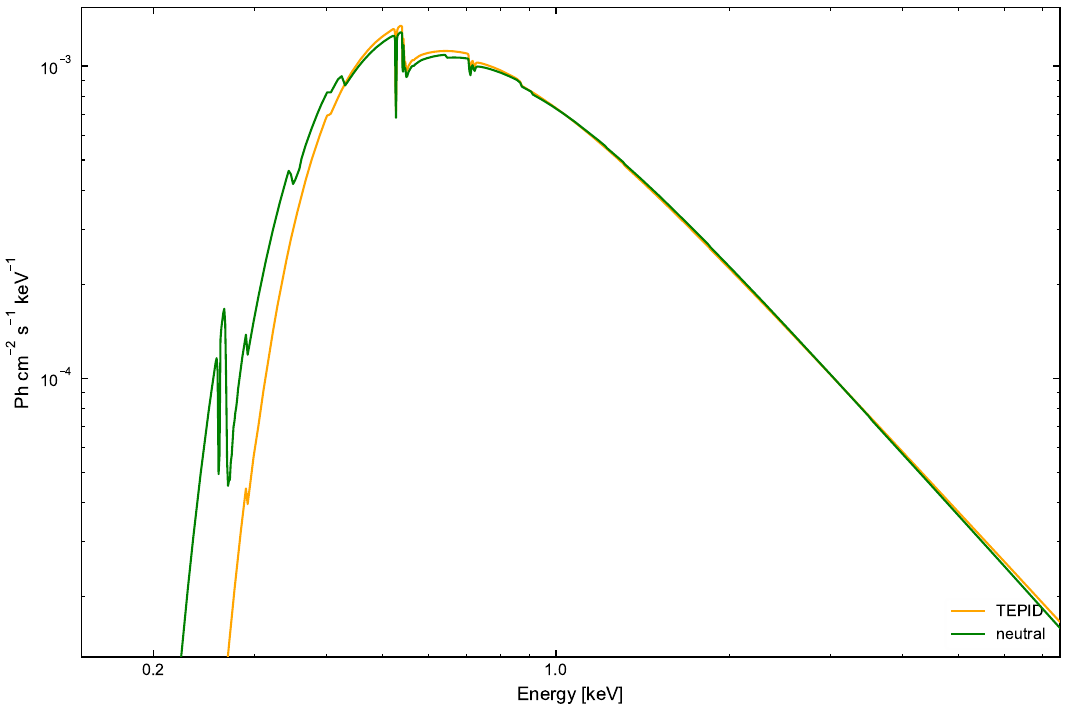}
    \includegraphics[width=0.7\textwidth]{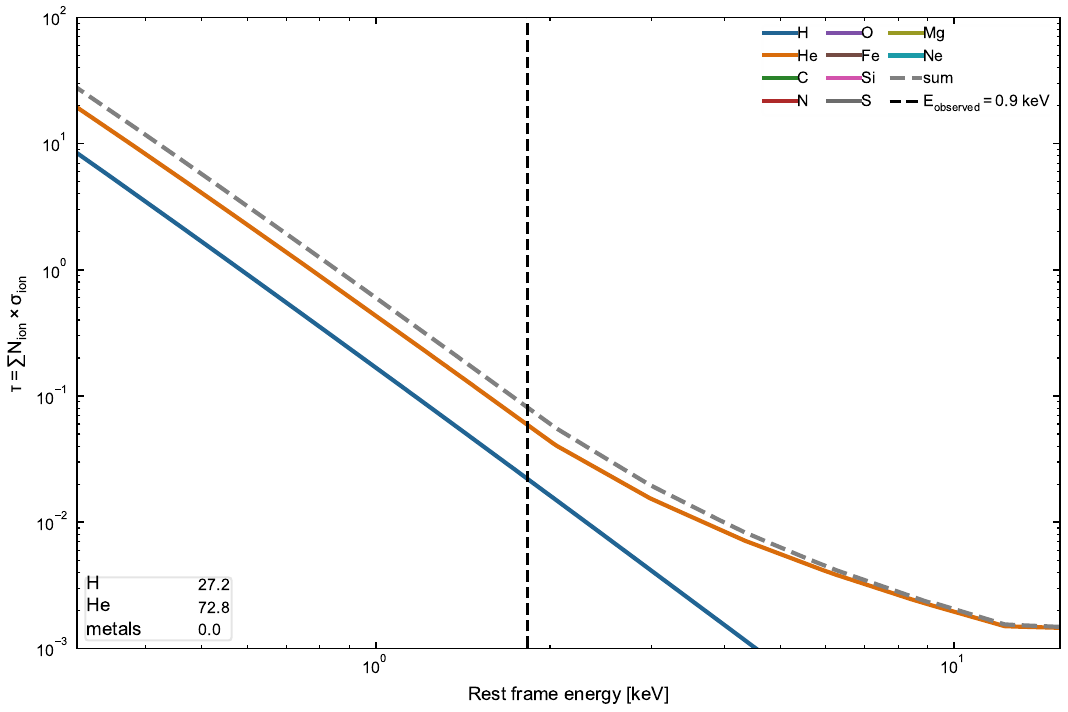}
    \caption{\textbf{Photon spectrum and element-wise opacity breakdown for the GRB 080411 TEPID-only best-fit} \textit{Top panel}: The TEPID best-fit model photon spectrum is plotted in orange as a function of observed energy. For comparison, the best-fit neutral model is plotted in green. \textit{Bottom panel}: Opacity contributions per element (summed over all ions) as a function of rest-frame energy. At $E_{rest}< 1$ keV, H and He are contributing to the absorption, instead at $E_{rest} >$ 1 keV metal contributions start to become significant. The vertical dashed line marks the E$_{obs} = 0.9$ keV, with percent contributions of H, He and metals reported in the bottom left.}
    \label{fig:phot_spec4}
\end{figure}
\clearpage

\begin{figure}
    \centering
    \includegraphics[width=0.7\textwidth]{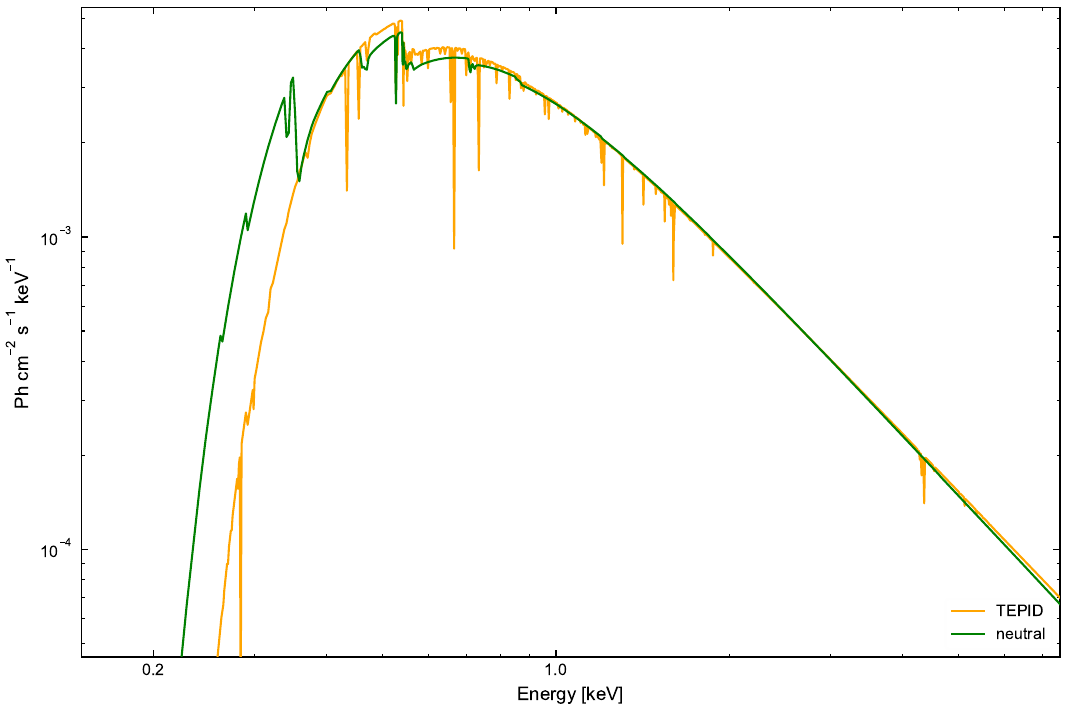}
    \includegraphics[width=0.7\textwidth]{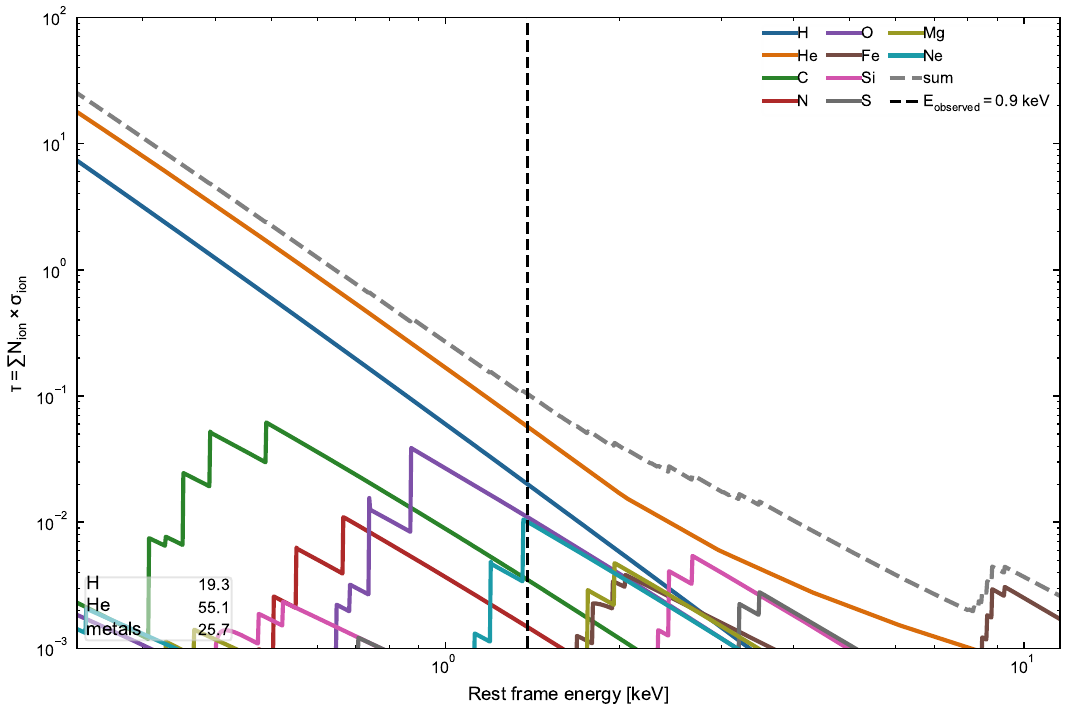}
    \caption{\textbf{Photon spectrum and element-wise opacity breakdown for the GRB 090618 TEPID-only best-fit} \textit{Top panel}: The TEPID best-fit model photon spectrum is plotted in orange as a function of observed energy. For comparison, the best-fit neutral model is plotted in green. \textit{Bottom panel}: Opacity contributions per element (summed over all ions) as a function of rest-frame energy. At $E_{rest}< 1$ keV, H and He are contributing to the absorption, instead at $E_{rest} >$ 1 keV metal contributions start to become significant. The vertical dashed line marks the E$_{obs} = 0.9$ keV, with percent contributions of H, He and metals reported in the bottom left.}
    \label{fig:phot_spec5}
\end{figure}
\clearpage

\begin{figure}
    \centering
    \includegraphics[width=0.7\textwidth]{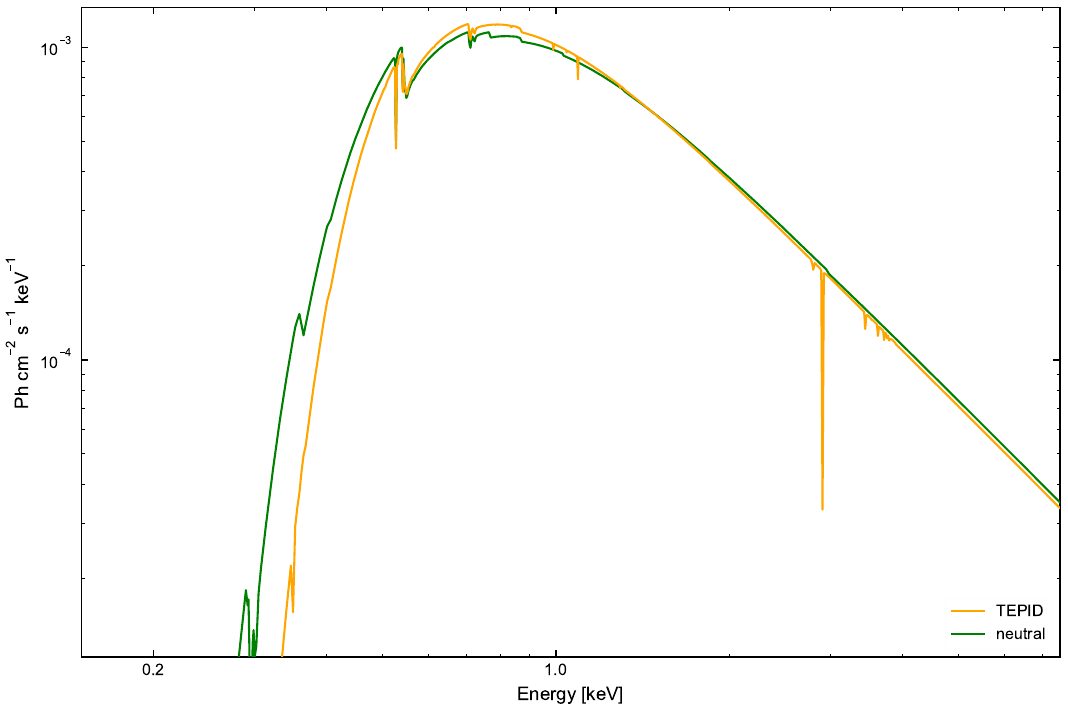}
    \includegraphics[width=0.7\textwidth]{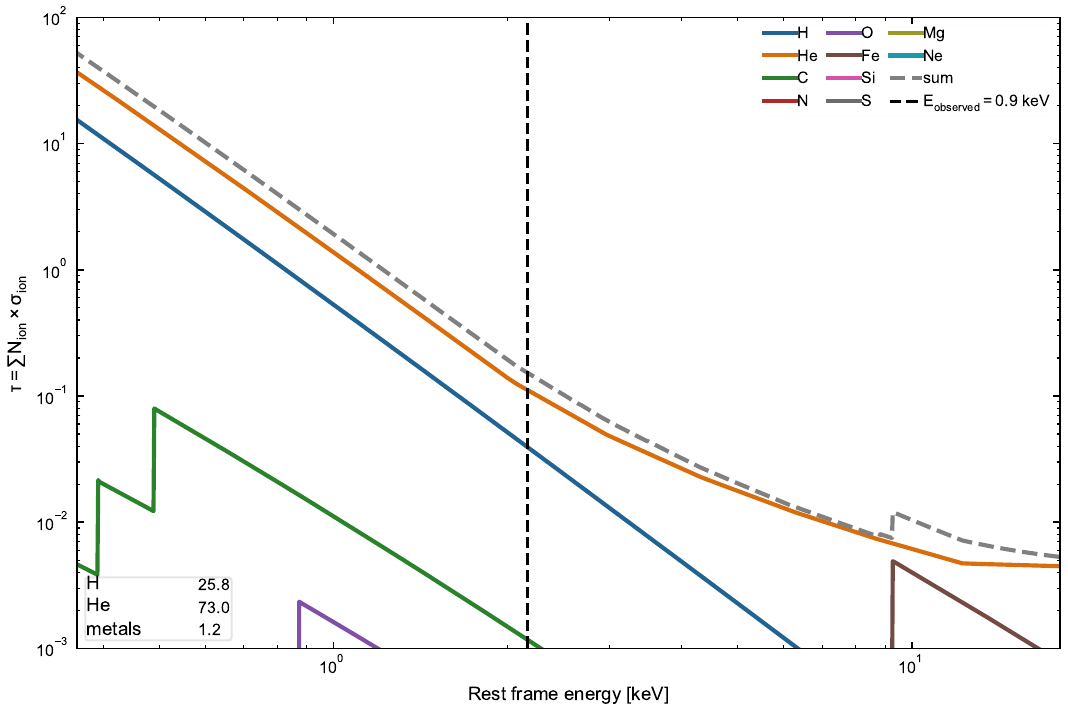}
    \caption{\textbf{Photon spectrum and element-wise opacity breakdown for the GRB 120711A TEPID-only best-fit} \textit{Top panel}: The TEPID best-fit model photon spectrum is plotted in orange as a function of observed energy. For comparison, the best-fit neutral model is plotted in green. \textit{Bottom panel}: Opacity contributions per element (summed over all ions) as a function of rest-frame energy. At $E_{rest}< 1$ keV, H and He are contributing to the absorption, instead at $E_{rest} >$ 1 keV metal contributions start to become significant. The vertical dashed line marks the E$_{obs} = 0.9$ keV, with percent contributions of H, He and metals reported in the bottom left.}
    \label{fig:phot_spec6}
\end{figure}
\clearpage

\begin{figure}
    \centering
    \includegraphics[width=0.7\textwidth]{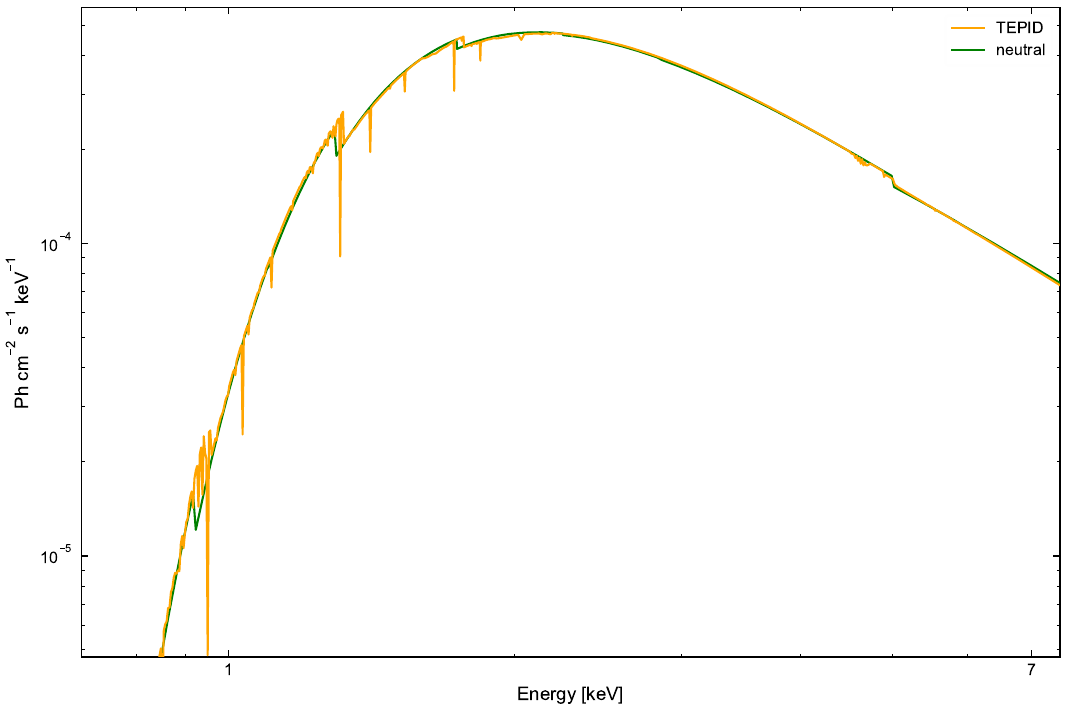}
    \includegraphics[width=0.7\textwidth]{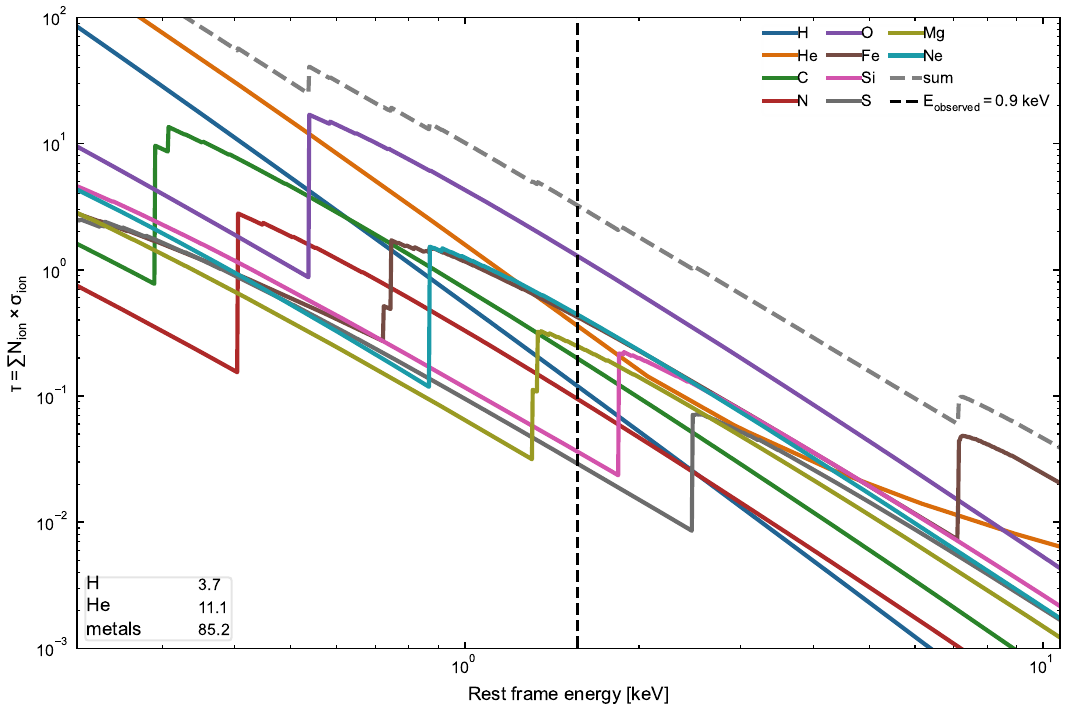}
    \caption{\textbf{Photon spectrum and element-wise opacity breakdown for the GRB 190114C TEPID-only best-fit} \textit{Top panel}: The TEPID best-fit model photon spectrum is plotted in orange as a function of observed energy. For comparison, the best-fit neutral model is plotted in green. \textit{Bottom panel}: Opacity contributions per element (summed over all ions) as a function of rest-frame energy. Given that we fit the spectrum above 0.9 keV due to strong host absorption, the vertical dashed line marks $E_{obs} = 1.1$ keV for this GRB with percent contributions of H, He and metals reported in the bottom left.}
    \label{fig:phot_spec7}
\end{figure}
\clearpage

\begin{figure}
    \centering
    \includegraphics[width=0.7\textwidth]{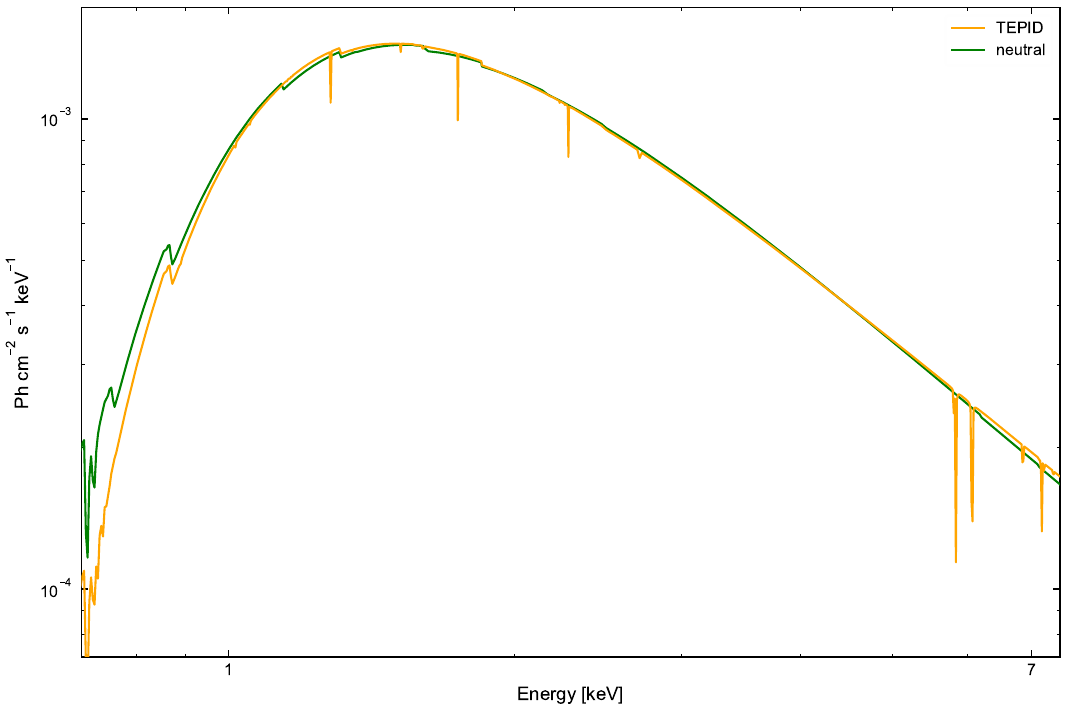}
    \includegraphics[width=0.7\textwidth]{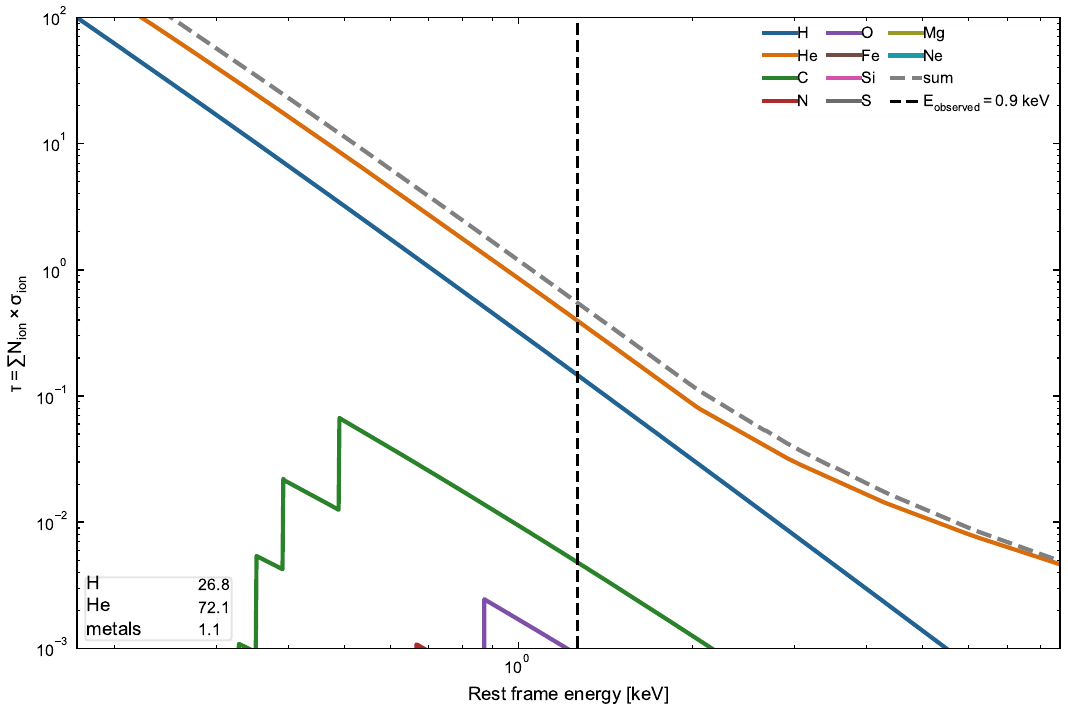}
    \caption{\textbf{Photon spectrum and element-wise opacity breakdown for the GRB 221009A TEPID-only best-fit} \textit{Top panel}: The TEPID best-fit model photon spectrum is plotted in orange as a function of observed energy. For comparison, the best-fit neutral model is plotted in green. \textit{Bottom panel}: Opacity contributions per element (summed over all ions) as a function of rest-frame energy. Given that we fit the spectrum above 0.9 keV due to strong Galactic absorption, the vertical dashed line marks $E_{obs} = 1.1$ keV for this GRB with percent contributions of H, He and metals reported in the bottom left.}
    \label{fig:phot_spec8}
\end{figure}
\clearpage

\begin{figure}
    \centering
    \includegraphics[width=0.9\textwidth]{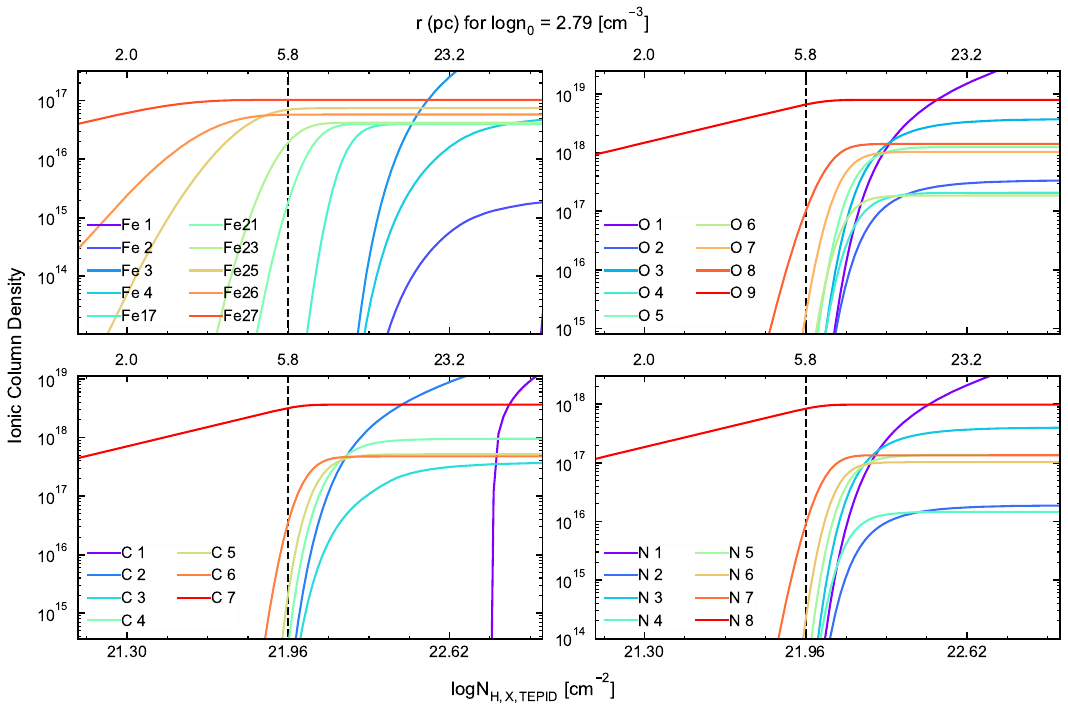}
    \caption{\textbf{Ionic column densities as a function of the hydrogen equivalent
column density for the best-fitting TEPID model of GRB 060729}. The top axis plots the distance converted assuming the best-fit number density. The elements in clockwise order from the top-left panel are iron, oxygen (top-right), nitrogen (bottom-right) and carbon (bottom-left). The black dashed line marks the TEPID best-fit column density.}
    \label{fig:cols}
\end{figure}
\clearpage

\begin{figure}
    \centering
    \includegraphics[width=0.9\textwidth]{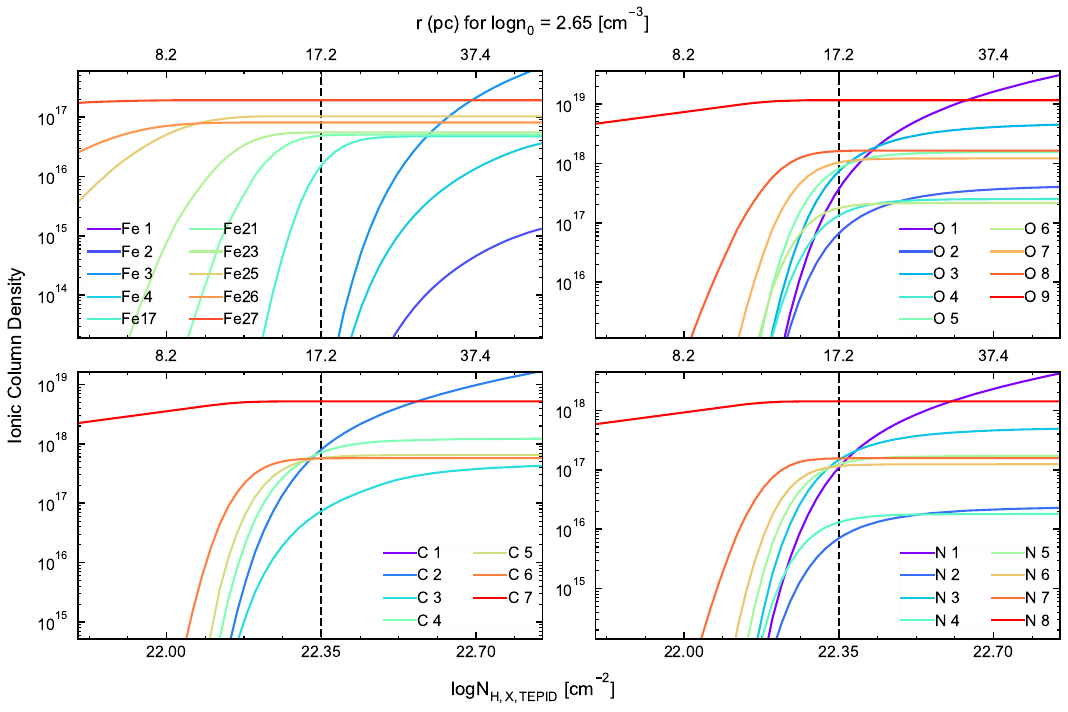}
    \caption{\textbf{Ionic column densities as a function of the hydrogen equivalent
column density for the best-fitting TEPID model of GRB 061121}. The top axis plots the distance converted assuming the best-fit number density. The elements in clockwise order from the top-left panel are iron, oxygen (top-right), nitrogen (bottom-right) and carbon (bottom-left). The black dashed line marks the TEPID best-fit column density.}
    \label{fig:cols_1}
\end{figure}
\clearpage

\begin{figure}
    \centering
    \includegraphics[width=0.9\textwidth]{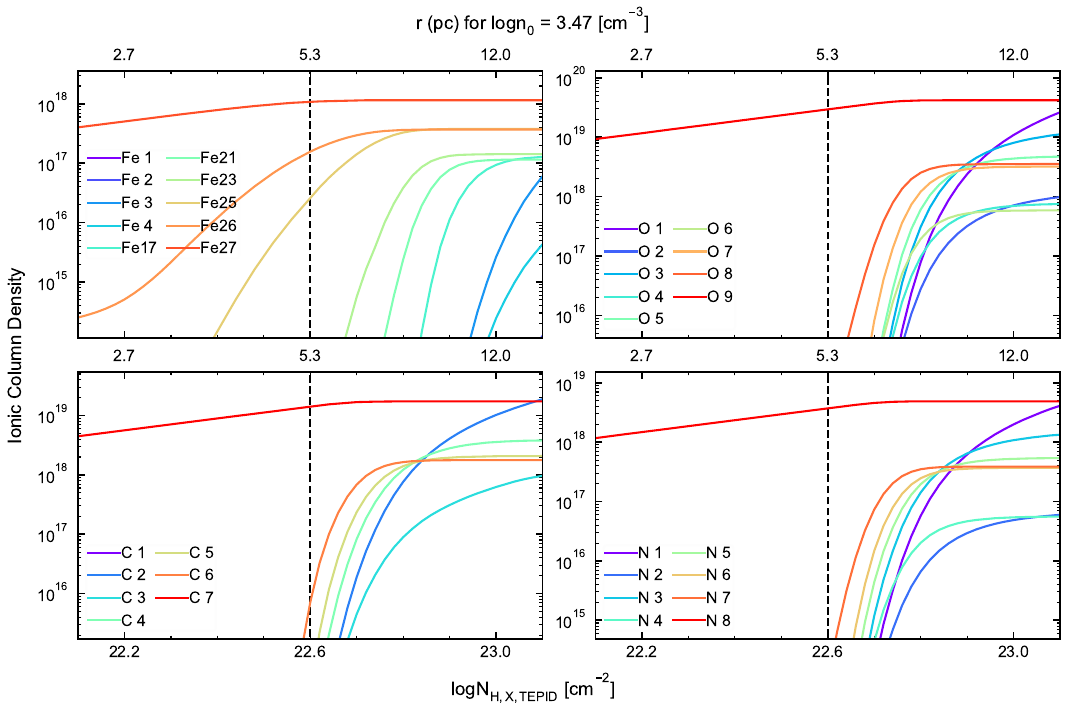}
    \caption{\textbf{Ionic column densities as a function of the hydrogen equivalent column density for the best-fitting TEPID+optical model of GRB 061121}. The top axis plots the distance converted assuming the best-fit number density. The elements in clockwise order from the top-left panel are iron, oxygen (top-right), nitrogen (bottom-right) and carbon (bottom-left). The black dashed line marks the TEPID best-fit column density.}
    \label{fig:cols_2}
\end{figure}
\clearpage

\begin{figure}
    \centering
    \includegraphics[width=0.9\textwidth]{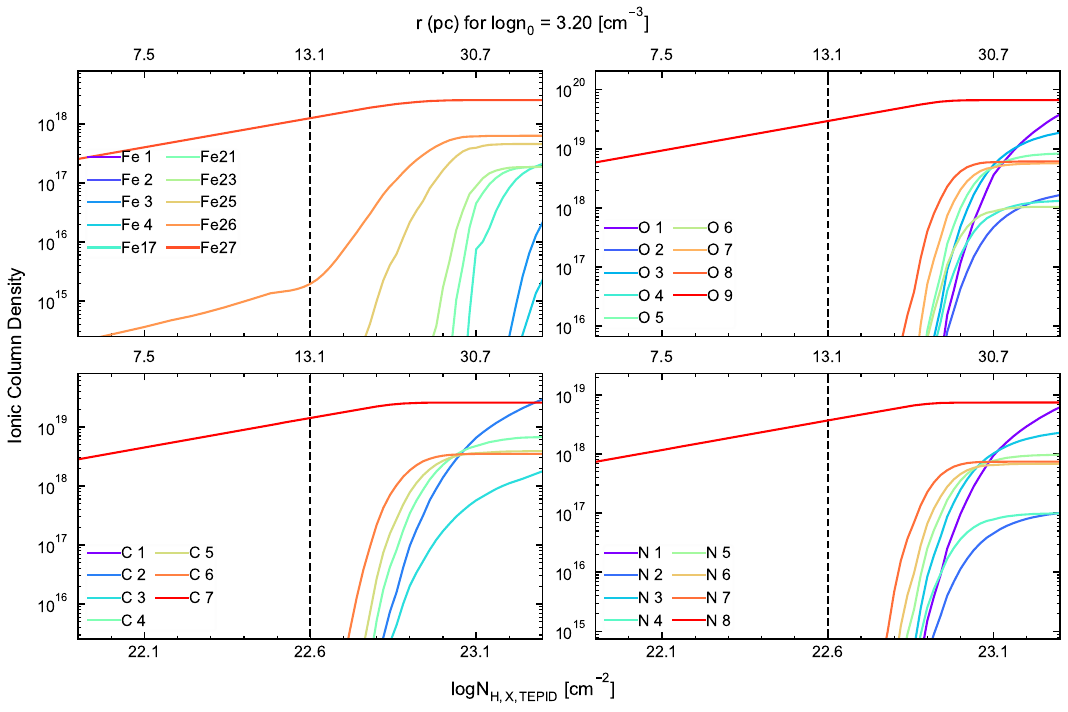}
    \caption{\textbf{Ionic column densities as a function of the hydrogen equivalent column density for the best-fitting TEPID model of GRB 080411}. The top axis plots the distance converted assuming the best-fit number density. The elements in clockwise order from the top-left panel are iron, oxygen (top-right), nitrogen (bottom-right) and carbon (bottom-left). The black dashed line marks the TEPID best-fit column density.}
    \label{fig:cols_3}
\end{figure}
\clearpage

\begin{figure}
    \centering
    \includegraphics[width=0.9\textwidth]{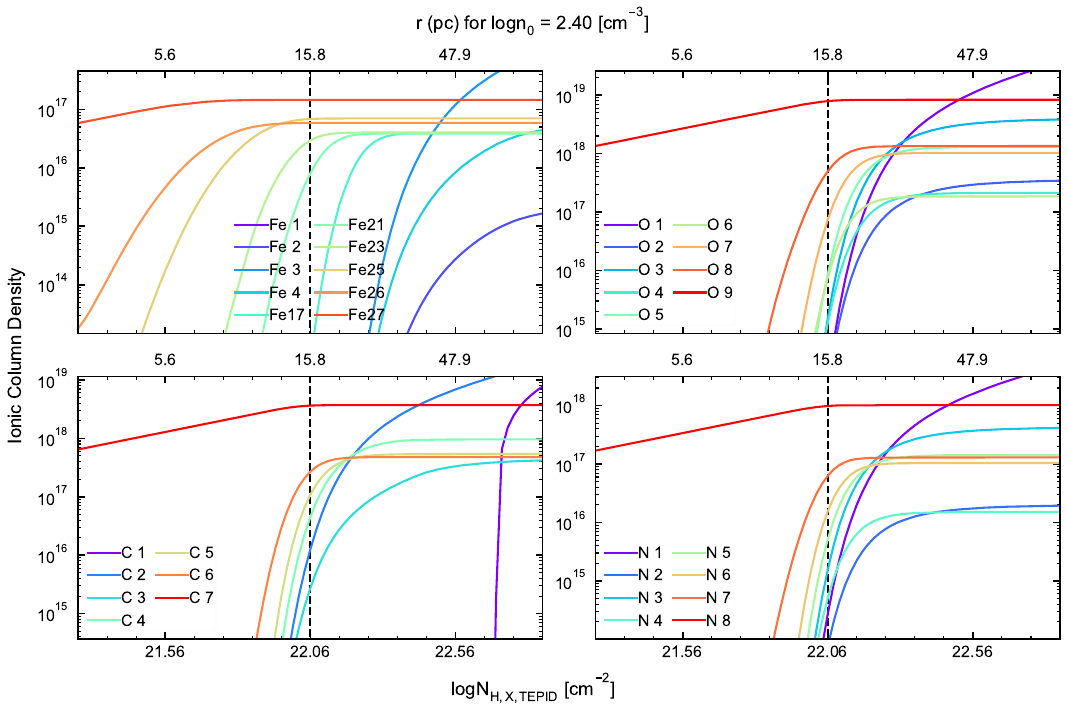}
    \caption{\textbf{Ionic column densities as a function of the hydrogen equivalent column density for the best-fitting TEPID model of GRB 090618}. The top axis plots the distance converted assuming the best-fit number density. The elements in clockwise order from the top-left panel are iron, oxygen (top-right), nitrogen (bottom-right) and carbon (bottom-left). The black dashed line marks the TEPID best-fit column density.}
    \label{fig:cols_4}
\end{figure}
\clearpage

\begin{figure}
    \centering
    \includegraphics[width=0.9\textwidth]{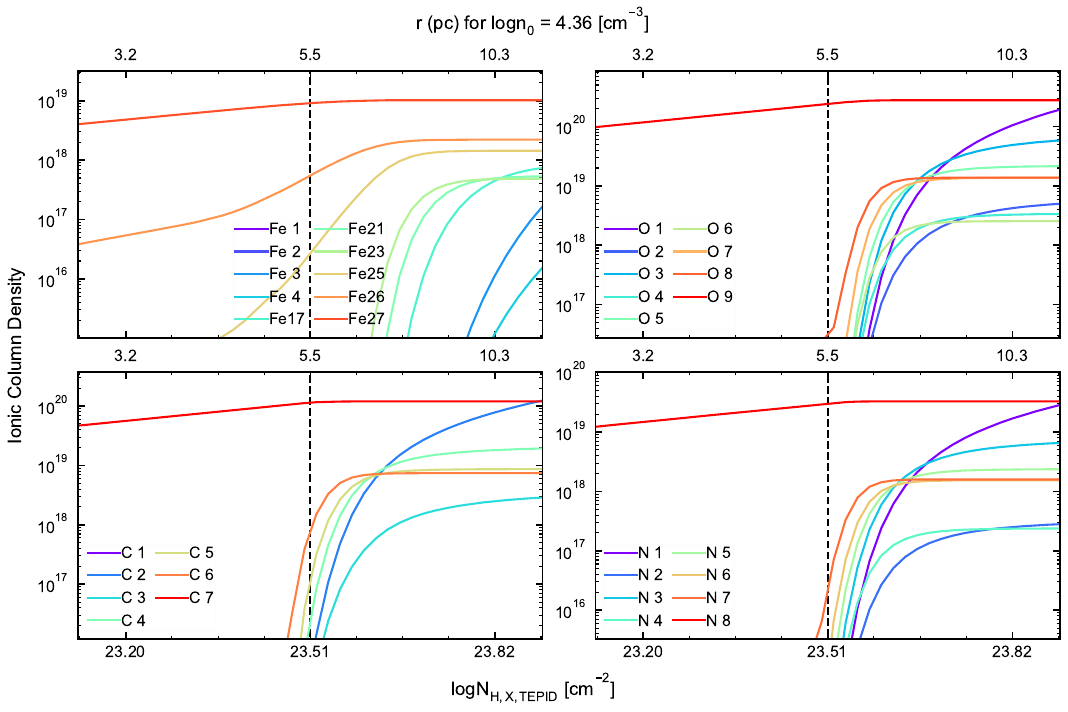}
    \caption{\textbf{Ionic column densities as a function of the hydrogen equivalent column density for the best-fitting TEPID model of GRB 120711A}. The top axis plots the distance converted assuming the best-fit number density. The elements in clockwise order from the top-left panel are iron, oxygen (top-right), nitrogen (bottom-right) and carbon (bottom-left). The black dashed line marks the TEPID best-fit column density.}
    \label{fig:cols_5}
\end{figure}
\clearpage

\begin{figure}
    \centering
    \includegraphics[width=0.9\textwidth]{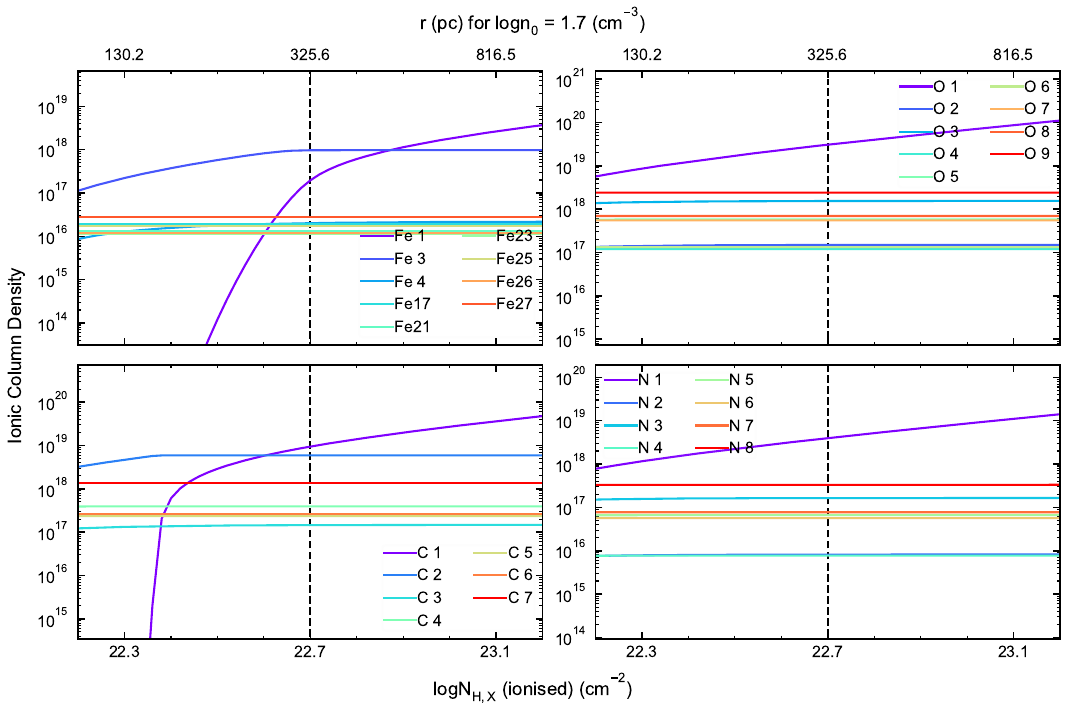}
    \caption{\textbf{Ionic column densities as a function of the hydrogen equivalent column density for the best-fitting TEPID model of GRB 190114C.} The top axis plots the distance converted assuming the best-fit number density. The elements in clockwise order from the top-left panel are iron, oxygen (top-right), nitrogen (bottom-right) and carbon (bottom-left). The black dashed line marks the TEPID best-fit column density.}
    \label{fig:cols_6}
\end{figure}
\clearpage

\begin{figure}
    \centering
    \includegraphics[width=0.9\textwidth]{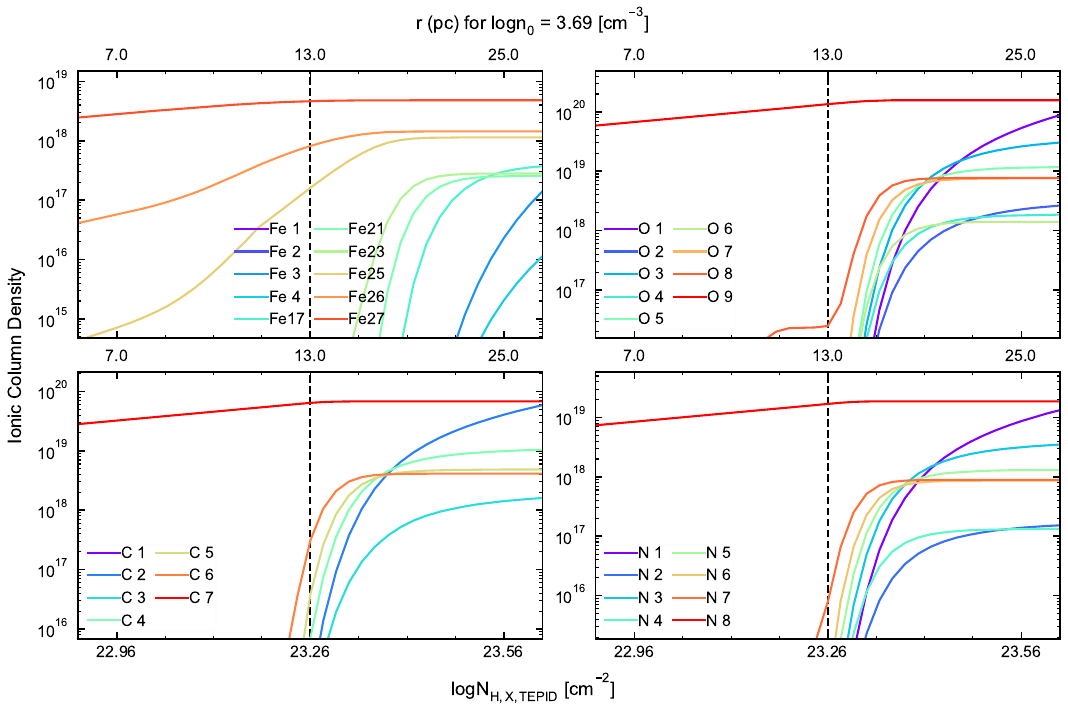}
    \caption{\textbf{Ionic column densities as a function of the hydrogen equivalent
column density for the best-fitting TEPID model of GRB 221009A.} The top axis plots the distance converted assuming the best-fit number density. The elements in clockwise order from the top-left panel are iron, oxygen (top-right), nitrogen (bottom-right) and carbon (bottom-left). The black dashed line marks the TEPID best-fit column density.}
    \label{fig:cols_7}
\end{figure}
\clearpage

\begin{figure}
    \centering
    \includegraphics[width=0.7\textwidth]{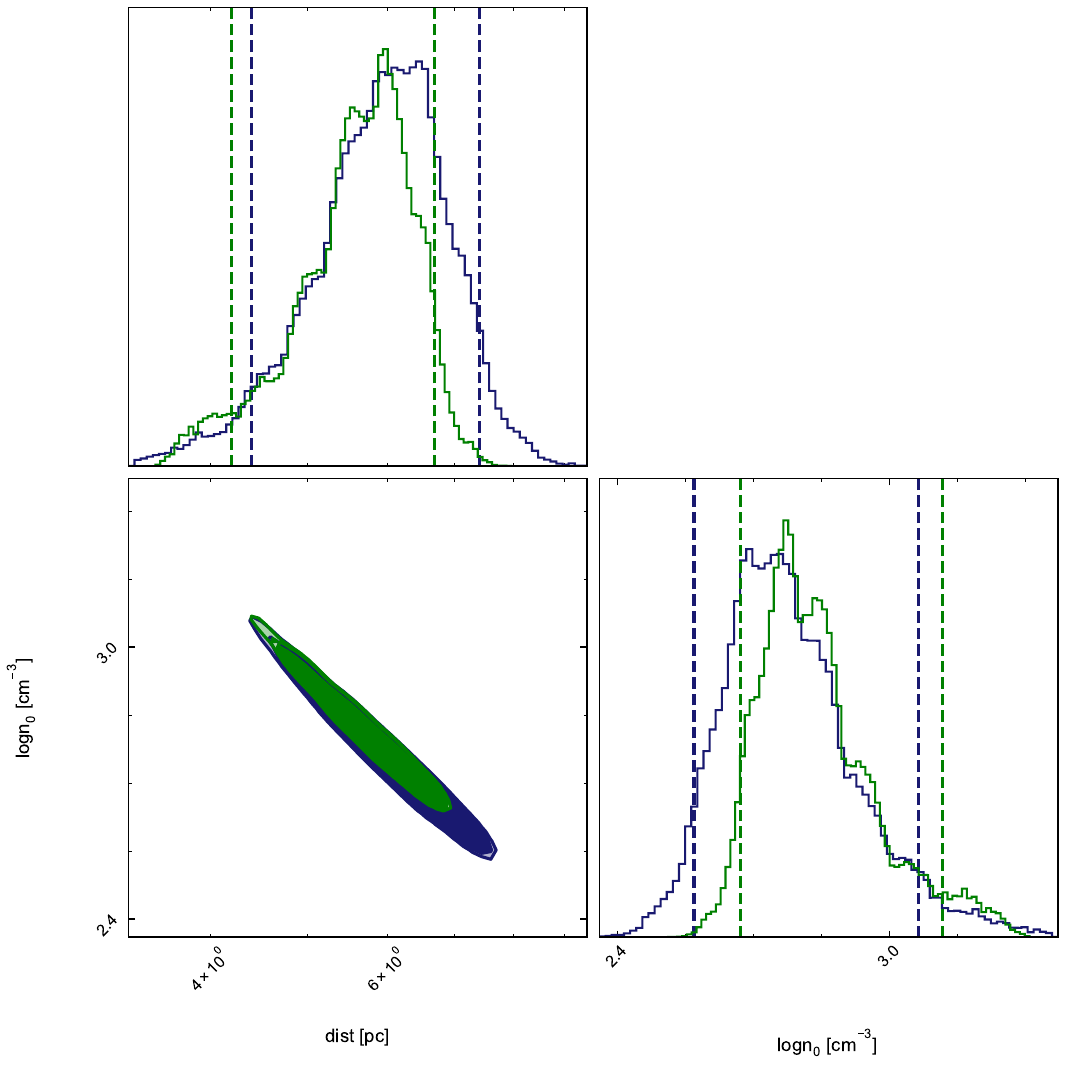}
    \includegraphics[width=0.7\textwidth]{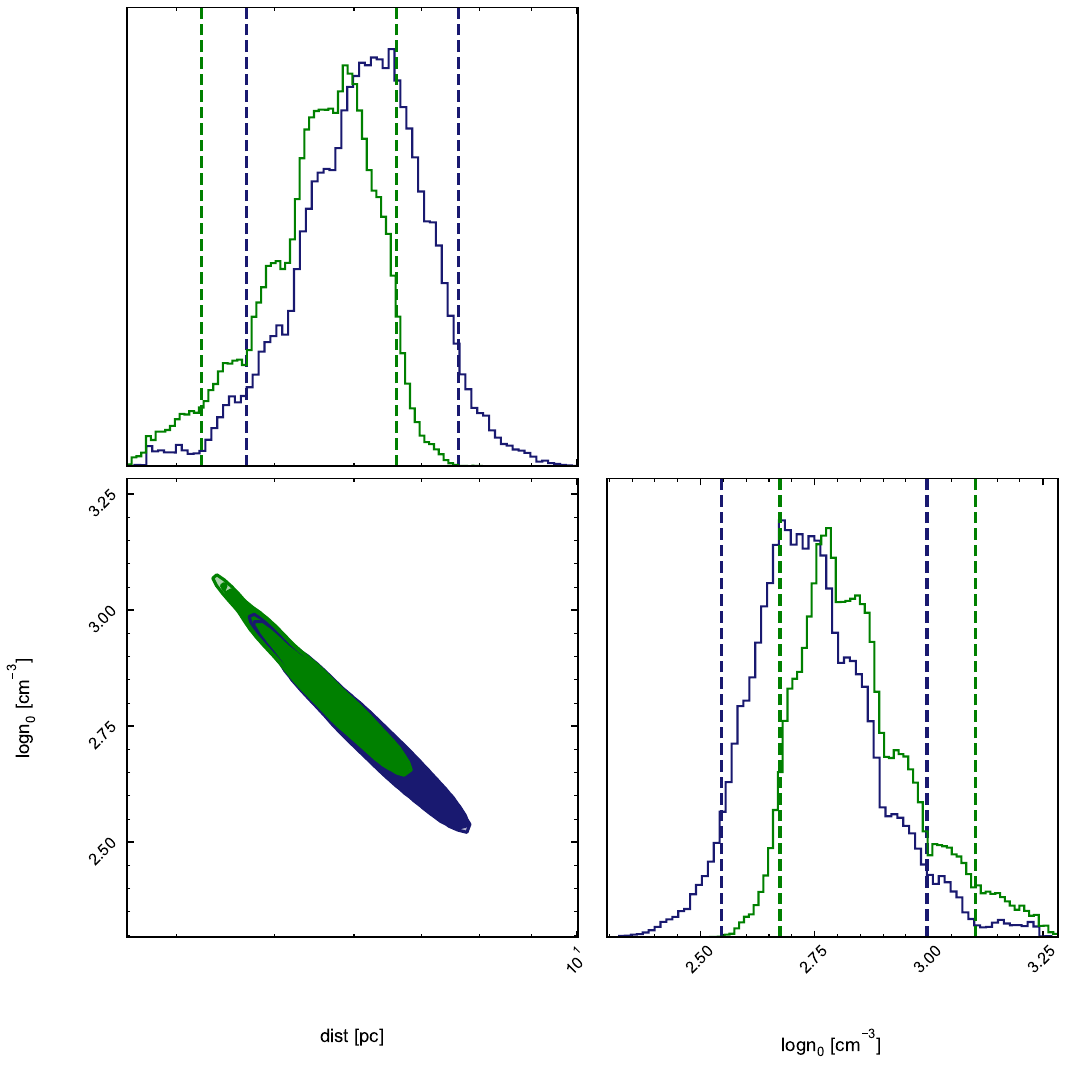}
    \caption{\textbf{MCMC contour plots for the TEPID best-fits for GRB 060729.} \textit{Top panel}: TEPID-only and \textit{Bottom panel}: TEPID+optical model fits, with both plotted at the 90\% confidence level. For both panels, the blue contours are for the fit with free metallicity, whereas the green contours are for the fit with fixed solar metallicity.}
    \label{fig:contour1}
\end{figure}

\begin{figure}
    \centering
    \includegraphics[width=0.7\textwidth]{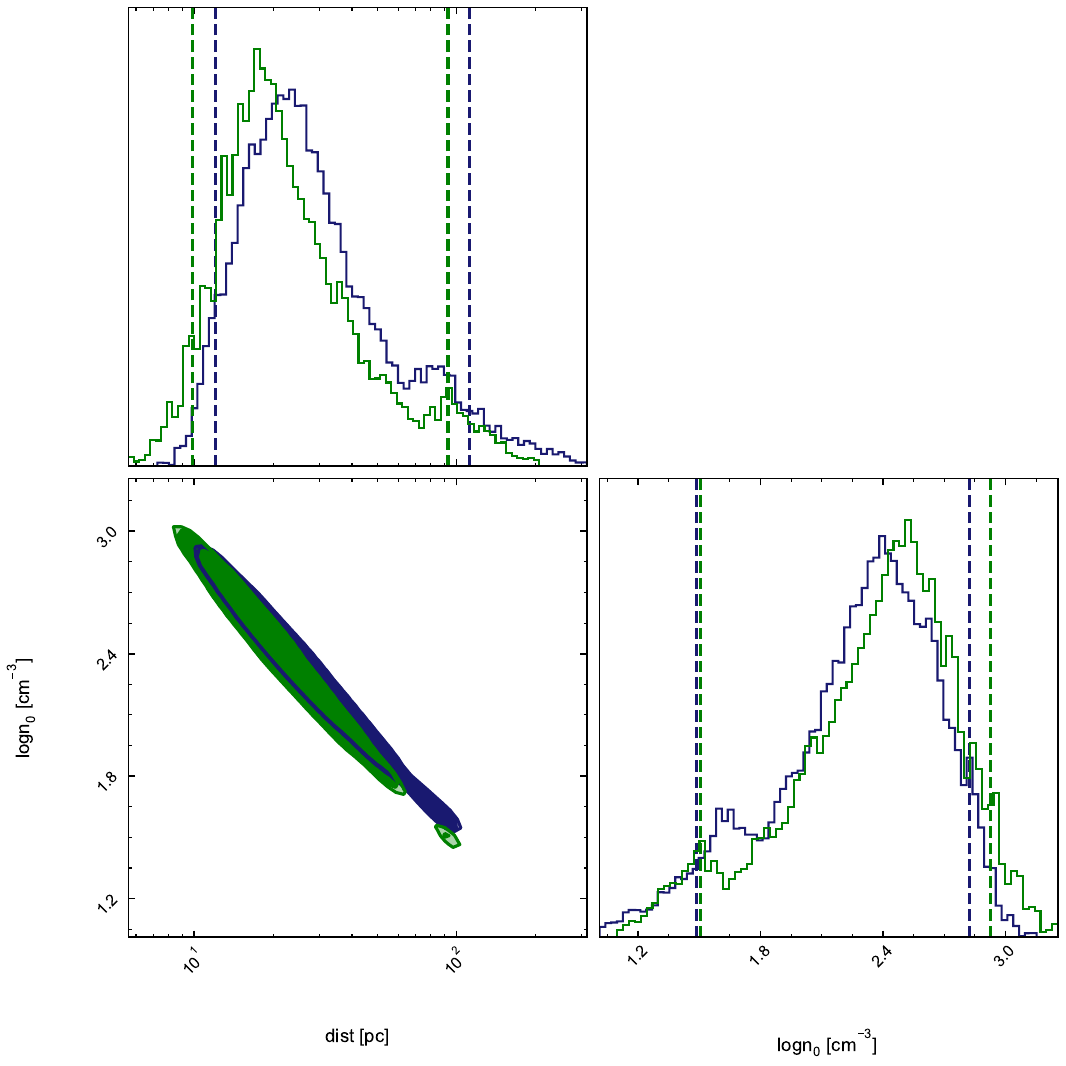}
    \includegraphics[width=0.7\textwidth]{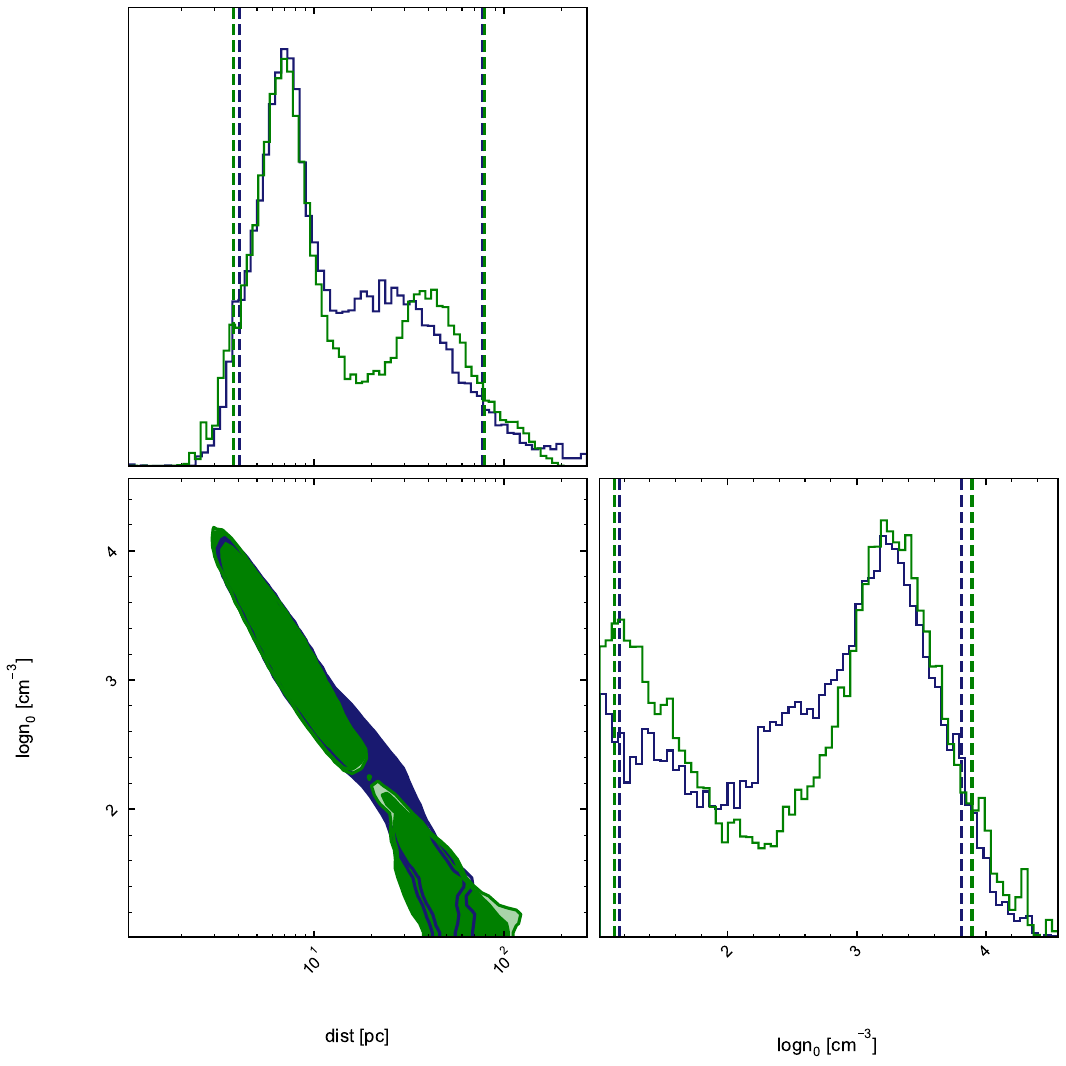}
    \caption{\textbf{MCMC contour plots for the TEPID best-fits for GRB 061121.} \textit{Top panel}: TEPID-only and \textit{Bottom panel}: TEPID+optical model fits, with both plotted at the 90\% confidence level. For both panels, the blue contours are for the fit with free metallicity, whereas the green contours are for the fit with fixed solar metallicity.}
    \label{fig:contour2}
\end{figure}

\begin{figure}
    \centering
    \includegraphics[width=0.7\textwidth]{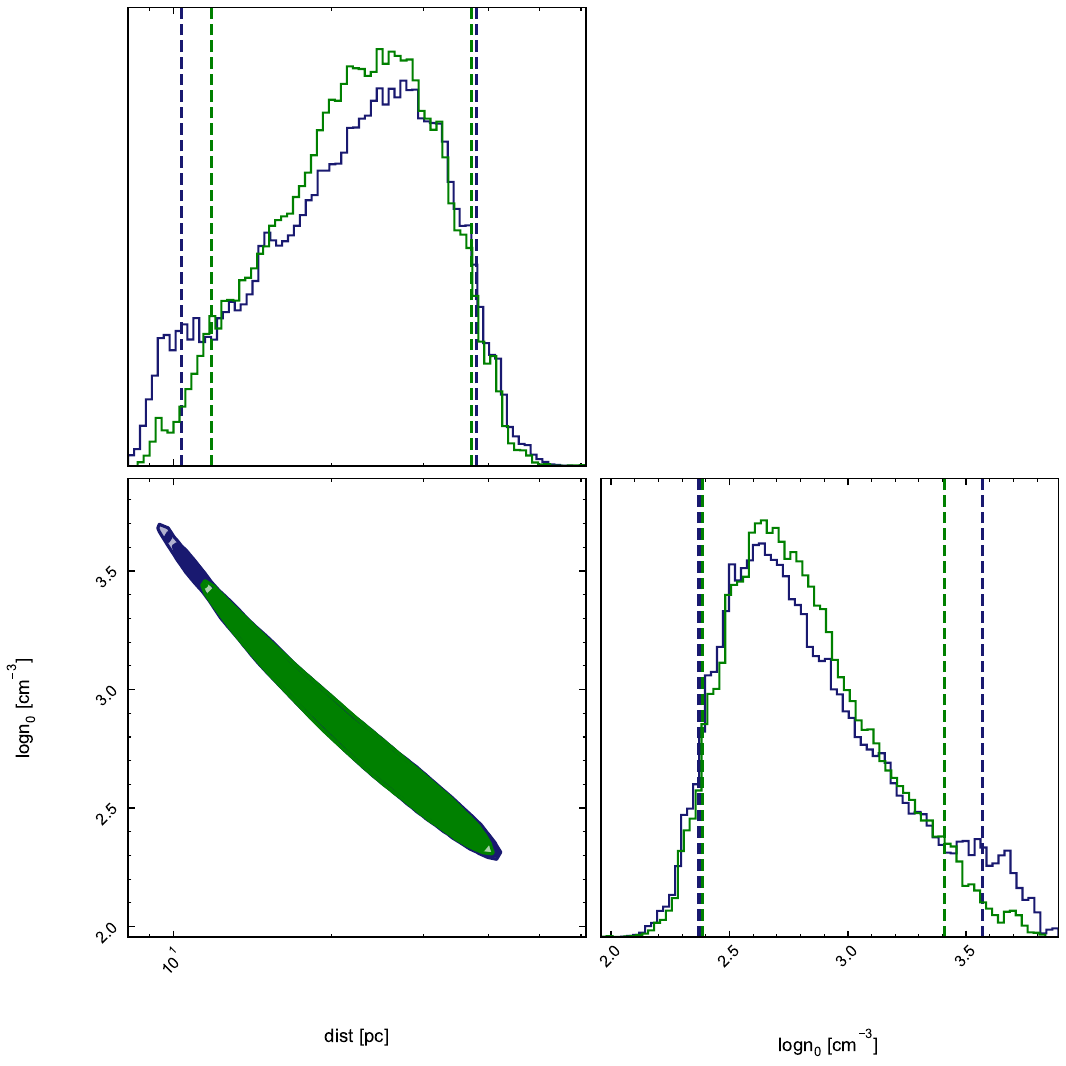}
    \includegraphics[width=0.7\textwidth]{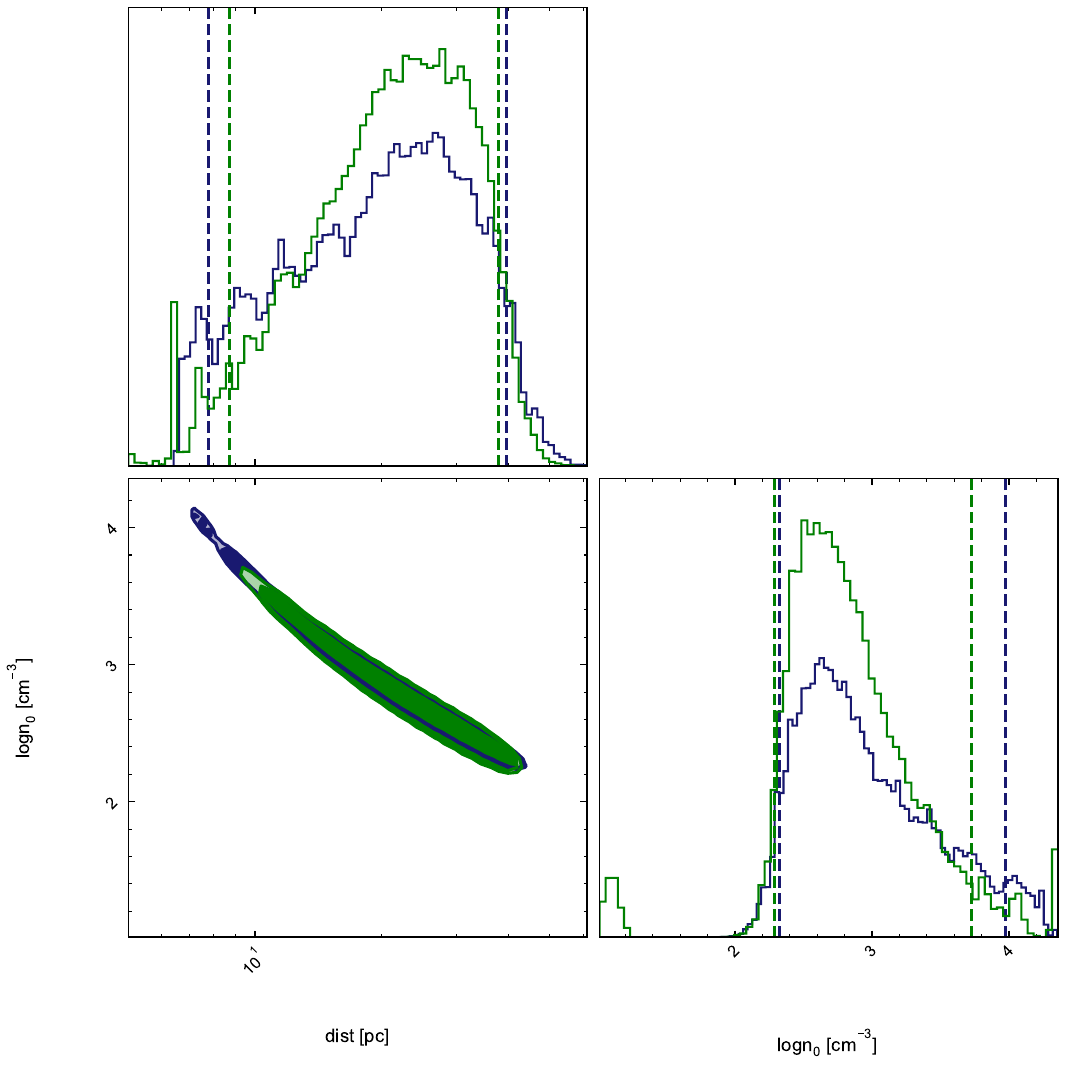}
    \caption{\textbf{MCMC contour plots for the TEPID best-fits for GRB 080411.} \textit{Top panel}: TEPID-only and \textit{Bottom panel}: TEPID+optical model fits, with both plotted at the 90\% confidence level. For both panels, the blue contours are for the fit with free metallicity, whereas the green contours are for the fit with fixed solar metallicity.}
    \label{fig:contour3}
\end{figure}

\begin{figure}
    \centering
    \includegraphics[width=0.7\textwidth]{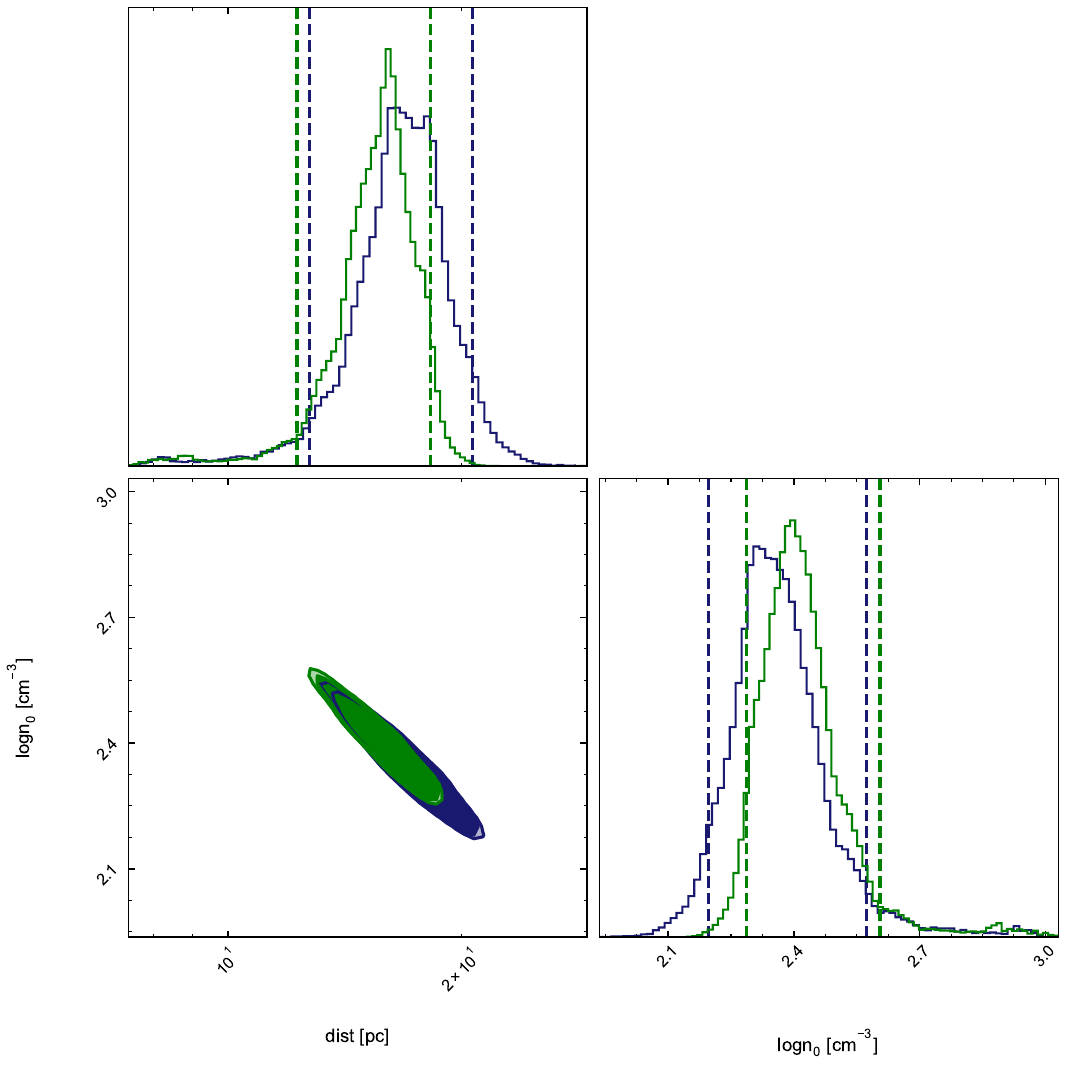}
    \includegraphics[width=0.7\textwidth]{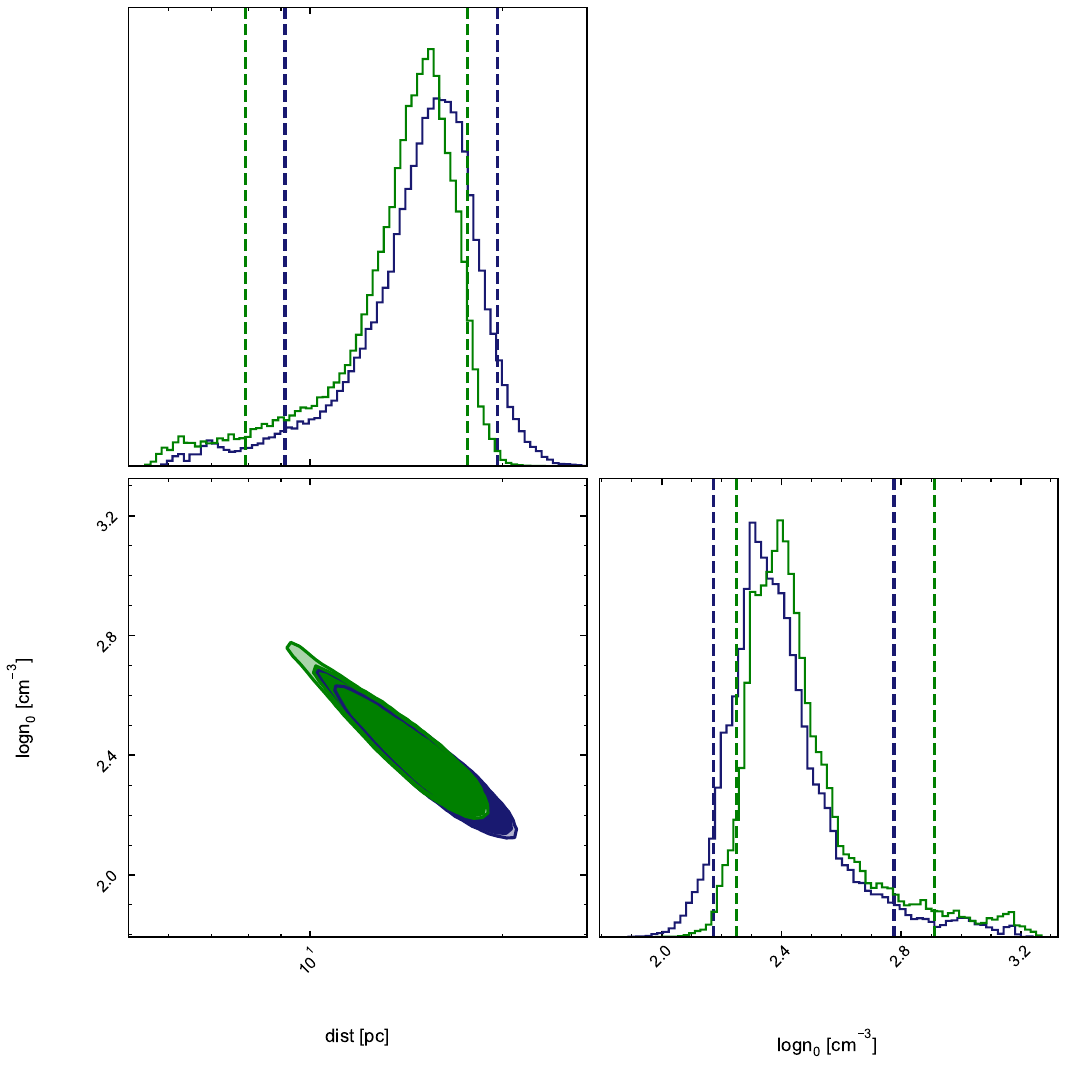}
    \caption{\textbf{MCMC contour plots for the TEPID best-fits for GRB 090618.} \textit{Top panel}: TEPID-only and \textit{Bottom panel}: TEPID+optical model fits, with both plotted at the 90\% confidence level. For both panels, the blue contours are for the fit with free metallicity, whereas the green contours are for the fit with fixed solar metallicity.}
    \label{fig:contour4}
\end{figure}

\begin{figure}
    \centering
    \includegraphics[width=0.7\textwidth]{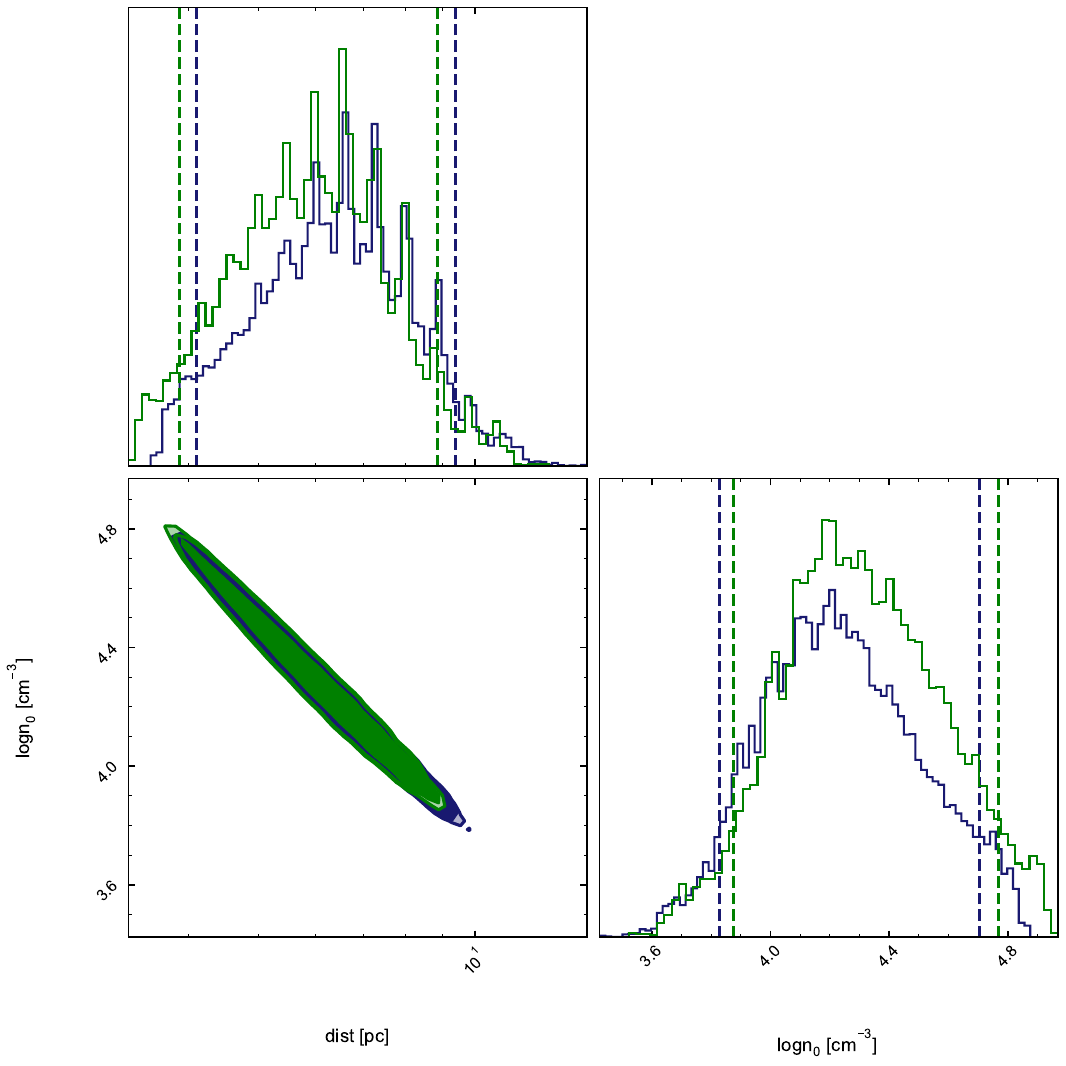}
    \includegraphics[width=0.7\textwidth]{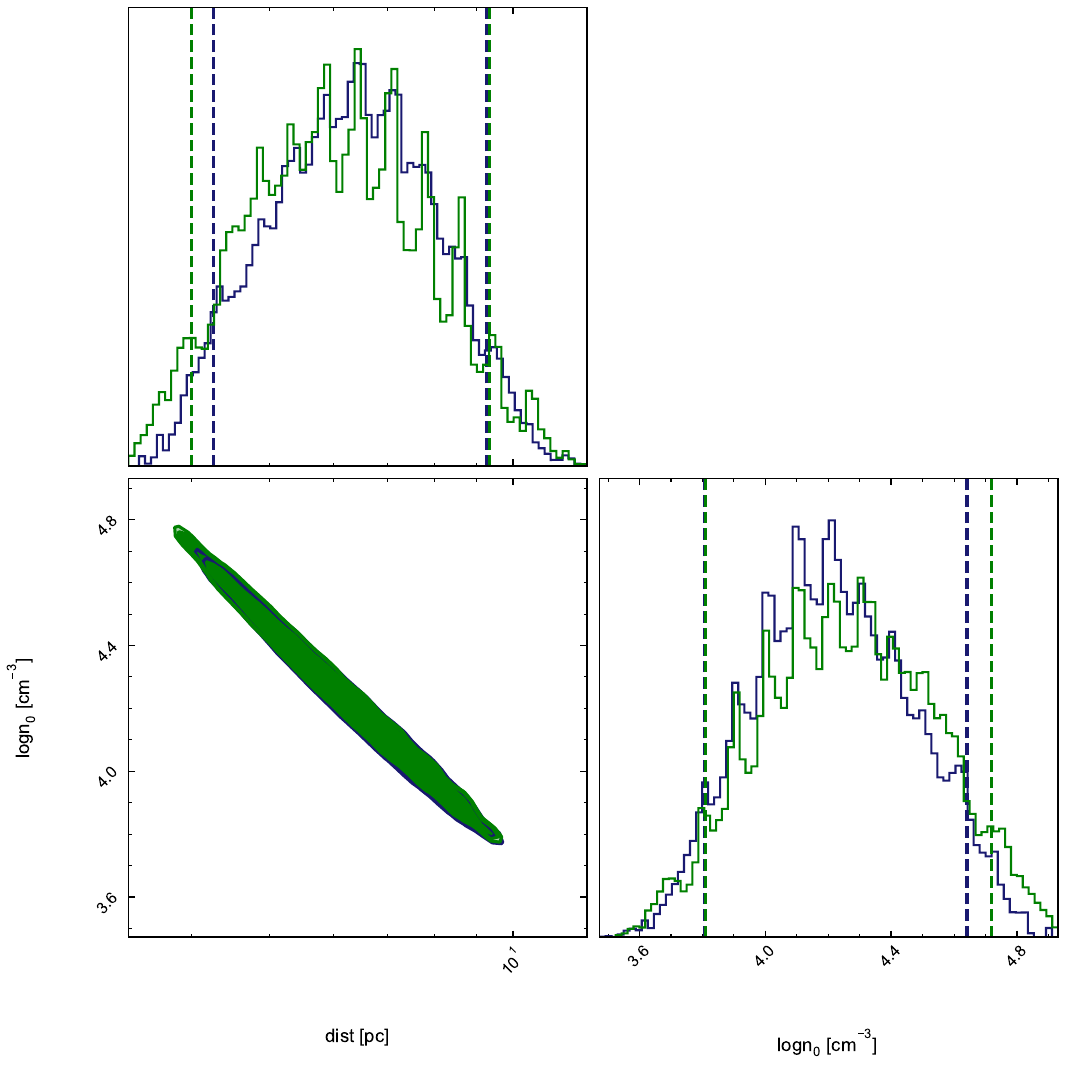}
    \caption{\textbf{MCMC contour plots for the TEPID best-fits for GRB 120711A.} \textit{Top panel}: TEPID-only and \textit{Bottom panel}: TEPID+optical model fits, with both plotted at the 90\% confidence level. For both panels, the blue contours are for the fit with free metallicity, whereas the green contours are for the fit with fixed solar metallicity.}
    \label{fig:contour5}
\end{figure}

\begin{figure}
    \centering
    \includegraphics[width=0.7\textwidth]{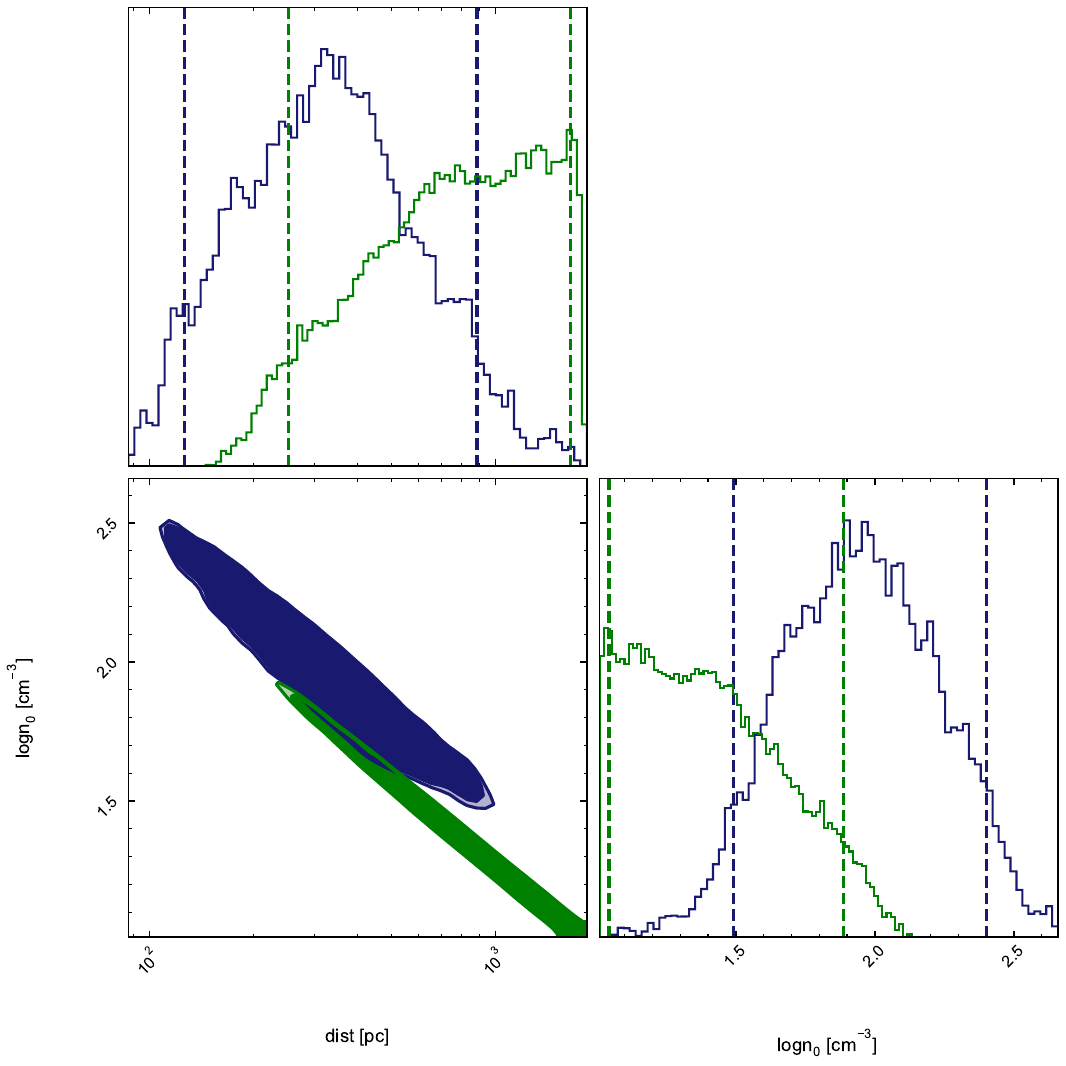}
    \includegraphics[width=0.7\textwidth]{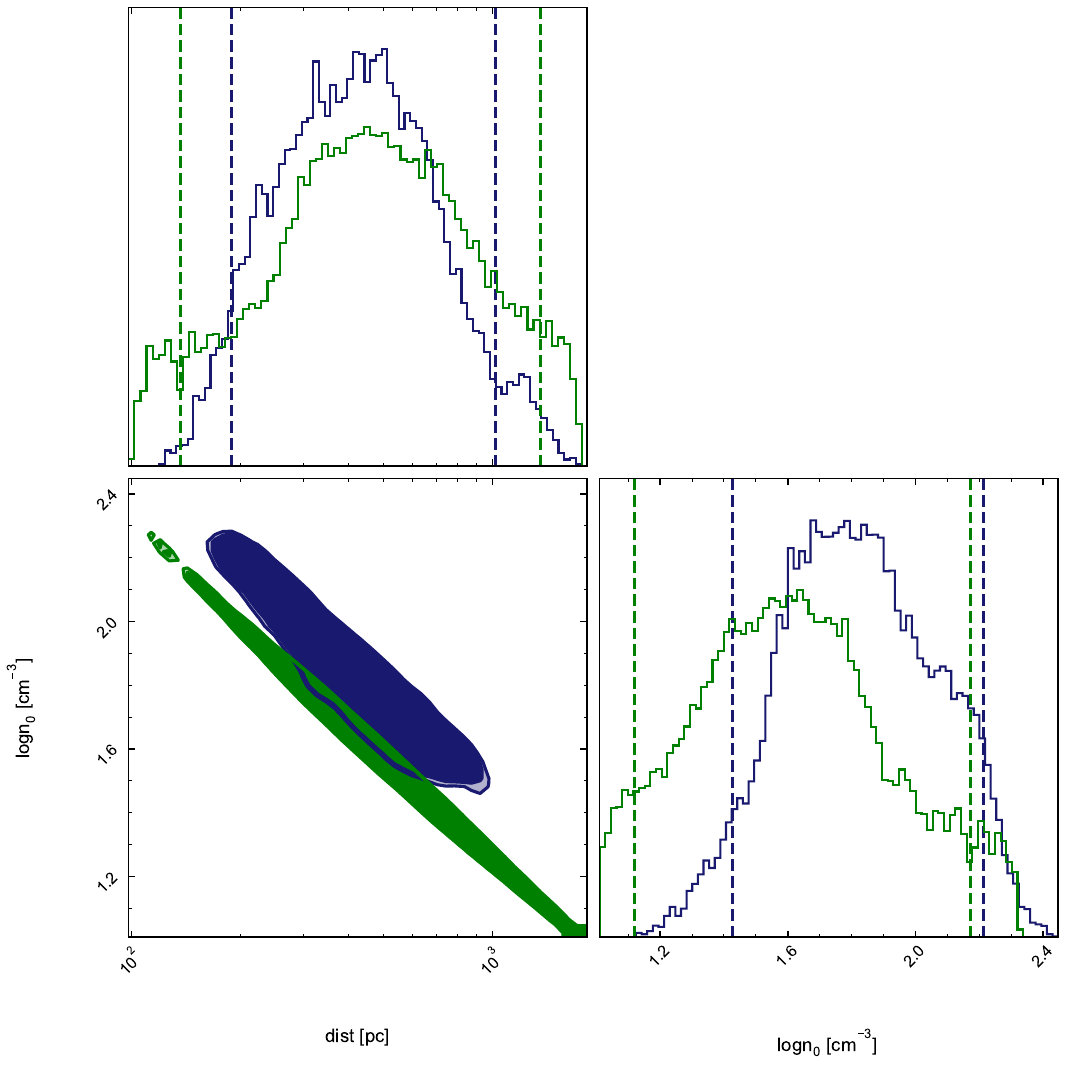}
    \caption{\textbf{MCMC contour plots for the TEPID best-fits for GRB 190114C.} \textit{Top panel}: TEPID-only and \textit{Bottom panel}: TEPID+optical model fits, with both plotted at the 90\% confidence level. For both panels, the blue contours are for the fit with free metallicity, whereas the green contours are for the fit with fixed solar metallicity.}
    \label{fig:contour6}
\end{figure}

\clearpage

\begin{figure}
    \centering
    \includegraphics[width=0.7\textwidth]{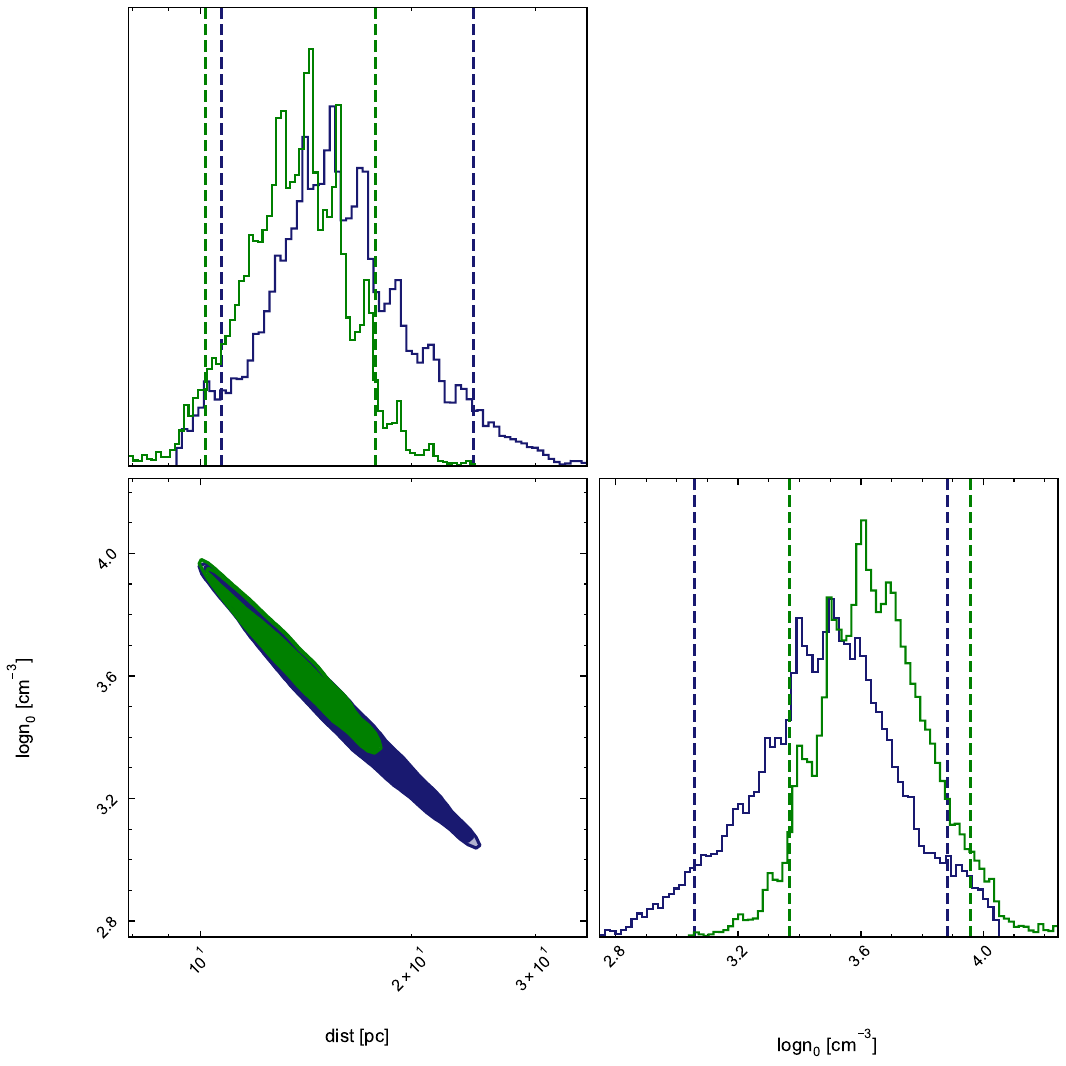}
    \caption{\textbf{MCMC contour plots for the TEPID-only best-fits for GRB 221009A.} The blue contours are for the fit with free metallicity, whereas the green contours are for the fit with fixed solar metallicity.}
    \label{fig:contour7}
\end{figure}

\subsection{Compilation of properties supporting a collapsar origin}

The prompt phase information for GRB 060729 have been compiled from \cite{2007ApJ...662..443G, Tsvetkova_2021} \& \citep{Gehrels2006}; for GRB 061121 from \cite{2007ApJ...663.1125P}, for GRB 080411 from \cite{2008GCN..7591....1S, 2017ApJ...837..119A} \& \cite{2012MNRAS.419..614U}, for GRB 090618 from \cite{09T90} \& \cite{2012MNRAS.419..614U}; for GRB 120711A from \cite{1207t90} \& \cite{Tsvetkova_2017}; for GRB 190114C from \cite{2019GCN.23724....1K, Ursi_2020} \& \cite{Du_2021} and for GRB 221009A from \cite{Lesage_2023} \& \cite{Lan_2023}. 

The supernova information for GRB 060729 \& GRB 090618 are taken from \cite{2011MNRAS.413..669C}, for GRB 190114C from \cite{2022A&A...659A..39M} and for GRB 221009A from \cite{2024NatAs...8..774B}. Information on star-formation in the host galaxy for GRB 060729 is taken from \cite{2011MNRAS.413..669C}, for GRB 061121 from \citep{2019A&A...623A..26P}, for GRB 190114C from \cite{2020A&A...633A..68D} and for GRB 221009A from \cite{2024NatAs...8..774B}. Offsets have been taken for GRB 060729, GRB 090618 and GRB 120711A from \cite{2016ApJ...817..144B}, for GRB 190114C from \cite{2020A&A...633A..68D} and for GRB 221009A from \cite{Levan_2023}. For GRB 061121, we have made an estimate for the offset based on the \textit{Keck} image of the host available online from the catalogue of \cite{2013EAS....61..391P}. 



\end{document}